\documentclass[11pt,a4paper]{article}

\usepackage{amsfonts}
\usepackage{multirow}
\usepackage{color}
\usepackage{cite}
\usepackage{graphicx}
\usepackage{float}
\usepackage{amsmath}
\usepackage[colorlinks=true,linkcolor=blue,citecolor=blue]{hyperref}%
\usepackage{subcaption}
\usepackage{comment}
\usepackage{adjustbox} 
\usepackage{tikz}
\usetikzlibrary{shapes.geometric, arrows}
\usepackage[absolute,overlay]{textpos}
\tikzstyle{block1} = [rectangle, minimum width=3cm, minimum height=1cm, text centered, draw=black, fill=red!30]
\tikzstyle{block2} = [rectangle, minimum width=3cm, minimum height=1cm, text centered, draw=black, fill=green!30]
\tikzstyle{block3} = [rectangle, minimum width=3cm, minimum height=1cm, text centered, draw=black, fill=yellow!30]
\tikzstyle{block4} = [rectangle, minimum width=2.5cm, minimum height=1cm, text centered, draw=black, fill=blue!30]
\tikzstyle{block5} = [rectangle, minimum width=2.5cm, minimum height=1cm, text centered, draw=black, fill=orange!30]
\tikzstyle{arrow} = [thick,->,>=stealth]
\tikzstyle{arrow} = [thick,->,>=stealth]

\title{\textbf{Spectral structure of fluctuations around $n$-vortices in the Abelian-Higgs model}}

\author{
A. Alonso Izquierdo$^{(a,b)}$, D. Migu\'elez-Caballero$^{(c)}$, and J. Queiruga$^{(a,b)}$
\\ {\normalsize {\it $^{(a)}$ Departamento de Matematica
Aplicada, Universidad de Salamanca,}} \\{\normalsize{\it Casas del Parque 2, 37008, Salamanca, Spain}}
\\ {\normalsize {\it $^{(b)}$ IUFFyM, Universidad de Salamanca, Plaza de la Merced 1, 37008, Salamanca, Spain}}
\\ {\normalsize {\it $^{(c)}$ Departamento de Fisica Teorica, Atomica y Optica,}} \\ {\normalsize {\it Universidad de Valladolid, 47011 Valladolid, Spain}}
}

\date{}

\begin{document}

\maketitle

\begin{abstract}
We study in detail the internal structure of rotationally invariant higher-charge Abelian-Higgs vortices. The symmetry of the normal modes close to the critical regime for type II vortices determines the possible disintegration channels. The full internal mode structure is discussed, finding modes that conserve the vortex symmetry (Derrick type modes) and modes whose symmetry differs (multipolar modes). 
\end{abstract}


\section{Introduction}

Vortices are two dimensional solitons whose charge is exponentially localized in $\mathbb{R}^2$. Their topological charge is determined by the homotopy group $\pi_1(\mathbb{S}_1)=\mathbb{Z}$, and as a consequence, they may exist in multivortex configurations with any (positive or negative) charge. In their local version, they are described by a complex field (usually called  Higgs field) and a $U(1)$ gauge field responsible for the local gauge symmetry. Usually, the vortex solutions of a theory with this physical content are called Abelian-Higgs vortices and they are fundamental in many areas of physics from condensed matter \cite{Abrikosov1957, superfluid-vortex, Nielsen1973} to cosmology \cite{Vilenkin2000}. In the usual formulation of the Abelian-Higgs model, the functional form of the self-interacting potential of the Higgs field determines the vacuum structure of the model, and therefore, the topological structure of the solitons. But more interestingly, the strength of the potential, characterized by a self-coupling constant, $\lambda$, has important consequences for higher charge solitons. There is a special value of $\lambda=\lambda_c$ (usually called critical, BPS or self-dual) at which individual charge one vortices do not interact statically \cite{Manton2002, Jaffe1980}. This means that, at this value, the static intersolitonic forces vanish. Below $\lambda_c$ individual vortices attract. As a consequence, at sufficiently large times, a configuration of $N$ vortices located at different points in space will eventually coalesce and form a charge $N$ vortex. On the other hand, for $\lambda>\lambda_c$ and at sufficiently large times, the field configuration will consist of a set of charge one vortices moving at (possibly) different velocities. 

The BPS Abelian-Higgs model deserves special attention. At $\lambda=\lambda_c$ vortices satisfy at set of first order static equations, the so-called BPS equations which saturate the energy in each topological sector \cite{Bogomolny1976, Prasad1975, Taubes1980, Taubes1980b}. In addition, the low velocity dynamics is dictated by the kinetic terms in the action. This gives rise to an effective manifold, called the moduli space, whose geodesics correspond to trajectories of the interacting vortices \cite{manton1981, Samols1992}. The vortex scattering has been thoroughly studied in the literature \cite{Ruback1988, Shellard1988, Matzner1981,Moriarty1988,Arthur199} and it has become clear recently that the moduli space picture can be improved by adding possible internal excitations \cite{Krusch2024, AlonsoIzquierdo2024d, AlonsoIzquierdo2024b}. Internal excitations are linear perturbations on top of the vortex, and, for the charge one case, they have been studied in the literature \cite{AlonsoIzquierdo2016,AlonsoIzquierdo, AlonsoIzquierdo2016b, Goodband1995, Arodz1991, Arodz1996, Kojo2007, AlonsoIzquierdo2024}. In an excited vortex scattering the internal modes may move through the spectrum changing their frequency. This generates effective forces that may lead to a relevant deviation form the moduli space trajectories. In the non-BPS case, the internal modes may play also a relevant role, in this case as competitors of the intersolitonic forces.  Recently, it has been suggested that the vibrational modes of cosmic strings—spatially extended counterparts of the vortices studied in this work—could act as sources of gravitational waves \cite{Blanco-Pillado2024}. Although such applications go beyond the immediate scope of our paper, they highlight the broader relevance of a detailed understanding of the fluctuation spectrum in the Abelian Higgs model.

This paper constitutes a natural continuation of two previously published studies \cite{AlonsoIzquierdo2016, AlonsoIzquierdo2016b}, which investigate the Abelian Higgs model at critical coupling ($\lambda = 1$). At this special point, powerful supersymmetric techniques allow for an elegant and efficient spectral analysis. In contrast, the present work addresses the general (non-BPS) case $\lambda \neq 1$, where such techniques are no longer applicable, and the spectral problem must be approached from first principles. The study of the excitation spectrum of vortex solitons in the Abelian Higgs model has a rich history, beginning with the foundational work of Goodband and Hindmarsh in 1995 \cite{Goodband1995}, who offered only a partial spectrum. Other subsequent investigations \cite{Arodz1991, Arodz1996, Kojo2007} addressed specific limits or configurations. To the best of our knowledge, our contribution is the first complete and systematic investigation of the full excitation spectrum of vortices away from the BPS limit. Our results yield a comprehensive classification of the eigenmodes and their properties for arbitrary values of the coupling constant, resolving a decades-old problem. In this sense, the present article concludes the research program initiated in the aforementioned works. 

This paper is organized as follows: in Sec. \ref{sec:intro} we present the model and compute the spectral equations. In Sec. \ref{bps} the review the main properties of the spectrum of self-dual vortices. In Sec. \ref{non-bps} we analyzed in detail the spectral structure of non-BPS vortices. In Sec. \ref{num_spec} we compute numerically the frequencies and profiles of the first internal modes. Finally, Sec. \ref{conclusions} is devoted to our conclusions and further discussion. We also add one appendix (\ref{appen}) with numerical details. 


\section{The Abelian-Higgs model}\label{sec:intro}

The Abelian-Higgs model in $\mathbb{R}^{1,2}$ describes the minimal coupling between a $U(1)$-gauge field, $A_\mu(x)=(A_0(x),A_1(x),A_2(x))$, and a complex scalar field, $\phi(x)=\phi_1(x)+i\phi_2(x)$. After a trivial rescaling of the field and coordinates, the action can be written as  
\begin{equation}
S[\phi,A]=\int d^3 x \left[ -\frac{1}{4} F_{\mu\nu}F^{\mu \nu} + \frac{1}{2} \overline{D_\mu \phi}\, D^\mu \phi -U(\phi,\overline{\phi}) \right]
\label{action1} ,
\end{equation}
where the self-interaction potential term 
\[
U(\phi,\overline{\phi}) = \frac{\lambda}{8} (\overline{\phi}\, \phi-1)^2  .
\]
 The covariant derivative has the form $D_\mu \phi(x) = (\partial_\mu -i A_\mu(x))\phi(x)$ and the electromagnetic field tensor is $F_{\mu\nu}(x)=\partial_\mu A_\nu(x) - \partial_\nu A_\mu(x)$. We choose the mostly minus convention for the metric tensor in Minkowski space,  $\eta_{\mu\nu}={\rm diag}(1,-1,-1),$ with $\mu,\nu=0,1,2$ and use the Einstein index convention. In the temporal gauge $A_0=0$, the second order differential equations for the complex field $\phi$ and the spatial components of the vector field $A_\mu$ are given by 	
\begin{eqnarray}
	&& \frac{1}{2}\partial_0^2 \phi - \frac{1}{2}D_jD_j\phi = - \frac{\partial U }{\partial \overline{\phi}} \hspace{0.0cm}, \label{genedo1} \\
	&& \partial_0^2 A_j  - \partial_k F_{kj} = - \frac{i}{2} \Big[ \overline{\phi} \, D_j \phi -  \overline{D_j \phi} \,\phi \Big] \hspace{0.0cm}. \label{genedo2}
\end{eqnarray}
The equations (\ref{genedo1}) and (\ref{genedo2}) must be complemented with the Gauss law
\begin{equation}
\partial_{0i}A_i = - \frac{i}{2} \Big[ \overline{\phi} \, \partial_0 \phi -  \overline{\partial_0 \phi} \,\phi \Big] \hspace{0.0cm}, \label{GaussLaw}
\end{equation}
which is automatically satisfied for static solutions. Static rotationally invariant $n$-vortex solutions can be found by imposing the radial gauge condition $A_r=0$ and the ansatz
\begin{equation}
\phi(r,\theta) = f_n(r) e^{in\theta} \hspace{0.0cm},\hspace{0.5cm} r A_\theta(r,\theta) = n \beta_n(r), \label{ansatz}
\end{equation}
where we use the convention $A_1=A_r \cos \theta - A_\theta \sin \theta$ and $A_2=A_r \sin \theta + A_\theta \cos \theta$. The radial profile functions $f_n(r)$ and $\beta_n(r)$ must satisfy the differential equations
\begin{eqnarray}
	&& \frac{d^2 f_n}{dr^2} + \frac{1}{r} \frac{df_n}{dr} - \frac{n^2 (1-\beta_n)^2 f_n}{r^2} + \frac{\lambda}{2} f_n (1-f_n^2) = 0, \label{edof} \\
	&& \frac{d^2 \beta_n}{dr^2} - \frac{1}{r} \frac{d\beta_n}{dr} + (1-\beta_n) f_n^2 = 0. \label{edobeta}
\end{eqnarray}
Regularity of the solutions at the origin imposes the following condition for the profile functions
\[
f_n(0) = 0 \hspace{0.0cm},\hspace{0.5cm} \beta_n(0)=0  .
\]
It can be shown by using (\ref{edof}) and (\ref{edobeta}) that the power expansion of $n$-vortices around $r=0$ has the form
\[
f_n(r) = r^n \sum_{i=0}^\infty d_i r^i \hspace{0.0cm},\hspace{0.5cm} \beta_n(r) = r^2 \sum_{i=0}^\infty c_i r^i  ,
\]
where $d_{2i+1}=c_{2i+1}=0$ with $i\in \mathbb{N}$ and $c_2=c_4=\dots = c_{2n-2}=0$, resulting into the following expressions
\begin{equation}
    f_n(r) = r^n \sum_{i=0}^\infty d_{2i} \, r^{2i} \hspace{0.0cm},\hspace{0.5cm} \beta_n(r) = r^2\Big( c_0+\sum_{i=n}^\infty c_{2i} r^{2i} \Big). \label{VortexOrigin}
\end{equation}
The rest of the coefficients $c_i$ can be obtained using recurrence equations from the values of $c_0$ and $d_0$, see \cite{Manton2002}. On the other hand, the asymptotic conditions for these solutions
\begin{equation}
\lim_{r\rightarrow \infty} f_n(r)= 1 \hspace{0.0cm},\hspace{0.5cm} \lim_{r\rightarrow \infty} \beta_n(r) = 1,  \label{asymptotic}
\end{equation} 
guarantee the usual requirement of finite energy and magnetic flux quantization $\Phi=\frac{1}{2\pi} \int_{\mathbb{R}^2} d^2 x F_{12} =n$. For intermediate values of $r$, equations (\ref{edof}) and (\ref{edobeta}) must be solved numerically. For a static rotationally invariant $n$-vortex solution we will denote the scalar and vector components as
\[
\psi(\vec{x};n)=\psi_1(\vec{x};n) + i \, \psi_2(\vec{x};n) \hspace{00cm},\hspace{0.5cm} V(\vec{x};n)=(V_1(\vec{x};n),V_2(\vec{x};n)) \hspace{0.5cm}\mbox{with}\hspace{0.5cm} \vec{x}=(x_1,x_2) ,
\]
which will be assembled into a four real component column $\Psi(r,\theta) \in {\cal C}^\infty(\mathbb{R}^2) \oplus \mathbb{R}^4$ in the form
\begin{equation}
\Psi(r,\theta)=\left(\begin{array}{c} V_1(\vec{x};n) \\ V_2(\vec{x};n) \\ \psi_1(\vec{x};n) \\ \psi_2(\vec{x};n) \end{array}  \right) =  \left(\begin{array}{c}- \frac{n\beta_n(r)}{r} \sin \theta \\ \frac{n\beta_n(r)}{r} \cos \theta  \\ f_n(r) \cos (n\theta) \\ f_n(r) \sin (n\theta) \end{array}  \right) \label{vortexcolumn}.
\end{equation}
At this point, we will define certain quantities that contain information about the angular dependence of the vector and scalar components functions $F(r,\theta)\in {\cal C}^\infty(\mathbb{R}^2) \oplus \mathbb{R}^4$ of interest, such as solutions, fluctuations, etc., as well as their radial behavior near the origin. The latter is particularly relevant as it allows us to identify the presence of singularities by simply analyzing the form of these quantities. For this reason, we will call them \textit{characteristics of the function} under study:

\vspace{0.2cm}

\noindent \textbf{Definition:} (\textit{Angular and radial characteristics}.) The \textit{angular characteristic} ${\rm ch}_\theta[F]$ of a function $F(r,\theta)\in {\cal C}^\infty(\mathbb{R}^2) \oplus \mathbb{R}^4$ is defined as
\[
{\rm ch}_\theta[F]  = {\alpha_1;\dots;\alpha_q \choose \beta_1;\dots;\beta_q },
\]
where the set of numbers in the rows specifies the angular dependence of the function $F$. Specifically, the vector components of $F$ depend only on a linear combination of sine and cosine functions with arguments $\alpha_1 \theta, \dots, \alpha_q\theta$ while the scalar component depends only on the angles $\beta_1 \theta, \dots, \beta_q\theta$. The \textit{angular characteristics number} is defined as $\#{\rm ch}_\theta[F] = \max\{ p,q\}$. 

On the other hand, the \textit{radial characteristic} ${\rm ch}_r[F]$,
\[
{\rm ch}_r[F]  = {\ell_1 \choose \ell_2} ,
\]
determines the radial dependence of the function in the neighborhood of $r=0$. In particular, the vector components of the function $F$ behave as $r^{\ell_1}$, whereas the scalar components behave as $r^{\ell_2}$.

\vspace{0.2cm}

For example, the angular and radial characteristics of the $n$-vortex solution $\Psi(x,\theta)$ are
\begin{equation}
{\rm ch}_\theta[\Psi] = {1\choose n} \hspace{0.5cm}\mbox{and}\hspace{0.5cm} {\rm ch}_r[\Psi] = {2\choose n} \label{chvortex},
\end{equation}
which is directly obtained from (\ref{vortexcolumn}) and the power expansion around the origin \eqref{VortexOrigin} .

\vspace{0.2cm}

The main goal of this paper is the study of the spectrum of normal modes associated with $n$-vortices. In order to achieve this, the dynamics of small fluctuations $\xi(\vec{x})$ around the static $n$-vortex (\ref{vortexcolumn}) will be analyzed. Consequently, the perturbed $n$-vortex, denoted collectively by $\widetilde{\Psi}(\vec{x},n)$, is expressed as
\begin{equation}
\widetilde{\Psi}(\vec{x},n) = \Psi(\vec{x},n) + \epsilon\, \xi(\vec{x})=\left(\begin{array}{c} V_1(\vec{x};n) \\ V_2(\vec{x};n) \\ \psi_1(\vec{x};n) \\ \psi_2(\vec{x};n)\end{array}  \right) + \epsilon\,  \left(\begin{array}{c} a_1(\vec{x}) \\ a_2(\vec{x})  \\ \varphi_1(\vec{x}) \\ \varphi_2(\vec{x}) \end{array}  \right) \label{perturvortexcolumn},
\end{equation}
where 
\[
\xi(\vec{x})=\left( \begin{array}{c c c c}a_1(\vec{x}) & a_2(\vec{x}) & \varphi_1(\vec{x}) & \varphi_2(\vec{x}) \end{array} \right)^t ,
\]
denote the fluctuation column. To discard pure (non-physical) gauge fluctuations, the so called \textit{background gauge}
\begin{equation}
	\partial_1 a_1( \vec{x}) + \partial_2 a_2( \vec{x}) - (\,\psi_1( \vec{x})\, \varphi_2( \vec{x})-\psi_2( \vec{x})\,\varphi_1( \vec{x})\,)=0  , 
	\label{backgroundgauge}
\end{equation}
is imposed as the gauge fixing condition on the fluctuation modes. With this set-up the normal modes of an $n$-vortex solution $\Psi(\vec{x},n)$ are determined by the spectral condition 
\begin{equation}
{\cal H}^+ \xi_\nu(\vec{x}) =\omega_\nu^2 \, \xi_\nu(\vec{x}) , \label{spectralproblem}
\end{equation}
where $\nu$ is a label in either the discrete or the continuous spectrum useful to enumerate the eigenfunctions and eigenvalues, and ${\cal H}^+$ is the second-order small fluctuation operator
{\footnotesize\begin{equation}
	{\cal H}^+= \left( \begin{array}{cccc}
	-\Delta + |\psi|^2 & 0 & -2 \widetilde{D}_1 \psi_2 & 2 \widetilde{D}_1 \psi_1 \\
	0 & -\Delta +|\psi|^2 & -2 \widetilde{D}_2 \psi_2 & 2 \widetilde{D}_2 \psi_1 \\
	-2 \widetilde{D}_1 \psi_2 & -2 \widetilde{D}_2\psi_2 & -\Delta + \frac{3\lambda}{2} \psi_1^2 + (1+\frac{\lambda}{2})\psi_2^2 -\frac{\lambda}{2}  +V_kV_k & -2 V_k \partial_k + (\lambda-1)\psi_1\psi_2\\
	2\widetilde{D}_1\psi_1 & 2 \widetilde{D}_2 \psi_1 & 2V_k \partial_k + (\lambda-1)\psi_1\psi_2 & -\Delta + (1+\frac{\lambda}{2})\psi_1^2 +\frac{3\lambda}{2} \psi_2^2  -\frac{\lambda}{2} + V_kV_k
	\end{array} \right)\label{hessianoperator}
   ,
	\end{equation}}
where $\widetilde{D}_i\psi_j = \partial_i\psi_j+\epsilon^{ik} V_i \psi_k$. The Sturm-Liouville eigenvalue problem (\ref{spectralproblem}) is obtained by linearizing the field equations (in the background gauge) around the vortices.

The fluctuation eigenfunctions $\xi(\vec{x})$ belong in general to a rigged Hilbert space, such that there exist square integrable eigenfunctions $\xi_\nu(\vec{x})\in L^2(\mathbb{R}^2)\oplus \mathbb{R}^4$ belonging to the discrete spectrum, for which the norm $\|\xi(\vec{x})\|$ is bounded:
\begin{equation}
	\|\xi(\vec{x})\|^2  = \int_{\mathbb{R}^2} d^2x \Big[ (a_1(\vec{x}))^2 + (a_2(\vec{x}))^2 + (\varphi_1(\vec{x}))^2 + (\varphi_2(\vec{x}))^2 \Big] < +\infty \, ,
	\label{normalization}
\end{equation}
together with scattering (unbounded) eigenfunctions $\xi_\nu(\vec{x})$ with $\nu$ ranging in a dense set. 

Note that the operator (\ref{hessianoperator}) resulting for the asymptotic values (\ref{asymptotic}) of the vortices exhibits a triply degenerate spectrum emerging at the threshold value $\omega_c^2 = 1$ together with once degenerate spectrum on the value $\omega_c^2= \lambda$. As a consequence, the continuous spectrum of the operator (\ref{hessianoperator}) begins at $\omega_c^2=\lambda$ when $\lambda <1$, and at the value $\omega_c^2=1$ when $\lambda \geq 1$. From a technical perspective, eigenfunctions corresponding to eigenvalues in the range $\lambda < \omega_\nu^2 < 1$ consist of a localized gauge field component and a non-localized scalar field component (which behave as a scattering mode). These type of modes is usually called quasi-bound modes. This behavior is reversed in the range $1<\omega_\nu^2 < \lambda$.


\section{Spectral structure for the fluctuations around self-dual vortices}

\label{bps}

In this paper, our objective is to determine the complete structure of the normal modes for rotationally invariant $n$-vortices in the Abelian-Higgs model for arbitrary values of the coupling constant $\lambda$. This analysis is entirely new, with the exception of the specific case $\lambda=1$ (self-dual vortices), which was thoroughly investigated in \cite{AlonsoIzquierdo2016, AlonsoIzquierdo2016b}. In this section, we briefly summarize the results presented in those works. The motivation for this summary is twofold: (i) The general results obtained in this paper must reproduce those of the aforementioned works when the coupling constant takes the specific value $\lambda=1$; and (ii) it will become evident that the mathematical techniques employed in \cite{AlonsoIzquierdo2016, AlonsoIzquierdo2016b} are applicable exclusively to self-dual vortices, corresponding to the special case $\lambda=1$. This limitation underscores the necessity of developing new methods to extend the analysis beyond the self-dual regime. This new approach is presented in Sec. \ref{non-bps} of the present work.

As previously mentioned, we will describe the spectral structure of self-dual vortices, which satisfy the BPS first-order differential equations
\begin{equation}
\widetilde{D}_1\psi_1 \mp \, \widetilde{D}_2 \psi_2 =0, \hspace{0.5cm} \hspace{0.5cm} \widetilde{D}_1\psi_2 \pm \, \widetilde{D}_2 \psi_1 =0, \hspace{0.5cm} \hspace{0.5cm} F_{12}\mp \frac{1}{2} (1-|\psi|^2 )=0 .\label{bpsequation}
\end{equation}
Equations (\ref{bpsequation}) are derived using a Bogomolny decomposition of the energy functional. Substituting the ansatz (\ref{ansatz}) into the first-order PDE system (\ref{bpsequation}) leads to the following system of first order ODEs
\begin{equation}
\frac{df_n(r)}{d r}=\frac{n}{r} f_n(r) [1-\beta_n(r)]\hspace{0.0cm},\hspace{0.5cm} \frac{d\beta_n(r)}{d r}=\frac{r}{2n}[1-f_n^2(r)] \, . \label{ode1}
\end{equation}
In the aforementioned papers the spectral problem in the self-dual/BPS case was solved by exploiting a hidden supersymmetric structure involving the second order vortex fluctuation operator (\ref{hessianoperator}), which can be factorized as ${\cal H}^+= {\cal D}^\dagger \, {\cal D}$, where ${\cal D}$ is the first order differential matrix operator 
\[
{\cal D} = \left( \begin{array}{cccc}
-\partial_2 & \partial_1 & \psi_1 & \psi_2 \\
-\partial_1 & -\partial_2 & -\psi_2 & \psi_1 \\
\psi_1 & -\psi_2 & -\partial_2 + V_1 & -\partial_1 -V_2 \\
\psi_2 & \psi_1 & \partial_1+V_2 & -\partial_2 + V_1
\end{array} \right) ,
\]
obtained from the BPS equations (\ref{bpsequation}) and the background gauge condition (\ref{backgroundgauge}). It can be shown that the intertwined supersymmetric partner operator ${\cal H}^-= {\cal D} \, {\cal D}^\dagger$ has a block diagonal structure. This property can be exploited to investigate the fluctuation spectral problem of the BPS vortex. Recall that our goal in this paper is to study this problem for general (non-BPS) vortices for arbitrary values of $\lambda$, where this supersymmetric approach is no longer applicable. 

\vspace{0.1cm}

The spectral structure identified in the BPS limit reveals that two apparently different types of eigenfunctions emerge in the spectral problem for self-dual vortices:

\begin{itemize}
	\item \textbf{Zero modes}: It is well known that at the critical value $\lambda=1$, vortices can be located at any point without experiencing any forces between them. For charge-$n$ vortices, this freedom in the positions of the vortex centers suggests the existence of $2|n|$ linearly independent zero modes (modes with vanishing eigenvalue $\omega^2_\mu=0$). This was proved by E. Weinberg in \cite{Weinberg1979} using a generalization of the index theorem for elliptic operators and the supersymmetric structure mentioned earlier. Following the notation used in \cite{Weinberg1979}, these $2|n|$ zero modes associated with the rotationally invariant BPS $n$-vortex have the structure
\begin{eqnarray}
\xi_0(\vec{x},n,k)&=& r^{n-k-1} \left( \begin{array}{c} h_{nk}(r) \, \sin[(n-k-1)\theta] \\ h_{nk}(r) \, \cos[(n-k-1)\theta] \\  -\frac{h_{nk}'(r)}{f_n(r)} \, \cos(k\theta) \\ - \frac{h_{nk}'(r)}{f_n(r)} \, \sin(k\theta) \end{array} \right)  , \label{bpszeromode4}
\\
\xi_0^\perp(\vec{x},n,k)&=& r^{n-k-1} \left( \begin{array}{c} h_{nk}(r) \, \cos[(n-k-1)\theta] \\ -h_{nk}(r) \, \sin[(n-k-1)\theta] \\  - \frac{h_{nk}'(r)}{f_n(r)} \, \sin(k\theta) \\  \frac{h_{nk}'(r)}{f_n(r)} \, \cos(k\theta)  \end{array} \right)  , \label{bpszeromode41}
\end{eqnarray}
where $k=0,1,2,\dots,n-1$ and the radial form factor $h_n(r)$ satisfies the second-order ODE
\begin{equation}
-r \, h_{nk}''(r)+[1+2k-2n\,\beta_n(r)]\,h_{nk}'(r) + r \,f_n^2(r)\, h_{nk}(r)=0, \label{ode5}
\end{equation}
with the boundary conditions $h_{nk}(0)\neq 0$ and $\lim_{r\rightarrow \infty} h_{nk}(r) =0$. The eigenfunctions (\ref{bpszeromode4}) and (\ref{bpszeromode41}) are degenerate, although they are mutually orthogonal with respect to the interior product defined by the norm \eqref{normalization}.

\vspace{0.1cm}

The angular characteristic of the zero modes is
\begin{equation}
{\rm ch}_\theta[\xi_0(\vec{x},n,k)] = {\rm ch}_\theta[\xi_0^\perp(\vec{x},n,k)] = {n-k-1\choose k}, \label{angchbpszeromode}
\end{equation}
which means that only one trigonometric function enters in the expression of the components, that is,
$\# {\rm ch}_\theta[\xi_0(\vec{x},n,k)] =  \# {\rm ch}_\theta[\xi_0^\perp(\vec{x},n,k)] = 1$.

Since $ h_{nk}(r) = c_0^{(n,k)} + c_{2k+2}^{(n,k)} r^{2k+2} + \dots$ near $r=0$, 
\begin{equation}
{\rm ch}_r[\xi_0(\vec{x},n,k)] = {\rm ch}_r[\xi_0^\perp(\vec{x},n,k)] = {n-k-1\choose k}.\label{radchbpszeromode}
\end{equation}
Note that the \textit{radial characteristic} ${\rm ch}_r[\xi_0(\vec{x},n,k)]$ directly provides the constraints on the values of $k$. Regularity conditions require the components of this characteristic to be non-negative, otherwise a singularity would emerge at the origin. This means in this case that $k\in \{0,1,\dots, n-1\}$, as previously indicated. 

In addition to these $2 \vert n \vert $ orthogonal zero modes, a continuous spectrum of scattering modes emerges for $\omega^2_\mu\geq1$.

\item \textbf{Shape modes}: Eigenfunctions belonging to the strictly positive spectrum of ${\cal H}^+$ arise with the form:
\begin{eqnarray}
\xi_\lambda(\vec{x},n,k)&=& \left( \begin{array}{c} \sin \theta \cos (k\theta) \frac{d v_{nk}(r)}{d r} - \frac{k}{r} \, v_{nk}(r) \cos \theta \sin(k\theta) \\ -\cos \theta \cos (k\theta) \frac{d v_{nk}(r)}{d r} - \frac{k}{r} \, v_{nk}(r) \, \sin \theta \sin(k\theta) \\ f_n(r) \, v_{nk}(r)\, \cos(n\theta)\, \cos(k\theta) \\ f_n(r)\, v_{nk}(r)\, \sin(n\theta) \,\cos(k\theta)
 \end{array} \right) \hspace{0.0cm} , \hspace{0.3cm} k=0,1,2,\dots\, , \hspace{0.2cm} \label{bpsexcitedmode1}
\end{eqnarray}
\begin{eqnarray}
\chi_\lambda(\vec{x},n,k)&=& \left( \begin{array}{c} \sin \theta \sin (k\theta) \frac{d v_{nk}(r)}{d r} + \frac{k}{r} \, v_{nk}(r) \cos \theta \cos(k\theta) \\ -\cos \theta \sin (k\theta) \frac{d v_{nk}(r)}{d r} + \frac{k}{r} \, v_{nk}(r) \, \sin \theta \cos(k\theta) \\ f_n(r) \, v_{nk}(r)\, \cos(n\theta)\, \sin(k\theta) \\ f_n(r)\, v_{nk}(r)\, \sin(n\theta) \,\sin(k\theta)
 \end{array} \right) \hspace{0.0cm}, \hspace{0.3cm} k=1,2,\dots \, ,\hspace{0.2cm} \label{bpsexcitedmode2}
\end{eqnarray}
where the radial form factors $v_{nk}(r)$ can be determined as a solution of the 1D Sturm-Liouville problem
\begin{equation}
-\frac{d^2 v_{nk}(r)}{d r^2} -\frac{1}{r} \frac{d v_{nk}(r)}{d r} + \Big[ f_n^2(r)-\omega_\lambda^2 + \frac{k^2}{r^2} \Big] v_{nk}(r)=0  .
\label{ode55}
\end{equation}
For $k\neq 0$ the eigenfunctions $\xi_\lambda(\vec{x},n,k)$ and $\chi_\lambda(\vec{x},n,k)$ are degenerate, although linearly independent. The number of shape modes is non trivially related to the vortex charge $n$ (contrary to the zero modes) and depends on the strength of the potential well arising in the spectral problem (\ref{ode55}).

Since $v_{nk}(r) = v_0^{(n,k)} r^k + v_2^{(n,k)} r^{k+2} + \dots$ near $r=0$, the angular characteristics of the shape modes (\ref{bpsexcitedmode1}) and (\ref{bpsexcitedmode2}) become
\begin{equation}
{\rm ch}_\theta[\xi(\vec{x},n,k)] = {\rm ch}_\theta[\chi(\vec{x},n,k)] = {k-1;k+1\choose n-k;n+k} \hspace{0.5cm} \mbox{for} \hspace{0.5cm} k=1,2,\dots, \label{angchbpsshapemode}
\end{equation}
while the radial characteristics are
\begin{equation}
{\rm ch}_r[\xi(\vec{x},n,k)] = {\rm ch}_r[\chi(\vec{x},n,k)] = {k-1\choose n+k}  \hspace{0.5cm} \mbox{for} \hspace{0.5cm} k=1,2,\dots.  \label{radchbpsshapemode}
\end{equation}
The expression (\ref{angchbpsshapemode}) implies that the eigenfunctions describing these shape modes exhibit a more complex angular dependence than the zero modes, which can be described by the value of the angular characteristic number, $\# {\rm ch}_\theta[\xi(\vec{x},n,k)]  = \# {\rm ch}_\theta[\chi(\vec{x},n,k)] = 2$. In general, the vector and scalar contribution of the shape modes has  angular dependence of the forms $(k\pm 1)\theta$ and $(n\pm k)\theta$ respectively.

The double angular dependence is broken for $k=0$, where we find 
\begin{equation}
{\rm ch}_\theta[\xi(\vec{x},n,0)]  = {1\choose n} \hspace{0.5cm}\mbox{and} \hspace{0.5cm} {\rm ch}_r [\xi(\vec{x},n,0)] = {1\choose n}. \label{chbpsshapemode0}
\end{equation}
In this case we recover the dependence on only one angle for the components of the eigenfunction, as in the case of the zero modes. It is worth mentioning that for $k=0$ the angular characteristics (\ref{chbpsshapemode0}) of the shape mode (\ref{bpsexcitedmode1}) with $k=0$ coincides with that of the charge-$n$ vortex (\ref{chvortex}). This implies in particular that the $k=0$ modes are the so-called Derrick type modes, i.e., the $k=0$ modes only affect the vortex size but not its symmetry. Note that this is the only shape mode that is not doubly degenerate, see (\ref{bpsexcitedmode1}) and (\ref{bpsexcitedmode2}).
\end{itemize}

In summary, the spectral structure for rotationally invariant $n$-vortices in the self-dual case comprises a non-degenerate Derrick type mode with positive eigenvalue (responsible for size oscillations of the $n$-vortex); $2 \vert n \vert$ zero modes (which changes the vortex centers) and a charge-dependent number of doubly degenerate shape modes with positive eigenvalues (describing normal modes with non-trivial multipolar structure with respect to the $n$-vortex solution). 

\section{The spectral structure for the fluctuations around non-self-dual vortices}

\label{non-bps}

In this section we aim to describe the spectral structure of the non-BPS vortices. At the same time, we seek to develop a unified understanding of the different types of normal modes of vibration analyzed in Sec. \ref{bps}. The first step is to identify the angular dependence of the eigenfunctions, that is, their angular characteristics. One could guess that beyond the self-dual case the angular dependence of the eigenfunctions would become more complex than in the case of BPS vortices. However, as we show in the following lemma, the angular characteristics number is still 2:

\vspace{0.1cm}

\noindent \textbf{Lemma:} The spectral problem (\ref{spectralproblem}) associated to the non self-dual rotationally invariant $n$-vortex admits Derrick type eigenfunctions together with generic eigenmodes $\xi_\nu(\vec{x})$ whose angular characteristics are
\begin{equation}
{\rm ch}_\theta [\xi_\nu(\vec{x},n)] = {\overline{k}-1 \,;\,\overline{k}+1 \choose n-\overline{k}\,;\,n+\overline{k}} \hspace{0.5cm} \mbox{for} \hspace{0.5cm} \overline{k}=1,2,\dots \label{angthnonbpsvortex}
\end{equation}

\noindent \textbf{Proof:} The form of (\ref{angthnonbpsvortex}) indicates that the expression of the vector and scalar components of the ${\cal H}^+$-eigenfunctions depends only on two different angles. For this reason, it is enough to show that the ansatz 
\begin{equation}
    \xi_\nu(r,\theta) = \left( \begin{array}{c} \overline{g}_1(r) \, \sin (a \,\theta)  + \overline{g}_2(r) \sin (b \,\theta)  \\ \overline{g}_1(r) \, \cos (a \,\theta)  + \overline{g}_2(r) \cos (b \,\theta)  \\  \overline{t}_1(r) \, \cos (c \,\theta)  + \overline{t}_2(r) \cos (d \,\theta) \\  \overline{t}_1(r) \, \sin (c \,\theta)  + \overline{t}_2(r) \sin (d \,\theta) \end{array} \right), \label{genericform01}
\end{equation}
is compatible with the spectral equation ${\cal H}^+ \xi_\nu = \omega_\nu^2 \xi_\nu$ associated to the operator (\ref{hessianoperator}) for some choice of the parameters $a,b,c,d$ in (\ref{genericform01}). Because of the continuity of the eigenfunctions these parameters must be integers, that is, $a,b,c,d \in \mathbb{Z}$. If we substitute (\ref{genericform01}) into (\ref{spectralproblem}) the angular characteristics of both sides of the spectral equation can be obtained. The left-hand side verifies that
\[
{\rm ch}_\theta [{\cal H}^+ \xi_\nu] = { a \, ; \, b \,  ; \, 1+c-n\, ; \, 1-c+n\, ;\, 1+d-n\, ;\, 1-d+n \choose c\, ;\, d\, ;\, 1+a-n\, ;\, 1+a+n\, ;\, 1+b-n\, ;\, 1+b+n\, ;\, c-2n\, ;\, d-2n},
\]
while the right-hand side leads to
\[
{\rm ch}_\theta [\omega_\nu^2 \xi_\nu] = { a\, ;\,  b \choose c\, ;\, d}.
\]
This means that we end up with four equations to determine the radial profiles $\overline{g}_i(r)$ and $\overline{t}_i(r)$ with $i=1,2$, which involve sums of trigonometric functions with different angular dependencies. Specifically, the first and second equation include the angles $a\theta$, $b\theta$, $(1+c-n)\theta$, $(1-c+n)\theta$, $(1+d-n)\theta$, $(1-d+n)\theta$ while the third and the fourth ones involve the angles $c\theta$, $d\theta$, $(1+a-n)\theta$, $(1+a+n)\theta$, $(1+b-n)\theta$, $(1+b+n)\theta$, $(c-2n\theta)$ and $(d-2n)\theta$. Thus, $\#{\rm ch}_\theta [{\cal H}^+ \xi_\nu]=8$. Fortunately, there is a particular choice of the parameters
\begin{equation}
b=-2-a \hspace{0.0cm},\hspace{0.5cm} c=n-1-a  \hspace{0.0cm},\hspace{0.5cm} d= 1+n+a ,\label{constraint01}
\end{equation}
which reduces these dependencies to only two independent trigonometric functions for each equation, leading then to $\#{\rm ch}_\theta [{\cal H}^+ \xi_\nu]=2$ \footnote{Indeed, the choice of parameters  $b=-2-a ,\, d=n-1-a, \,c= 1+n+a$ is also a valid choice to solve the spectral problem, but leads to the same results as the choice \eqref{constraint01}.}. For the sake of simplicity, it is convenient to choose $a=\overline{k}-1$, such that the relation (\ref{constraint01}) implies that the fluctuations (\ref{genericform01}) follows the form
\begin{equation}
    \xi_\nu(\vec{x},n,\overline{k}) = \left( \begin{array}{c} -\overline{g}_1(r) \, \sin [(1-\overline{k}) \,\theta]  - \overline{g}_2(r) \sin [(1+\overline{k}) \,\theta]   \\  \overline{g}_1(r) \, \cos [(1-\overline{k}) \,\theta]  + \overline{g}_2(r) \cos [(1+\overline{k}) \,\theta]  \\  \overline{t}_1(r) \, \cos [(n-\overline{k}) \,\theta]  + \overline{t}_2(r) \cos [(n+\overline{k}) \,\theta]\\ \overline{t}_1(r) \, \sin [(n-\overline{k}) \,\theta]  + \overline{t}_2(r) \sin [(n+\overline{k}) \,\theta] \end{array} \right) \label{genericform02}.
\end{equation}
In this way, it is direct to prove that these eigenfunctions exhibit the symmetry
\[
\overline{k} \rightarrow - \overline{k},
\]
in such a way that we can restrict our study to the case $\overline{k} \geq 0$. This justifies the relation (\ref{angthnonbpsvortex}) introduced in the Lemma. The expression (\ref{genericform02}) when plugged into the spectral equation leads to four ordinary differential equations for the four radial functions $\overline{g}_i(r)$ and $\overline{t}_i(r)$ with $i=1,2$. The explicit form of these differential equations will be shown below. Recall that (\ref{genericform02}) provides the admissible eigenfunctions, but new requirements will be imposed on them, for example, to guarantee the regularity conditions at the origin or the fulfillment of the \textit{background gauge condition} in order to remove non-physical degrees of freedom. 

Note also that $\overline{k}=0$ is a special case, because the two trigonometric functions that enter each of the fluctuation components of (\ref{genericform02}) are equal, and the angular characteristic number is reduced to 1, that is, $\# {\rm ch}_\theta[\xi_\nu(\vec{x},n,0) ] = 1$. In this case, the eigenfunction (\ref{genericform02}) becomes 
\begin{equation}
    \xi_\nu(\vec{x},n,0) = \left( \begin{array}{c} -\widetilde{g}(r) \, \sin [\theta] \\  \widetilde{g}(r) \, \cos [\theta]  \\  \widetilde{t}(r) \, \cos [n \,\theta] \\ \widetilde{t}(r) \, \sin [n \,\theta] \end{array} \right), \label{genericform03}
\end{equation}
where
$$
\widetilde{g}(r)=\overline{g}_1(r)+ \overline{g}_2(r), \qquad \widetilde{t}(r)=t_1(r)+t_2(r).
$$
These eigenfunctions preserve the angular form of $n$-vortices, which means that they are Derrick type modes.

The same argument employed in this proof can be applied to a different choice of (\ref{genericform01}), which leads to the generic eigenfunctions
\begin{equation}
    \chi_\nu(\vec{x},n,\overline{k}) = \left( \begin{array}{c} -\overline{g}_1(r) \, \cos [(1-\overline{k}) \,\theta] + \overline{g}_2(r) \cos [(1+\overline{k}) \,\theta]   \\  -\overline{g}_1(r) \, \sin [(1-\overline{k}) \,\theta]  + \overline{g}_2(r) \sin [(1+\overline{k}) \,\theta]  \\  -\overline{t}_2(r) \, \sin [(n-\overline{k}) \,\theta]  + \overline{t}_1(r) \sin [(n+\overline{k}) \,\theta]\\ -\overline{t}_2(r) \, \cos [(n-\overline{k}) \,\theta]  - \overline{t}_1(r) \cos [(n+\overline{k}) \,\theta] \end{array} \right), \label{genericform02b}
\end{equation}
which are orthogonal to (\ref{genericform02}). The radial functions $\overline{g}_i(r)$ and $\overline{t}_i(r)$ with $i=1,2$ in (\ref{genericform02b}) verify the same differential equations as those included in (\ref{genericform02}). $\Box$

\vspace{0.2cm}

One important remark is that the generic form (\ref{genericform02}) and  (\ref{genericform02b}) unifies the two different types of eigenmodes described in the self-dual regime in Section \ref{bps}: 
\begin{enumerate}
\item The zero modes follow the pattern (\ref{genericform02}) (or (\ref{genericform02b})) by taking $\overline{k}=n-k$. With this choice, the angular momentum parameter $\overline{k}$ used in (\ref{genericform02}) ranges the values $\overline{k}=1,2,\dots,n-1,n$. Notice that the zero modes in the self-dual case have an \textit{angular characteristics number} equals to 1, $\#{\rm ch}_\theta[\xi_0(\vec{x})]=1$, but this is no longer possible for the general case. These modes acquire a new angular dependence when $\lambda\neq 1$.

\item On the other hand, the shape modes are recovered by simply taking $\overline{k}=k$ with $\overline{k}=0,1,\dots$, with $\overline{k}=0$ designating the Derrick type mode. Comparing  (\ref{genericform02}) with (\ref{bpsexcitedmode1}) the relation between the radial functions is given by $\overline{g}_1(r) = -\frac{1}{2} (v_{nk}'(r) + \frac{k v_{nk}(r)}{r})$, $\overline{g}_2(r)= -\frac{1}{2} (v_{nk}'(r) - \frac{k v_{nk}(r)}{r})$ and $\overline{t}_1(r)=\overline{t}_2(r)=\frac{1}{2} f_n(r) v_{nk}(r)$.
\end{enumerate}

As mentioned previously, by substituting the expressions (\ref{genericform02}) or (\ref{genericform03}) into the general problem (\ref{spectralproblem}) we end up with four second-order ordinary differential equations for the radial profile functions $\overline{g}_i(r)$ and $\overline{t}_i(r)$, $i=1,2$. We distinguish two different cases, $\overline{k}=0$ and $\overline{k}>0$:

\vspace{0.2cm}

\noindent \textsc{Derrick-type modes:} In the first case, $\overline{k}=0$, the reduced 1D eigenvalue problem is given by
\begin{eqnarray}
	&& - \frac{d^2 \widetilde{g}(r)}{dr^2} - \frac{1}{r}\frac{d\widetilde{g}(r)}{dr} + \Big( \frac{1}{r^2} + f_n^2(r) - \omega_n^2 \Big) \widetilde{g}(r) - \frac{2n}{r} (1-\beta_n(r)) f_n(r) \widetilde{t}(r) =0 ,\label{radialedo01} \\
	&& - \frac{d^2 \widetilde{t}(r)}{dr^2} - \frac{1}{r} \frac{d\widetilde{t}(r)}{dr} + \Big( \frac{n^2 (1-\beta_n(r))^2}{r^2} + \frac{3 \lambda}{2} f_n^2(r) - \frac{\lambda}{2} - \omega_n^2 \Big) \widetilde{t}(r) - \frac{2n}{r} (1-\beta_n(r)) f_n(r) \widetilde{g}(r) =0 .\nonumber
\end{eqnarray}

\vspace{0.2cm}

\noindent \textsc{Multipolar modes:} In the general case, $\overline{k}>0$ the radial profiles of the eigenfunctions (\ref{genericform02}) are determined by
{\small \begin{eqnarray}
	&& - \frac{d^2 \overline{g}_1(r)}{dr^2} - \frac{1}{r} \frac{d\overline{g}_1(r)}{dr}+ \Big[ \frac{(\overline{k}-1)^2}{r^2} +f_n^2(r) \Big] \overline{g}_1(r) - \Big[ \frac{n (1-\beta_n(r))f_n(r)}{r} + f_n'(r) \Big] \overline{t}_1(r) + \nonumber\\
	&&  \hspace{0.5cm} + \Big[ - \frac{n(1-\beta_n(r))f_n(r)}{r} + f_n'(r) \Big] \overline{t}_2(r) = \omega_n^2 \overline{g}_1(r) \hspace{0.0cm}, \nonumber \\
	&& - \frac{d^2 \overline{g}_2(r)}{dr^2} - \frac{1}{r} \frac{d\overline{g}_2(r)}{dr}+ \Big[ \frac{(\overline{k}+1)^2}{r^2} +f_n^2(r) \Big] \overline{g}_2(r) + \Big[ -\frac{n (1-\beta_n(r))f_n(r)}{r} + f_n'(r) \Big] \overline{t}_1(r) - \nonumber \\
	&&  \hspace{0.5cm} - \Big[ \frac{n(1-\beta_n(r))f_n(r)}{r} + f_n'(r) \Big] \overline{t}_2(r) = \omega_n^2 \overline{g}_2(r) \hspace{0.0cm}, \label{radialedo02} \\
	&& - \frac{d^2 \overline{t}_1(r)}{dr^2} - \frac{1}{r} \frac{d\overline{t}_1(r)}{dr}+ \Big[ \frac{(\overline{k}-n(1-\beta_n(r)))^2}{r^2} -\frac{\lambda}{2} + (\lambda+\frac{1}{2})f_n^2(r) \Big] \overline{t}_1(r) - \Big[ \frac{n (1-\beta_n(r))f_n(r)}{r} + f_n'(r) \Big] \overline{g}_1(r) + \nonumber \\
	&&  \hspace{0.5cm} + \Big[ - \frac{n(1-\beta_n(r))f_n(r)}{r} + f_n'(r) \Big] \overline{g}_2(r) + \frac{1}{2} (\lambda -1) f_n^2(r) \overline{t}_2(r) = \omega_n^2 \overline{t}_1 (r)\hspace{0.0cm}, \nonumber \\
	&& - \frac{d^2 \overline{t}_2(r)}{dr^2} - \frac{1}{r} \frac{d\overline{t}_2(r)}{dr}+ \Big[ \frac{(\overline{k}+n(1-\beta_n(r)))^2}{r^2} -\frac{\lambda}{2} + (\lambda+\frac{1}{2})f_n^2(r) \Big] \overline{t}_2(r) + \Big[- \frac{n (1-\beta_n(r))f_n(r)}{r} + f_n'(r) \Big] \overline{g}_1(r) - \nonumber \\
	&&  \hspace{0.5cm} - \Big[ \frac{n(1-\beta_n(r))f_n(r)}{r} + f_n'(r) \Big] \overline{g}_2(r) + \frac{1}{2} (\lambda -1) f_n^2(r) \overline{t}_1(r) = \omega_n^2 \overline{t}_2(r) \hspace{0.0cm}. \nonumber
\end{eqnarray}}

At this point, we restrict the admissible eigenfunctions (\ref{genericform02}) to those satisfying the \textit{background gauge condition}, removing all the non-physical degrees of freedom. This is a very delicate point because it may add a new condition on the radial spectral problems (\ref{radialedo01}) and (\ref{radialedo02}), which could lead to inconsistencies. Here, is not the case as we prove in the following proposition:

\vspace{0.2cm}

\noindent \textbf{Proposition:} The admissible eigenfunctions of the fluctuation spectral problem (\ref{spectralproblem}) associated with the non self-dual rotationally invariant $n$-vortices without superfluous gauge degrees of freedom are characterized as:

\vspace{0.1cm}
\begin{enumerate}
\item The Derrick type modes arise in the form
\begin{equation}
	\xi_\nu(\vec{x},n,0) = \left( \begin{array}{c} \widetilde{v}(r) \, \sin \theta  \\ - \widetilde{v}(r) \,\cos \theta  \\ \widetilde{u}(r) \, \cos(n\theta) \\ \widetilde{u}(r) \, \sin(n\theta) 
	\end{array} \right),  \label{genericform03b}
\end{equation}
where the radial functions $\zeta_\nu (r)= (\widetilde{v}(r),\widetilde{u}(r) )$ satisfy the reduced spectral problem $\overline{\cal H}_0 \zeta_\nu (x) = \omega_\nu^2 \zeta_\nu(x)$ where
\begin{equation}
\overline{\cal H}_0  = \left( \begin{array}{cc} - \frac{d^2}{dr^2} - \frac{1}{r} \frac{d}{dr} + \frac{1}{r^2} + f_n^2(r)  &  \frac{2n}{r} (1-\beta_n(r)) f_n(r) \\ \frac{2n}{r} (1-\beta_n(r)) f_n(r) & - \frac{d^2}{dr^2}- \frac{1}{r} \frac{d}{dr} + \frac{n^2 (1-\beta_n(r))^2}{r^2} + \frac{3 \lambda}{2} f_n^2(r) - \frac{\lambda}{2} 
\end{array} \right). \label{operatorhbar0}
\end{equation}

\item The multipolar eigenfunctions have the form
\begin{equation}
\xi_\nu(\vec{x},n,\overline{k}) = \left( \begin{array}{c} \cos(\overline{k} \theta) \sin\theta [v'(r) -r f_n(r) w(r)] - \sin (\overline{k}\theta) \cos \theta \frac{\overline{k} v(r)}{r}  \\
- \cos(\overline{k} \theta) \cos\theta [v'(r) -r f_n(r) w(r)] - \sin (\overline{k}\theta) \sin \theta \frac{\overline{k} v(r)}{r} \\
\cos(\overline{k}\theta)\cos(n\theta) u(r) + \overline{k} \sin(\overline{k}\theta) \sin(n\theta) w(r) \\
\cos(\overline{k}\theta)\sin(n\theta) u(r) - \overline{k} \sin(\overline{k}\theta) \cos(n\theta) w(r) \end{array} \right),\hspace{0.5cm}\overline{k}=1,2,\dots ,\label{genericform04b}  
\end{equation}
where the radial functions $\zeta_\nu(r)= (v(r),u(r),w(r))^t$ are determined by the reduced eigenvalue problem $\overline{\cal H} \, \zeta_\nu(x)= \omega_\nu^2 \, \zeta_\nu(x)$, 
where 
{\small 
\begin{equation}
\overline{\cal H}_k =\left( \begin{array}{ccc} - \frac{d^2}{dr^2} - \frac{1}{r} \frac{d}{dr}+  \frac{\overline{k}^2}{r^2} + f_n^2(r) & 0 & 2(f_n(r) +r f_n'(r)) \\  
\frac{2n(1-\beta_n(r))f_n(r)}{r} \frac{d}{dr} &  \begin{array}{c} - \frac{d^2}{dr^2} - \frac{1}{r} \frac{d}{dr}+ \frac{\overline{k}^2}{r^2} + \\ + \frac{n^2(1-\beta_n(r))^2}{r^2} + \frac{3\lambda}{2} f_n^2(r) - \frac{\lambda}{2} \end{array} & - \frac{2n(1-\beta_n(r))(\overline{k}^2 + r^2 f_n^2(r))}{r^2} \\
\frac{2f_n'(r)}{r} & - \frac{2n(1-\beta_n(r))}{r^2}  & \begin{array}{c} - \frac{d^2}{dr^2} - \frac{1}{r} \frac{d}{dr}+ \frac{\overline{k}^2}{r^2} + \\ + \frac{n^2(1-\beta_n(r))^2}{r^2} + f_n^2(r) + \frac{\lambda}{2} f_n^2(r) - \frac{\lambda}{2} \end{array} \end{array} \right) .\label{hbarrak}
\end{equation}
}

For $\overline{k}\geq 1$, a second type of eigenfunctions $\chi_\nu(\vec{x},n,\overline{k})$ (orthogonal to (\ref{genericform04b})) can be constructed simply by making the changes $\cos(\overline{k}\theta)\rightarrow \sin(\overline{k}\theta)$ and $\sin(\overline{k}\theta)\rightarrow -\cos(\overline{k}\theta)$ in (\ref{genericform04b}), which satisfies the same spectral problem. Therefore, eigenvalues with $\overline{k}\geq 1$ are doubly degenerate.
\end{enumerate}
\vspace{0.1cm}

\noindent \textbf{Proof:} We need to restrict the previous admissible eigenfunctions (\ref{genericform02}) and (\ref{genericform03}) such that they satisfy the background gauge conditions (\ref{backgroundgauge}) and check that no inconsistencies are introduced in the equations.

In the case of Derrick type modes, we notice that (\ref{genericform03b}) follows the same form as (\ref{genericform03}) simply changing $v(r)=-\widetilde{g}(r)$ and $u(r)=\widetilde{t}(r)$. This is done to get a better fitting between this situation and the general situation, explained below. In this case, the background gauge condition is automatically satisfied and the differential equations (\ref{radialedo01}) read now
\begin{eqnarray}
	&& - \widetilde{v}''(r) - \frac{1}{r} \widetilde{v}'(r) + \Big( \frac{1}{r^2} + f_n^2(r) - \omega_n^2 \Big) \widetilde{v}(r) + \frac{2n}{r} (1-\beta_n(r)) f_n(r) \widetilde{u}(r) =0, \label{edoDerrick} \\
	&& - \widetilde{u}''(r) - \frac{1}{r} \widetilde{u}'(r) + \Big( \frac{n^2 (1-\beta_n(r))^2}{r^2} + \frac{3 \lambda}{2} f_n^2(r) - \frac{\lambda}{2} - \omega_n^2 \Big) \widetilde{u}(r) + \frac{2n}{r} (1-\beta_n(r)) f_n(r) \widetilde{v}(r) =0, \nonumber
\end{eqnarray}
which define the spectral problem associated to the operator (\ref{operatorhbar0}). 

For the so called multipolar eigenfunctions, the situation is more complex. It is convenient to define the functions
\[
\overline{g}_+= -\overline{g}_1+\overline{g}_2 \hspace{0.0cm},\hspace{0.5cm}\overline{g}_-= -\overline{g}_1-\overline{g}_2 \hspace{0.cm},\hspace{0.5cm} \overline{t}_+= \overline{t}_1+\overline{t}_2 \hspace{0.cm},\hspace{0.5cm} \overline{t}_-= \overline{t}_1-\overline{t}_2,
\]
such that the equations (\ref{radialedo02}) now become
\begin{eqnarray}
	&& - \frac{d^2 \overline{g}_+(r)}{dr^2} - \frac{1}{r} \frac{d\overline{g}_+(r)}{dr}+ \Big[ \frac{\overline{k}^2+1}{r^2} +f_n^2(r) \Big] \overline{g}_+(r) - 
	\frac{2 \overline{k}}{r^2} \overline{g}_- (r)+2 f_n'(r) \, \overline{t}_-(r) = \omega_n^2 \overline{g}_+(r), \nonumber \\
	&& - \frac{d^2 \overline{g}_-(r)}{dr^2} - \frac{1}{r} \frac{d\overline{g}_-(r)}{dr}+ \Big[ \frac{\overline{k}^2+1}{r^2} +f_n^2(r) \Big] \overline{g}_-(r) - 
	\frac{2 \overline{k}}{r^2} \overline{g}_+ +\frac{2 n (1-\beta_n(r)) f_n(r)}{r} \overline{t}_+(r) = \omega_n^2 \overline{g}_-(r),\nonumber \\
	&& - \frac{d^2 \overline{t}_+(r)}{dr^2} - \frac{1}{r} \frac{d\overline{t}_+(r)}{dr}+ \Big[ \frac{\overline{k}^2+n^2(1-\beta_n(r))^2}{r^2} + \frac{3\lambda}{2} f_n^2(r) - \frac{\lambda}{2} \Big] \overline{t}_+(r) - 
	\frac{2 n \overline{k} (1-\beta_n(r))}{r^2} \overline{t}_- (r)\nonumber \\
	&&  \hspace{0.5cm} + \frac{2n(1-\beta_n(r))f_n(r)}{r} \, \overline{g}_-(r) = \omega_n^2 \overline{t}_+ (r),\label{radialedo03} \\
	&& - \frac{d^2 \overline{t}_-(r)}{dr^2} - \frac{1}{r} \frac{d\overline{t}_-(r)}{dr}+ \Big[ \frac{\overline{k}^2+n^2(1-\beta_n(r))^2}{r^2} + f_n^2(r)+ \frac{\lambda}{2} f_n^2(r) - \frac{\lambda}{2} \Big] \overline{t}_-(r) - 
	\frac{2 n \overline{k} (1-\beta_n(r))}{r^2} \overline{t}_+ (r)\nonumber \\
	&&  \hspace{0.5cm} + 2 f_n'(r) \, \overline{g}_+(r) = \omega_n^2 \overline{t}_-(r). \nonumber
\end{eqnarray}
The introduction of the functions $\overline{g}_\pm(r)$ and $\overline{t}_\pm(r)$ is motivated because now the background gauge condition (\ref{backgroundgauge}) is simply expressed as
\[
\frac{\partial \overline{g}_+(r)}{\partial r} + \frac{1}{r} \overline{g}_+(r) - \frac{\overline{k}}{r} \overline{g}_-(r) - f_n(r)\, \overline{t}_- (r)=0,
\]
which allows us to solve for $\overline{g}_-$
\begin{equation}
\overline{g}_-(r)=\frac{r}{\overline{k}} \Big( \frac{\partial \overline{g}_+(r)}{\partial r} + \frac{1}{r} \overline{g}_+(r) - f_n(r) \, \overline{t}_-(r) \Big). \label{gmenos}
\end{equation}
If (\ref{gmenos}) is substituted into (\ref{genericform02}) it is easy to see that the eigenfunctions of this class have the form
\begin{eqnarray}
	a_1(r,\theta) &=& \cos(\overline{k} \theta) \sin(\theta) [v'(r) -r f_n(r) w(r)] - \sin (\overline{k}\theta) \cos \theta \frac{\overline{k} v(r)}{r}, \nonumber  \\
	a_2(r,\theta) &=& - \cos(\overline{k} \theta) \cos(\theta) [v'(r) -r f_n(r) w(r)] - \sin (\overline{k}\theta) \sin \theta \frac{\overline{k} v(r)}{r} ,\nonumber \\
	\varphi_1(r,\theta) &=& \cos(\overline{k}\theta)\cos(n\theta) u(r) + \overline{k} \sin(\overline{k}\theta) \sin(n\theta) w(r), \label{genericform04c}  \\
	\varphi_2(r,\theta) &=& \cos(\overline{k}\theta)\sin(n\theta) u(r) - \overline{k} \sin(\overline{k}\theta) \cos(n\theta) w(r), \nonumber
\end{eqnarray}
where we have introduced $\overline{g}_+(r)= \frac{\overline{k}}{r} v(r)$, $\overline{t}_+(r) = u(r)$ and $\overline{t}_-(r)=\overline{k} w(r)$. Notice that by construction (\ref{genericform04c}) automatically satisfies  the gauge condition (\ref{backgroundgauge}). It remains to verify the consistency of the procedure. The first, third and fourth equations in (\ref{radialedo03}) provide the relations
\begin{eqnarray}
	&& - \frac{d^2 v(r)}{dr^2} - \frac{1}{r} \frac{dv(r)}{dr}+ \Big[ \frac{\overline{k}^2}{r^2} + f_n^2(r) \Big] v(r) + 2(f_n(r) +r f_n'(r)) w(r) = \omega_n^2 v(r), \nonumber \\
	&& - \frac{d^2 u(r)}{dr^2} - \frac{1}{r} \frac{du(r)}{dr}+ \Big[ \frac{\overline{k}^2}{r^2} + \frac{n^2(1-\beta_n(r))^2}{r^2} + \frac{3\lambda}{2} f_n^2(r) - \frac{\lambda}{2} \Big] u(r)- \frac{2n(1-\beta_n(r))(\overline{k}^2 + r^2 f_n^2(r))}{r^2} w(r) \nonumber\\
	&& \hspace{0.5cm}+ \frac{2n(1-\beta_n(r))f_n(r)}{r} \frac{dv(r)}{dr} = \omega_n^2 u(r), \label{genericform04d} \\
	&& - \frac{d^2 w(r)}{dr^2} - \frac{1}{r} \frac{dw(r)}{dr}+ \Big[ \frac{\overline{k}^2}{r^2} + \frac{n^2(1-\beta_n(r))^2}{r^2} + f_n^2(r) + \frac{\lambda}{2} f_n^2(r) - \frac{\lambda}{2} \Big] w(r)- \frac{2n(1-\beta_n(r))}{r^2} u(r) + \nonumber\\
	&& \hspace{0.5cm}+ \frac{2f_n'(r)}{r} v(r) = \omega_n^2 w(r), \nonumber
\end{eqnarray}
which can be used to determine the profiles of the three radial functions $v(r)$, $u(r)$ and $w(r)$. This establishes the spectral problem for the operator $\overline{H}_k$, as stated in the proposition. Notably, the second equation derived from (\ref{radialedo03}) corresponds to a linear combination of the remaining equations. Specifically, it can be expressed as the sum of $r/\overline{k}$ times the derivative of the first equation in (\ref{genericform04d}), $1/\overline{k}$ times the first equation itself, and $-r f_n(r)/\overline{k}$ times the fourth equation. Note that in this calculation the differential equation (\ref{edof}), which governs the scalar radial profile of the vortex, must be employed. This ensures the consistency of the procedure. $\Box$

\vspace{0.2cm}

Finally, all the previous information regarding the angular dependence of the eigenfunctions under the background gauge must be complemented by an analysis of their radial dependence at the origin and at infinity. This is necessary to determine, for instance, possible restrictions on the value of the parameter $k$. This analysis is summarized in the following result:

\vspace{0.2cm}

\noindent \textbf{Theorem:} The spectrum of the eigenvalue problem (\ref{spectralproblem}) associated with the fluctuation operator (\ref{hessianoperator}) for non self-dual rotationally invariant $n$-vortices comprises three types of eigenfunctions: 
\begin{enumerate}
    \item \textit{Derrick-type modes:} These modes are characterized by the form (\ref{genericform03b}), which depends on two radial functions $\widetilde{v}(r)$ and $\widetilde{u}(r)$. These functions are determined by the reduced spectral problem associated to the 1D $2\times 2$ matrix differential operator (\ref{operatorhbar0}). The number of these modes is only restricted by the strength of the potential well arising in the spectral problem.

    \item \textit{Multipolar modes:} They follow the expression (\ref{genericform04b}) and involve three radial functions $v(r)$, $u(r)$ and $w(r)$, which correspond to the eigenfunctions of the 1D $3\times 3$ matrix differential operator (\ref{hbarrak}). There are two classes of the multipolar modes:
    \begin{enumerate}
        \item \textit{Type A:} There are eigenfunctions $\xi_\nu^{\rm A}(\vec{x},n,\overline{k})$ with radial characteristic
\[
{\rm ch}_r[\xi_\nu^{\rm A}(\vec{x},n,\overline{k})] = {\overline{k}-1 \choose n-\overline{k}},
\]
which implies that the vector and scalar fields has opposite dependence on the parameter $\overline{k}$ in the power series with respect to $r$ near the origin. As a consequence, the number of eigenfunctions of this type are reduced to the range $\overline{k}\in \{1, \dots, n \}$.

{ \item \textit{Type B:} The second class of multipolar eigenfunctions $\xi_\nu^{\rm B}(\vec{x},n,\overline{k})$ has radial characteristic
\[
{\rm ch}_r[\xi_\nu^{\rm B}(\vec{x},n,\overline{k})] = {\overline{k}-1 \choose n+\overline{k}},
\]
such that the vector and scalar fields exhibit the same powers of $r$ with respect to the parameter $\overline{k}$ in the expansion series near the origin. Now, the number of the discrete modes is only fixed by the potential well strength of the corresponding eigenvalue problem.}

    \end{enumerate}
\end{enumerate}

\vspace{0.1cm}

\noindent \textbf{Proof:} It will be shown that the shape modes following the expressions (\ref{genericform03b}) and (\ref{genericform04b}) are square integrable eigenfunctions, $\xi_\nu (\vec{x})\in L^2(\mathbb{R}^2)\oplus \mathbb{R}^4$ of the operators (\ref{operatorhbar0}) and (\ref{hbarrak}). It only remains to show that the norm of these modes is well-defined. We distinguish two different scenarios:

\vspace{0.2cm}

\noindent 1. For Derrick type modes the norm of the eigenfunction is written as
\begin{equation}
\| \xi(\vec{x},n,0)\|^2 = 2 \pi \int_0^\infty r [\widetilde{v}^2(r) + \widetilde{u}^2(r) ] \, dr ,\label{norma0}
\end{equation}
in terms of the radial functions $\widetilde{v}(r)$ and $\widetilde{u}(r)$. These functions satisfy the ordinary differential equations (\ref{edoDerrick}). In order to have $\| \xi(\vec{x},n,0)\|< \infty$  regularity conditions must be imposed on the functions $v(r)$ and $u(r)$ together with a fast enough decay when $r\rightarrow \infty$. First, we shall study the behavior of these functions in the neighborhood of $r=0$. In order to do that, we plug the expansion 
\begin{equation}
\widetilde{v}(r) = r^s \sum_{i=0}^\infty \widetilde{v}_i \, r^i \hspace{0.0cm}, \hspace{0.5cm} \widetilde{u}(r) = r^t \sum_{i=0}^\infty \widetilde{u}_i r^i, \label{powerseries0}
\end{equation}
into the equations which characterize the spectral problem for (\ref{operatorhbar0}). By construction, the coefficients $\widetilde{v}_0$ and $\widetilde{u}_0$ do not vanish and $s$ and $t$ are the orders of the power series for $\widetilde{v}(r)$ and $\widetilde{u}(r)$ respectively. Explicitly, at lowest order we obtain the conditions
\begin{eqnarray}
    (1-s^2) \widetilde{v}_0 r^{s-2} + [1-(s+1)^2] \widetilde{v}_1 r^{s-1} + O(r^{n_1})  &=& 0 ,\label{order0a} \\
    (n^2-t^2)\widetilde{u}_0 r^{t-2} + [n^2-(t+1)^2]\widetilde{u}_1 r^{t-1} + O(r^{n_2})&=& 0 ,\label{order0b}
\end{eqnarray}
where $n_1=\min\{s,n+t-1\}$ and $n_2=\min\{t,n+s-1\}$.


The existence of a non-trivial solution demands that the order of the power series (\ref{powerseries0}) must be set as
\[
s=1 \hspace{0.5cm}, \hspace{0.5cm} t=n \hspace{0.5cm} \mbox{with} \hspace{0.5cm} \widetilde{v}_0,\widetilde{u}_0\in \mathbb{R}\hspace{0.5cm} \mbox{and} \hspace{0.5cm} \widetilde{v}_1,\widetilde{u}_1=0,
\]
which besides maintains the consistency in the order of the powers shown in (\ref{order0a}) and (\ref{order0b}). The previous condition implies that the radial characteristic of the Derrick type modes is fixed as
\[
{\rm ch}_r[\xi_\mu(\vec{x},n,0)] = {1\choose n},
\]

which means that this solution is regular at the origin. As a consequence, the integrand of the norm (\ref{norma0}) has a good behavior at $r=0$: $2\pi r[\widetilde{v}^2(r)+\widetilde{u}^2(r)] \sim 2\pi r^3$.

On the other hand, the asymptotic behavior of the eigenfunctions can be obtained form the differential equations
\[
-\frac{d^2 \widetilde{v}(r)}{d r^2} - \frac{1}{r} \frac{d \widetilde{v}(r)}{d r} + \Big[ \frac{1}{r^2} + 1- \omega_n^2 \Big] \widetilde{v}(r)= 0 \hspace{0.3cm},\hspace{0.3cm} -\frac{d^2 \widetilde{u}(r)}{d r^2} - \frac{1}{r} \frac{d \widetilde{u}(r)}{d r} + \Big[ \lambda - \omega_n^2 \Big] \widetilde{v}(r)= 0,
\]
which are derived from the general equations (\ref{edoDerrick}) in this approximation. It can be shown that
\[
\widetilde{v}(r) \stackrel{r\rightarrow \infty}{\longrightarrow} \left\{ \begin{array}{lll}  \widehat {C}_1^{(v)} \, I_1(\sqrt{1-\omega_n^2}\, r) + \widehat{C}_2^{(v)} \, K_1(\sqrt{1-\omega_n^2}\, r), & \mbox{ if } \omega_n^2 <1, \\ C_1^{(v)} \, J_1(\sqrt{\omega_n^2-1}\, r) + C_2^{(v)} \, Y_1(\sqrt{\omega_n^2-1}\, r), & \mbox{ if } \omega_n^2 \geq 1,  \end{array} \right.
\]
and
\[
\widetilde{u}(r) \stackrel{r\rightarrow \infty}{\longrightarrow} \left\{ \begin{array}{lll}  \widehat {C}_1^{(u)} \, I_0(\sqrt{\lambda-\omega_n^2}\, r) +\widehat{C}_2^{(u)} \, K_0(\sqrt{\lambda-\omega_n^2}\, r), & \mbox{ if } \omega_n^2 <\lambda ,\\ C_1^{(u)} \, J_0(\sqrt{\omega_n^2-\lambda}\, r) +  C_2^{(u)} \, Y_0(\sqrt{\omega_n^2-\lambda}\, r), & \mbox{ if } \omega_n^2 \geq \lambda,  \end{array} \right.
\]
where $J_i$, $Y_i$, $I_i$ and $K_i$ stand for the usual modified Bessel functions. The shape modes (belonging to the discrete spectrum) are attained when the values of $v_0$ and $u_0$ are suitably chosen such that the constants $\widehat {C}_1^{(v)}$ and $\widehat {C}_1^{(u)}$ vanish in the asymptotic behavior. Therefore, for these eigenfunctions the norm (\ref{norma0}) is well-defined. 

\vspace{0.2cm}

\noindent 2. The case of the multipolar modes is more complicated. Now, the norm for these modes (\ref{genericform04b}) is given by
\begin{equation}
\| \xi(\vec{x},n,\overline{k})\|^2 = \int_0^\infty \pi r \Big[ u^2(r) + \overline{k}^2 w^2(r) + \left(v'(r)- r f_n(r) w(r)\right)^2 + \frac{\overline{k}^2 v^2(r)}{r^2}  \Big] \, dr, \label{norma1}
\end{equation}
in terms of the radial functions $v(r)$, $u(r)$ and $w(r)$. These functions satisfy the ordinary differential equations (\ref{genericform04d}).  Now, we substitute the power expansion near $r=0$
\[
v(r)=\sum_{i=0}^\infty v_i r^{s+i} \hspace{0.cm},\hspace{0.5cm} u(r)=\sum_{i=0}^\infty u_i r^{t+i} \hspace{0.cm},\hspace{0.5cm} w(r)=\sum_{i=0}^\infty w_i r^{m+i} ,
\]
into the equations which characterize the spectral problem. The first equation that we find is 
\begin{equation}
(\overline{k}^2 - s^2) v_0 r^{s-2}  + 2 d_0 w_0 (1+n) r^{n+m} + [\overline{k}^2 - (s+1)^2] v_1 r^{s-1} + O(r^{\widetilde{n}_1})  = 0 .\label{indicial1a}
\end{equation}
with $\widetilde{n}_1= \min\{s,n+m+1\}$. Note that this relation involves the lowest power $r^{s-2}$ and $r^{n+m}$. Since $s$ and $m$ are at this point unknown, it cannot yet be determined which term dominates at the lowest order. The second equation in this process reads as 
{
\begin{eqnarray}
&&(-t^2+ \overline{k}^2+n^2) u_0 r^{t-2} - 2n \overline{k}^2 w_0 r^{m-2} +2 n s d_0 v_0 r^{n+s-2} + (-(t+1)^2+ \overline{k}^2+n^2) u_1 r^{t-1} - \nonumber \\
&& \hspace{1cm} - 2n \overline{k}^2 w_1 r^{m-1}+ 2 nd_0 v_1(1+s) r^{n+s-1}  + O(r^{\widetilde{n}_2})  = 0, \label{indicial1b}
\end{eqnarray}
}
whereas the third one becomes
\begin{eqnarray}
	&& (-m^2 + \overline{k}^2 + n^2) w_0 r^{m-2} - 2n u_0 r^{t-2}  + 2 n d_0 v_0 r^{n+s-2} + (-(m+1)^2 + \overline{k}^2 + n^2) w_1 r^{m-1} - \nonumber \\
    &&  \hspace{1cm} - 2n u_1 r^{t-1} + 2 n d_0 v_1 r^{n+s-1} + O(r^{\widetilde{n}_2})  =0 , \label{indicial1c}
\end{eqnarray}
being $\widetilde{n}_2=\min \{t,m,n+2\}$. Note that (\ref{indicial1b}) and (\ref{indicial1c}) involve the lowest power $r^{t-2}$, $r^{m-2}$ and $r^{n+s-2}$, leading to the same situation as before, where it is not predetermined which term dominates at the lowest order. For this reason, we have to distinguish two different cases:

\noindent \textbf{Type A:} A first choice of the orders of the series that allows for non-trivial solutions to the recurrence relations (\ref{indicial1a}), (\ref{indicial1b}) and (\ref{indicial1c}) is given by 
\begin{equation}
s=\overline{k} \hspace{0.cm} , \hspace{0.5cm} t=n-\overline{k} \hspace{0.cm} , \hspace{0.5cm} m=n-\overline{k}. \label{optionAA}
\end{equation}
With this choice, the lowest order term in (\ref{indicial1a}) vanishes, and the subsequent term leads to the condition $v_1=0$. The remaining terms  determine then the higher-order coefficients starting from $v_0\in \mathbb{R}$ (and also $w_0$ from the expansion of the function $w(r)$). On the other hand, the equations (\ref{indicial1a}) and (\ref{indicial1b}) now take the form 
\begin{eqnarray*}
2 \overline{k}n(u_0-\overline{k}\, w_0)r^{n-\overline{k}-2}+[(-(n-\overline{k}+1)^2+\overline{k}^2+n^2)u_1-2n \overline{k}^2 w_1]r^{n-\overline{k}-1}+O(r^{n-\overline{k}})  &=& 0, \\
	-2n(u_0-\overline{k}\, w_0)^{n-\overline{k}-2} + [(-(n-\overline{k}+1)^2+\overline{k}^2+n^2)w_1-2n \overline{k}^2 u_1] r^{n-\overline{k}-1} + O(r^{n-\overline{k}})  &=& 0.
\end{eqnarray*}
These equations yield non-trivial solutions under the conditions $v_0,w_0\in \mathbb{R}$ and $u_0= \overline{k} \, w_0$. Additionally, the next terms in the previous relation imply that $u_1=w_1=0$ and the following coefficients in the power series are determined by the values of $u_0$ and $w_0$ in the recurrence equations. 

As a consequence of the previous analysis, the eigenfunctions constructed with the choice (\ref{optionAA}) exhibit the following radial characteristic
\[
{\rm ch}_r[\xi_\nu^{\rm A}(\vec{x},n,\overline{k})] = {\overline{k}-1 \choose n- \overline{k}},
\]
Clearly, this implies that the value of the parameter $\overline{k}$ is restricted to $\overline{k}=1,\dots, n$, since otherwise, one of the components of the vortex solution would fail to be regular at the origin. Note that these Type A multipolar modes in the self-dual case give rise to zero modes. 

\vspace{0.2cm}

\noindent \textbf{Type B:} A second choice of the orders of the series leading to non-trivial solutions is given by
\begin{equation}
s=\overline{k} \hspace{0.cm} , \hspace{0.5cm} t=n+\overline{k} \hspace{0.cm} , \hspace{0.5cm} m=n+\overline{k}, \label{optionBB}
\end{equation}
In this case the lowest order term in (\ref{indicial1a}) automatically vanishes, and the following one leads to $v_1=0$. The rest of the terms determine, as before, the higher-order coefficients starting from $v_0,w_0\in \mathbb{R}$. On the other hand, (\ref{indicial1b}) and (\ref{indicial1c}) read now 
\begin{eqnarray*}
&&\hspace{-0.4cm} -2 n \overline{k}(u_0+\overline{k} w_0-d_0 v_0) r^{n+\overline{k}-2}+\left[ \left(-(n+\overline{k}+1)^2+\overline{k}^2+n^2\right)u_1-2n \overline{k}^2 w_1+2 n d_0 v_1(1+\overline{k})\right]  r^{n+\overline{k}-1}+ \\
&& + O(r^{n+\overline{k}}) = 0
\end{eqnarray*}
and
\begin{eqnarray*}
&&\hspace{-0.4cm} -2n(u_0 +\overline{k} w_0- d_0 v_0) r^{n+\overline{k}-2}+ \left[ \left(-(n+\overline{k}+1)^2+\overline{k}^2+n^2 \right) w_1-2 n u_1+2n d_0 v_1\right] r^{n+\overline{k}-1} + \\
&& + O(r^{n+\overline{k}})= 0.
\end{eqnarray*}
Non-trivial regular modes can be obtained if we fix $v_0,w_0\in \mathbb{R}$ under the restriction $u_0=d_0v_0-\overline{k}w_0$. In addition, it can be checked that $u_1,w_1=0$. This analysis shows that the radial characteristic of the Type B multipolar eigenmodes is given by
\[
{\rm ch}_r[\xi_\nu^{\rm B}(\vec{x},n,\overline{k})] = {\overline{k}-1\choose n+\overline{k}},
\]
which does not introduce any restriction on the value of $\overline{k}\geq 1$. In the self-dual case this type of mode reproduces the shape modes with positive eigenvalues presented in \cite{AlonsoIzquierdo2016}.

The asymptotic behavior of the multipolar eigenfunctions are governed by the differential equation
\begin{eqnarray*}
	&& - \frac{d^2 v}{dr^2} - \frac{1}{r} \frac{dv}{dr}+ \Big[ \frac{\overline{k}^2}{r^2} + 1 \Big] v(r) + 2  w(r) = \omega_n^2 v(r),   \\
	&& - \frac{d^2 u}{dr^2} - \frac{1}{r} \frac{du}{dr}+ \Big[ \frac{\overline{k}^2}{r^2} + \lambda \Big] u(r) = \omega_n^2 u(r),  \\
	&& - \frac{d^2 w}{dr^2} - \frac{1}{r} \frac{dw}{dr}+ \Big[ \frac{\overline{k}^2}{r^2} + 1 \Big] w(r) = \omega_n^2 w(r)  ,
\end{eqnarray*}
The second and the third equations leads to 
\[
u(r) \stackrel{r\rightarrow \infty}{\longrightarrow} \left\{ \begin{array}{lll}  \widehat {C}_1^{(u)} \, I_{\overline{k}}(\sqrt{\lambda-\omega_n^2}\, r) +\widehat{C}_2^{(u)} \, K_{\overline{k}}(\sqrt{\lambda-\omega_n^2}\, r), & \mbox{ if } \omega_n^2 <\lambda, \\ C_1^{(u)} \, J_{\overline{k}}(\sqrt{\omega_n^2-\lambda}\, r) +  C_2^{(u)} \, Y_{\overline{k}}(\sqrt{\omega_n^2-\lambda}\, r), & \mbox{ if } \omega_n^2 \geq \lambda,  \end{array} \right.
\]
and
\[
w(r) \stackrel{r\rightarrow \infty}{\longrightarrow} \left\{ \begin{array}{lll}  \widehat {C}_1^{(w)} \, I_{\overline{k}}(\sqrt{1-\omega_n^2}\, r) + \widehat{C}_2^{(w)} \, K_{\overline{k}}(\sqrt{1-\omega_n^2}\, r), & \mbox{ if } \omega_n^2 <1, \\ C_1^{(w)} \, J_{\overline{k}}(\sqrt{\omega_n^2-1}\, r) + C_2^{(w)} \, Y_{\overline{k}}(\sqrt{\omega_n^2-1}\, r), & \mbox{ if } \omega_n^2 \geq 1.  \end{array} \right.
\]
{
Finally, the first of the equations provides the expression
\[
v(r) \stackrel{r\rightarrow \infty}{\longrightarrow} \left\{ \begin{array}{lll}  \widehat {C}_1^{(v)} \, I_{\overline{k}}(\sqrt{1-\omega_n^2}\, r) + \widehat{C}_2^{(v)} \, K_{\overline{k}}(\sqrt{1-\omega_n^2}\, r) + v_p(r),& \mbox{ if } \omega_n^2 <1, \\ C_1^{(v)} \, J_{\overline{k}}(\sqrt{\omega_n^2-1}\, r) + C_2^{(v)} \, Y_{\overline{k}}(\sqrt{\omega_n^2-1}\, r) + v_p(r), & \mbox{ if } \omega_n^2 \geq 1 , \end{array} \right.
\]
where the particular solution $v_p(r)$ involves the presence of Meijer functions whose contribution is negligible compared to the contribution of the homogeneous solution.  It can be clearly seen from the previous expressions that this problem has two different mass thresholds, one starting at $\omega^2=1$ and the other one starting at $\omega^2=\lambda$. In other words, if $\omega_n^2>\lambda$ $u(r)$ becomes an oscillatory function at infinity. The same phenomenon occurs for $w(r)$ and $v(r)$ for $\omega_n^2>1$. 

} As before, it is possible to find values for the constants $v_0$ and $w_0$ such that the solution exhibits asymptotic behavior where the constants $\widehat{C}_1^{u,u,w}$ are zero, ensuring that the eigenfunctions vanish at infinity. $\Box$

At this stage, we have outlined the complete spectrum of fluctuations for the non-BPS vortices. Specifically, we have shown that the modes with $\overline{k}=0$ correspond to Derrick-type modes. Additionally, the multipolar modes with $\overline{k}=1$ have a clear interpretation as the translational modes. This interpretation is confirmed by observing that an infinitesimal gauge-invariant translation along the $x$- and $y$-directions generates the general form of the two fundamental translational modes \cite{Manton2002,Tong2014}
\begin{equation}\label{ZeroModes}
\xi_{0,x}(\vec{x}) = \left( 
\begin{array}{c} 0  \\
F_{12}(\vec{x})  \\
\widetilde{D}_1\psi_1(\vec{x})\\
\widetilde{D}_1\psi_2(\vec{x}) \end{array} \right),  \qquad
\xi_{0,y}(\vec{x}) = \left( 
\begin{array}{c} -F_{12}(\vec{x})  \\
0  \\
\widetilde{D}_2\psi_1(\vec{x})\\
\widetilde{D}_2\psi_2(\vec{x})  \end{array} \right). 
\end{equation}
Expressing \eqref{ZeroModes} in terms of the vortex profiles $f_n(r)$ and $\beta_n(r)$, these modes become
    \begin{equation}\label{ZeroModesx}
\xi_{0,x}(\vec{x}) = \left( 
\begin{array}{c} 0  \\
\frac{n}{r}\frac{d \beta_n(r)}{dr} \\
\frac{n f_n(r)}{r}(1-\beta_n(r))\sin(\theta)\sin(n \theta)+ \frac{d f_n(r)}{dr}\cos(\theta)\cos(n \theta)\\
-\frac{n f_n(r)}{r}(1-\beta_n(r))\sin(\theta)\cos(n \theta)+ \frac{d f_n(r)}{dr}\cos(\theta)\sin(n \theta)\end{array} \right),  \qquad
\end{equation}
    \begin{equation}\label{ZeroModesy}
\xi_{0,y}(\vec{x}) = \left( 
\begin{array}{c} -\frac{n}{r}\frac{d \beta_n(r)}{dr}  \\
 0\\
-\frac{n f_n(r)}{r}(1-\beta_n(r))\cos(\theta)\sin(n \theta)+ \frac{d f_n(r)}{dr}\sin(\theta)\cos(n \theta)\\
\frac{n f_n(r)}{r}(1-\beta_n(r))\cos(\theta)\cos(n \theta)+ \frac{d f_n(r)}{dr}\sin(\theta)\sin(n \theta)\end{array} \right).  \qquad
\end{equation}
The eigenfunction $\xi_{0,x}(\vec{x})$, given by \eqref{ZeroModesx}, can be written in the form \eqref{genericform04b} by simply choosing
\begin{equation}\label{uuww}
    \overline{k}=1, \quad v(r)=-n\frac{d \beta_n(r)}{dr}, \quad u(r)=\frac{d f_n (r)}{d r}, \quad w(r)=\frac{n f_n(r)}{r}(1-\beta_n(r)).
\end{equation}
On the other hand, the eigenmode $\xi_{0,y}(\vec{x})$, determined by \eqref{ZeroModesy}, can be identified with the orthogonal eigenfunction of (\ref{genericform04b}), which  arise when the changes  $\cos(\overline{k}\theta)\rightarrow \sin(\overline{k}\theta)$ and $\sin(\overline{k}\theta)\rightarrow -\cos(\overline{k}\theta)$ are considered in its expression. By direct substitution, it can be verified that the functions given by (\ref{uuww}) satisfy the spectral conditions \eqref{genericform04d}. To this end, it is necessary to use the differential equations \eqref{edof} and \eqref{edobeta}, which, for instance, lead to the relation $\frac{d v(r)}{d r}-r f_n(r) w(r)=\frac{v(r)}{r}$, which is useful for simplifying the resulting expressions in this calculation. It is worth mentioning that these two zero eigenmodes can be directly constructed from the vortex profiles $f_n(r)$ and $\beta_n(r)$ for any value of the coupling constant $\lambda$. 

Now, we explicitly show how the fluctuation spectrum of BPS $n$-vortices (in the self-dual regime $\lambda=1$) emerges from the general framework presented herein. 
\begin{itemize}
\item Firstly, it is clear that the $2n$ Type A multipolar modes become the $2n$ zero modes in the BPS case. Recall that these multipolar modes with $\overline{k}=1$ correspond to translational zero modes for any value of the coupling constant $\lambda$. The remaining $2n-2$ multipolar modes, whose eigenvalues are generally nonzero, become degenerate at zero when $\lambda=1$. This behavior is illustrated in the graphical representations of the spectra displayed in Figures \ref{fig:espectra1}-\ref{fig:espectra5}. The correspondence between the BPS and non-BPS formulations (\ref{bpszeromode4}) and (\ref{genericform04b}) is established by applying the following identifications: 
\[
k=n-\overline{k}, \hspace{0.2cm} v(r)=- \frac{r^{n-k} }{n-k} h_{nk}(r), \hspace{0.2cm} u(r)=- \frac{r^{n-k-1}}{f_n(r)}  h_{nk}'(r), \hspace{0.2cm}  w(r)=- \frac{r^{n-k-1}}{(n-k)f_n(r)}  h_{nk}'(r) \, .
\]
\item The non-degenerate Derrick type mode $\overline{k}=0$ arising in the general case corresponds to the shape mode (\ref{bpsexcitedmode1}) with $k=0$ in the BPS regime. To establish this correspondence, we must identify
\[
\widetilde{v}(r)=\frac{d v_{n k}(r)}{d r} \hspace{0.4cm} \text{and } \hspace{0.4cm} \widetilde{u}(r)=f_n(r) v_{n k}(r) \, .
\]

\item Finally, the Type B multipolar modes determine the degenerate shape modes with $k=1,2,\dots$ in the self-dual case. Now, we have
\[
k= \overline{k}, \hspace{0.4cm}v(r)= v_{n k}(r), \hspace{0.4cm} u(r)=f_n(r) v_{n k}(r) \hspace{0.4cm} \text{and } \hspace{0.4cm} w(r)=0 \, .
\]
    
\end{itemize}

\begin{figure}[ht]
    \centering
\begin{tikzpicture}[node distance=2.8cm, scale=0.8, transform shape]

    \node (top) [block1] {Internal Modes };
    \node (left) [block2, below left of=top,text width=3cm] {Derrick type:\\
    $\overline{k}=0$};
    \node (left2) [block2, below  of=left] {$\widetilde{v}(r)$, $\widetilde{u}(r)$: $\omega^2>0$};
    \node (right) [block2, below right of=top,text width=3cm] {Multipolar:
    \\ $\overline{k}=1,2,3\dots$};
    \node (right2) [block2, below  of=right] {$v(r)$, $u(r)$, $w(r)$};
    \node (rightA) [block4, below left of=right2] {Type A:  ($\overline{k}<n$)};
    \node (rightB) [block4, below right of=right2] {Type B: $\omega^2>0$};
    \node (rightAA) [block5, below left of=rightA, text width=3cm] {
        $\overline{k}=1$: $\omega^2=0$\\
        Translational modes
    };
    \node (rightBB) [block5, below right of=rightA] {
        $\overline{k}>1$
    };
    \node (rightBB1) [block3, below right of=rightBB, text width=1.5cm]{
    $\lambda>1$ $\omega^2<0$
    };
    \node (rightBB2) [block3, below left of=rightBB1, text width=1.5cm]{$\lambda=1$ $ \omega^2=0$};
    \node (rightBB3) [block3, below left  of=rightBB, text width=1.5cm]{$\lambda<1$ $\omega^2>0$};

    \draw [arrow] (top) -- (left);
    \draw [arrow] (top) -- (right);
    \draw [arrow] (right) -- (right2);
    \draw [arrow] (left) -- (left2);
    \draw [arrow] (right2) -- (rightA);
    \draw [arrow] (right2) -- (rightB);
    \draw [arrow] (rightA) -- (rightAA);
    \draw [arrow] (rightA) -- (rightBB);
    \draw [arrow] (rightBB) -- (rightBB1);
    \draw [arrow] (rightBB) -- (rightBB2);
    \draw [arrow] (rightBB) -- (rightBB3);
 
\end{tikzpicture}
\caption{Classification of the internal modes of the $n$-vortex in the Abelian-Higgs model.}
\label{Fig:Diagram}
\end{figure}

  Finally we summarize all the results of this section. First of all, all internal modes can be classified into two categories: Derrick type modes and multipolar modes. Every multipolar mode is double degenerate. Derrick type modes preserve the angular symmetry of the vortex and depend on two functions $\widetilde{u}(r)$ and $\widetilde{v}(r)$, while multipolar do not preserve the aforementioned symmetry and depend on three functions $v(r)$, $u(r)$ and $w(r)$.  Among multipolar modes, these can be of Type $A$ or Type $B$, the former ones are always associated with a positive eigenvalue. The Type $A$ eigenfunctions with $\overline{k}=1$ correspond to translational eigenmodes for any value of $\lambda$. The rest of the modes of this type corresponds to unstable modes for $\lambda>1$, zero modes for the BPS limit and stable modes for $\lambda<1$, see Fig. \ref{Fig:Diagram}.


\section{Vortex fluctuation spectra: numerical results}\label{num_spec}

In this section, we explicitly determine the small fluctuation spectrum for non-BPS vortices with the lowest vorticities. The method we employ is general and can be extended to higher vorticities. The analysis introduced in Section 4 reduces the problem to solving the eigenvalue equation associated with the operators (\ref{operatorhbar0}) and (\ref{hbarrak}). In fact, it can be verified that the operator \eqref{hbarrak} can also be used to obtain the spectrum of \eqref{operatorhbar0}. Specifically, by setting $\overline{k}=0$ in the first of these operators and applying the transformations $\widetilde{v}(r)=\frac{d v(r)}{dr}- r f_n(r) w(r)$ and $\widetilde{u}(r)= u(r)$, the spectrum of the second operator can be recovered. Consequently, the entire vortex fluctuation spectrum can be obtained by solving the eigenvalue problem for the operator \eqref{hbarrak}.

For the numerical analysis, we adopt the procedure described in \cite{AlonsoIzquierdo2016}, now adapted to account for the fact that the relevant operator in our case has a $3\times 3$ matrix structure. The details of the numerical algorithm used in this work are explicitly shown in Appendix \ref{appen}. It is worth noting that the approach presented in this article significantly reduces computational demands. For instance, using a $2000$-point mesh per dimension, a brute-force diagonalization of the original spectral problem (\ref{hessianoperator}) would require handling matrices of size $16,000,000\times 16,000,000$. However, our analysis reduces this to matrices of size $6000\times 6000$, which are computationally feasible on a conventional computer. The eigenvalue problem has been solved on the interval $[0,r_{max}]$ using $N=2000$ points, with $r_{max}=20$.

\begin{figure}[h!]
{\small\begin{tabular}{c}\includegraphics[height=4.2cm]{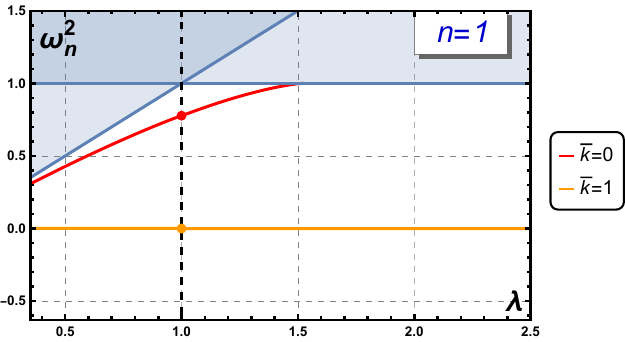} \end{tabular}\hspace{0.9cm} 
\begin{tabular}{|c|c|c|c|c|c|} \hline
\multicolumn{6}{|c|}{Case $n=1$} \\ \hline\hline
$\overline{k}$ & $\lambda=0.6$ & $\lambda=0.8$ & $\lambda=1.0$ & $\lambda=1.2$ & $\lambda=1.4$ \\ \hline\hline
$0$ & $0.50432$ & $0.64895$ & $0.77741$ & $0.88686$ & $0.97062$ \\ \hline
$1$ & $\sim 0.0$ & $\sim 0.0$ & $\sim 0.0$ & $\sim 0.0$ & $\sim 0.0$ \\ \hline
\end{tabular}}
\caption{Spectral structure for the rotationally invariant $1$-vortex as a function of the coupling constant $\lambda$: (a) Graphical representation and (b) Table of data for some values of $\lambda$. }  
\label{fig:espectra1}
\end{figure}

\begin{figure}[h!]
{\footnotesize\begin{tabular}{c}\includegraphics[height=4.2cm]{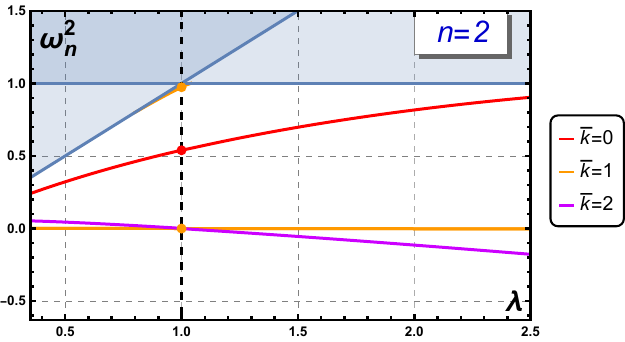} \end{tabular}\hspace{0.5cm} 
\begin{tabular}{|c|c|c|c|c|c|} \hline
\multicolumn{6}{|c|}{Case $n=2$} \\ \hline\hline
$\overline{k}$ & $\lambda=0.6$ & $\lambda=0.8$ & $\lambda=1.0$ & $\lambda=1.2$ & $\lambda=1.4$ \\ \hline\hline
$0$ &  $0.37130$ &  $0.46095$ & $0.53926$  & $0.60734$  &  $0.66932$  \\ \hline
$1$ & $\sim 0.0$ & $\sim 0.0$ & $\sim 0.0$ & $\sim 0.0$ & $\sim 0.0$ \\ 
    & - & $0.79905$ & $0.97338$ & - & - \\ \hline
$2$ & $0.037040$  & $0.020285$ &  $\sim 0.0$ &  $-0.021487$ & $-0.043283$  \\ \hline
\end{tabular}}
\caption{Spectral structure for the rotationally invariant $2$-vortex as a function of the coupling constant $\lambda$: (a) Graphical representation and (b) Table of data for some values of $\lambda$. }  
\label{fig:espectra2}
\end{figure}

\begin{figure}[h!]
{\footnotesize\begin{tabular}{c}\includegraphics[height=4.2cm]{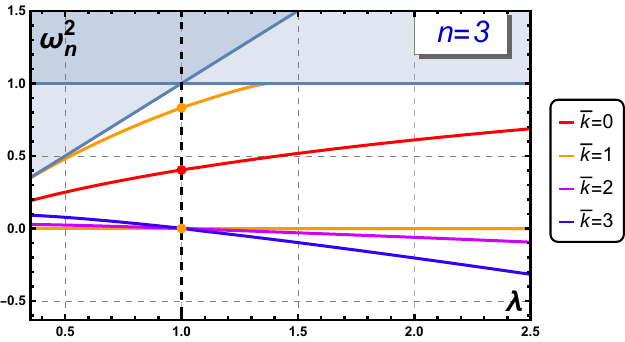} \end{tabular}\hspace{0.5cm} 
\begin{tabular}{|c|c|c|c|c|c|} \hline
\multicolumn{6}{|c|}{Case $n=3$} \\ \hline\hline
$\overline{k}$ & $\lambda=0.6$ & $\lambda=0.8$ & $\lambda=1.0$ & $\lambda=1.2$ & $\lambda=1.4$ \\ \hline\hline
$0$ & $0.28543$ & $0.34829$ &  $0.40373$  &  $0.45147$ & $0.49647$  \\ \hline
$1$ & $\sim 0.0$ & $\sim 0.0$ & $\sim 0.0$ & $\sim 0.0$ & $\sim 0.0$ 
\\ 
  & $0.56129$ & $0.70650$ & $0.83152$ & $0.93405$ &  - \\ \hline
$2$ & $0.019978$  & $0.010882$ & $\sim 0.0$ & $-0.011997$ & $-0.023537$ \\ \hline
$3$ & $0.064644$ & $0.035084$ & $\sim 0.0$ & $-0.037734$ & $-0.076565$ \\ \hline
\end{tabular}}
\caption{Spectral structure for the rotationally invariant $3$-vortex as a function of the coupling constant $\lambda$: (a) Graphical representation and (b) Table of data for some values of $\lambda$. }  
\label{fig:espectra3}
\end{figure}

\begin{figure}[h!]
{\footnotesize\begin{tabular}{c}\includegraphics[height=4.2cm]{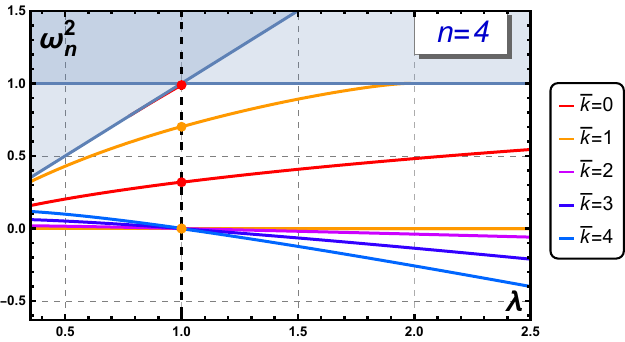} \end{tabular}\hspace{0.4cm} 
\begin{tabular}{|c|c|c|c|c|c|} \hline
\multicolumn{6}{|c|}{Case $n=4$} \\ \hline\hline
$\overline{k}$ & $\lambda=0.6$ & $\lambda=0.8$ & $\lambda=1.0$ & $\lambda=1.2$ & $\lambda=1.4$ \\ \hline\hline
$0$ & $0.22982$ & $0.27775$ & $0.31893$  & $0.35670$ &  $0.39156$ \\
    & - & $0.79934$ & $0.98820$ & - & -  \\ \hline
$1$ & $\sim 0.0$ & $\sim 0.0$ & $\sim 0.0$ & $\sim 0.0$ & $\sim 0.0$ 
\\ 
  & $0.49289$ & $0.60572$ & $0.70103$ & $0.78497$ &  $0.85835$ \\ \hline
$2$ & $0.013308$  & $0.0073629$ & $\sim 0.0$ & $-0.0076856$ & $-0.015048$ \\ \hline
$3$ & $0.043027$ & $0.023293$ & $\sim 0.0$ & $-0.025528$ & $-0.051394$ \\ \hline
$4$ & $0.082528$  & $0.044591$ & $\sim 0.0$ & $-0.048017$ & $-0.097455$ \\ \hline
\end{tabular}}
\caption{Spectral structure for the rotationally invariant $4$-vortex as a function of the coupling constant $\lambda$: (a) Graphical representation and (b) Table of data for some values of $\lambda$. }  
\label{fig:espectra4}
\end{figure}

\begin{figure}[h!]
{\scriptsize\begin{tabular}{c}\includegraphics[height=4.2cm]{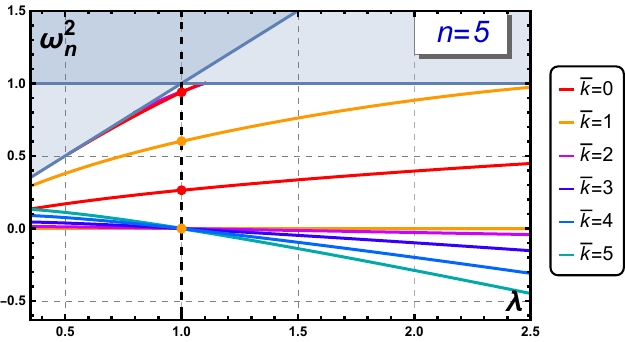} \end{tabular}\hspace{0.6cm} 
\begin{tabular}{|c|c|c|c|c|c|} \hline
\multicolumn{6}{|c|}{Case $n=5$} \\ \hline\hline
$\overline{k}$ & $\lambda=0.6$ & $\lambda=0.8$ & $\lambda=1.0$ & $\lambda=1.2$ & $\lambda=1.4$ \\ \hline\hline
$0$ & $0.19169$  & $0.23024$  &  $0.26394$  & $0.29389$  &  $0.32217$  \\
    & $0.59629$ & $0.77644$  & $0.93860$  & - & -  \\ \hline
$1$ & $\sim 0.0$ & $\sim 0.0$ & $\sim 0.0$ & $\sim 0.0$ & $\sim 0.0$ \\  
& $0.43131$  & $0.52295$  &  $0.60156$ &  $0.67009$ &  $0.73243$  \\ \hline
$2$ &  $0.0097114$ & $0.0053945$ & $\sim 0.0$ & $-0.0055940$ & $-0.010833$ \\
    & - & -  &  $0.94291$ & - &  - \\ \hline
$3$ &  $0.031756$ & $0.017279$ & $\sim 0.0$ &  $-0.018512$ &  $-0.037296$ \\ \hline
$4$ &  $0.063050$  & $0.034015$  & $\sim 0.0$ & $-0.037165$  & $-0.075204$  \\ \hline
$5$ &  $0.092745$  & $0.050039$  & $\sim 0.0$ &  $-0.053929$ & $-0.10943$  \\ \hline
\end{tabular}}
\caption{Spectral structure for the rotationally invariant $5$-vortex as a function of the coupling constant $\lambda$: (a) Graphical representation and (b) Table of data for some values of $\lambda$. }  
\label{fig:espectra5}
\end{figure}

The spectrum obtained for vortices with winding number $n=1,2,3,4,5$ can be found in Figures \ref{fig:espectra1}-\ref{fig:espectra5}. As expected, a vortex with $n=1$ does not posses any unstable eigenmode. Nevertheless, for $n>1$ and $\lambda>1$ every vortex has $n-1$ negative eigenvalues, which are responsible for their instability. It should be noted that for $\overline{k}>0$, every eigenvalue is doubly degenerate. Also, recall that for every value of $\lambda$ every vortex has two translational eigenmodes with $\overline{k}=1$.

Additionally, it can be seen that the unstable eigenmodes become zero modes in the BPS limit for every vortex with $\overline{k}>0$. It is also important to note that, as the vorticity grows, the potentials found in the spectral problem \eqref{numeric}-\eqref{numericbc} become deeper, which is the reason why it is possible to find more Type B eigenmodes for higher vorticities. For example, for $n=1$ only one Type B eigenmode can be found (see Figure \ref{fig:espectra1}). However, for $n=5$ five different Type B internal modes are present in the spectrum (see Figure \ref{fig:espectra5}) 

{ In the next subsections, the eigenfunctions corresponding to the spectra found will be displayed. The effects of these eigenfunctions on the vortex and its corresponding energy density will also be discussed. }


\subsection{Type A internal modes}

\begin{table}[h!]
    \centering
    \begin{tabular}{|c|c|c|}
 
      \hline
 \multirow{3}{0.3cm}{ \rotatebox{90}{$n=2$ \hspace{2cm}$\overline{k}=1$ } } 
              &
          \rotatebox{90}{\hspace{-0.5cm}$\lambda=0.6$} &
        \begin{minipage}{0.5\textwidth}
        \centering          
        \includegraphics[width=0.31\textwidth]{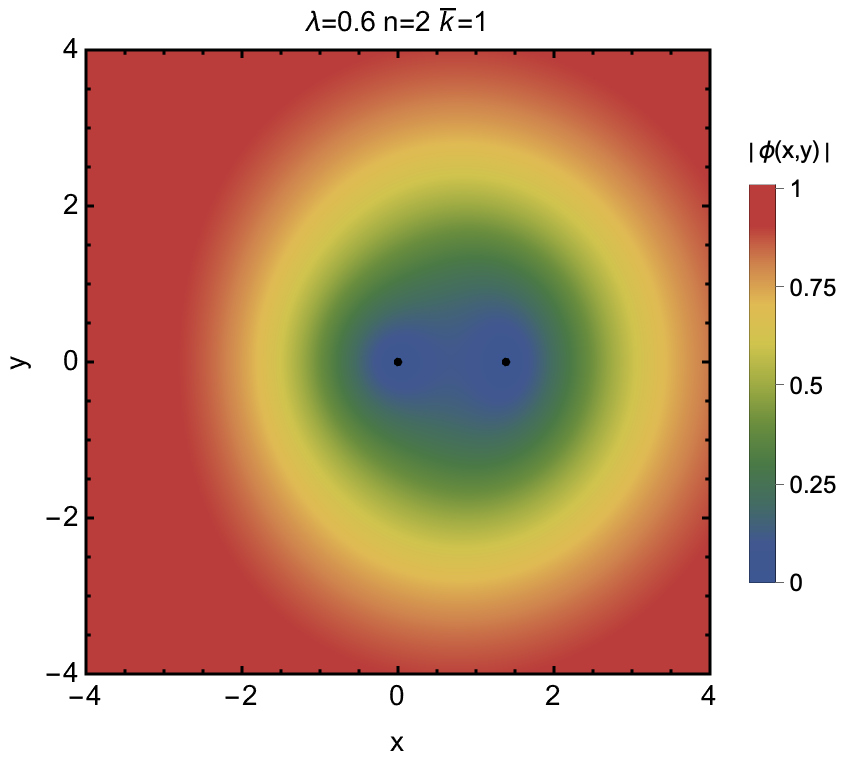}  \includegraphics[width=0.32\textwidth]{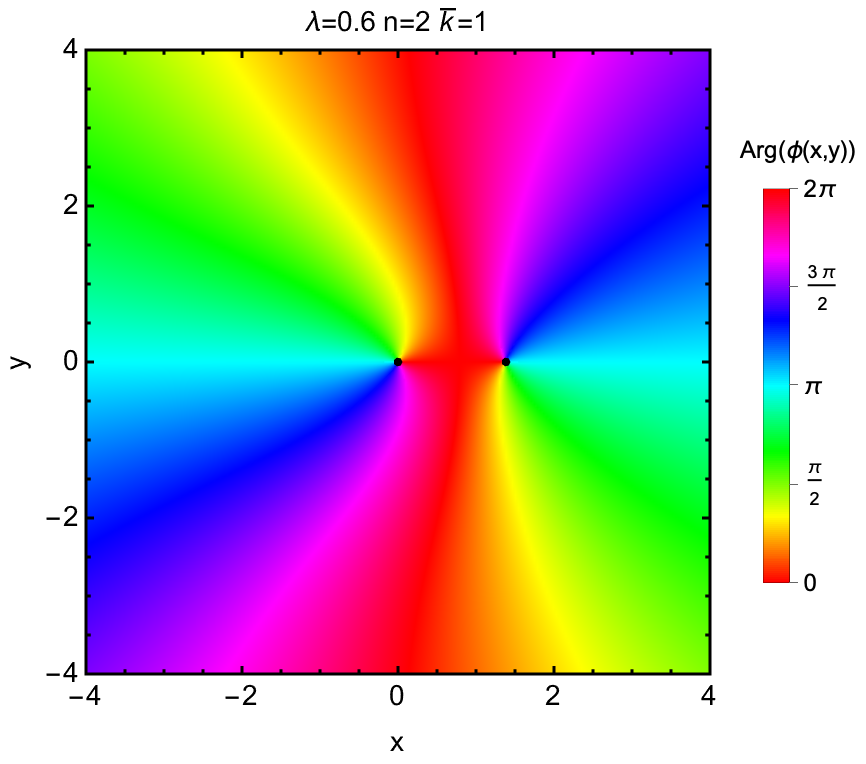}
        \raisebox{0.3\height}{\includegraphics[width=0.32\textwidth]{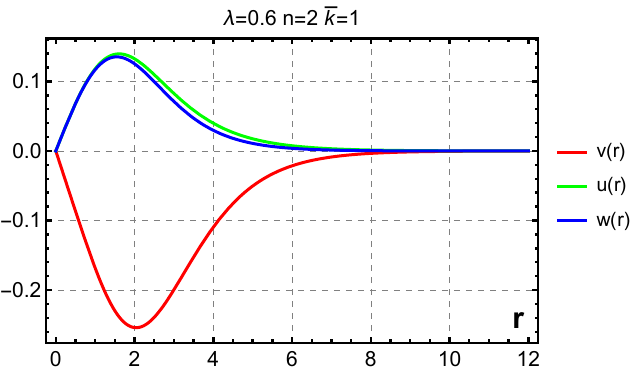}}
        \end{minipage}
     
\\
    &

          \rotatebox{90}{\hspace{-0.3cm}$\lambda=1$} &
        \begin{minipage}{0.5\textwidth}
        \centering                     
        \includegraphics[width=0.31\textwidth]{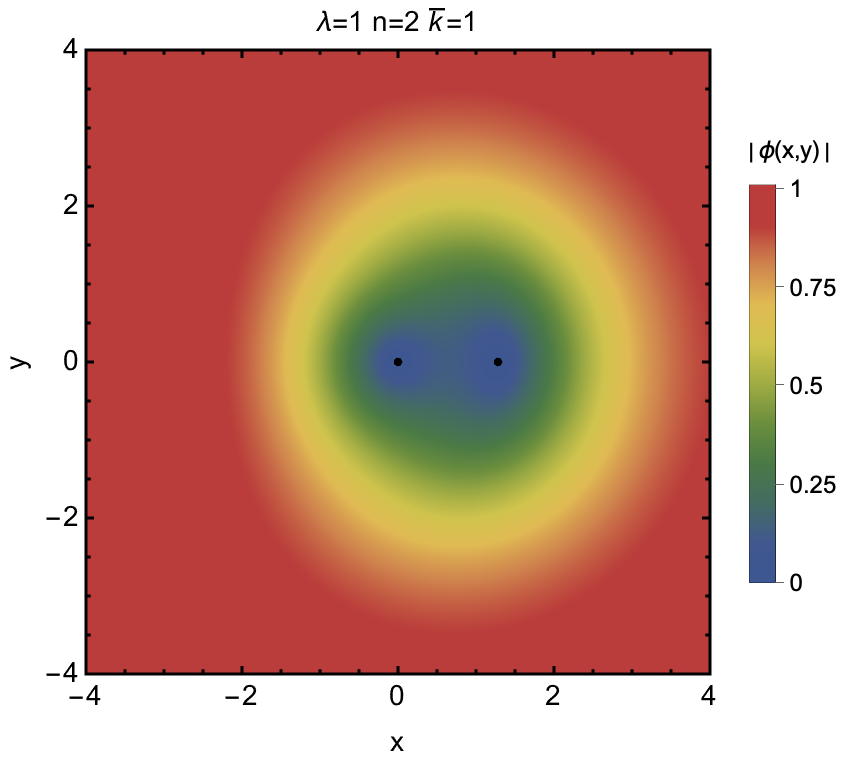}  \includegraphics[width=0.32\textwidth]{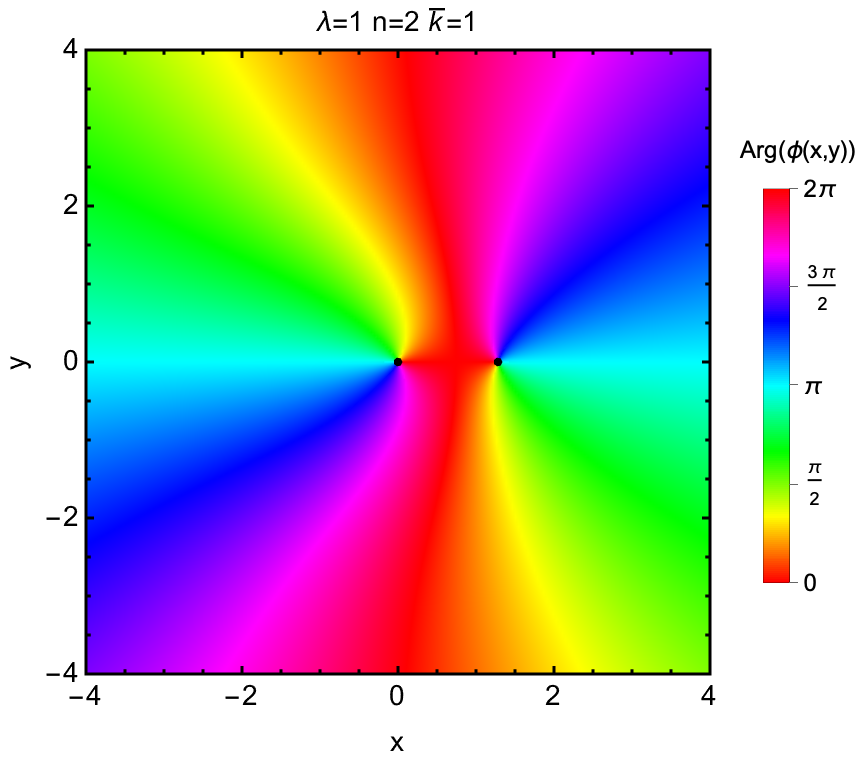}
        \raisebox{0.3\height}
       {\includegraphics[width=0.32\textwidth]{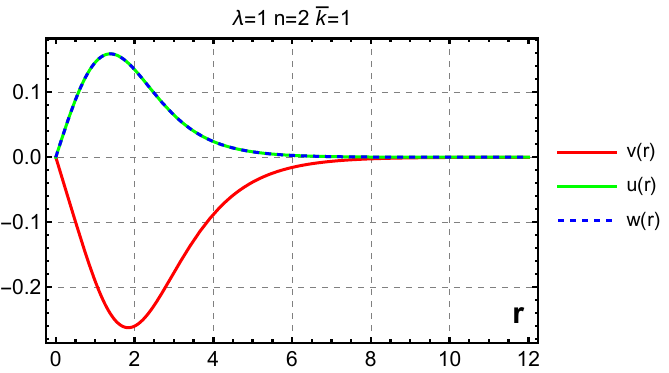}}
        \end{minipage}
 
\\
       &

          \rotatebox{90}{\hspace{-0.5cm}$\lambda=1.4$} &
        \begin{minipage}{0.5\textwidth}
        \centering                         
        \includegraphics[width=0.31\textwidth]{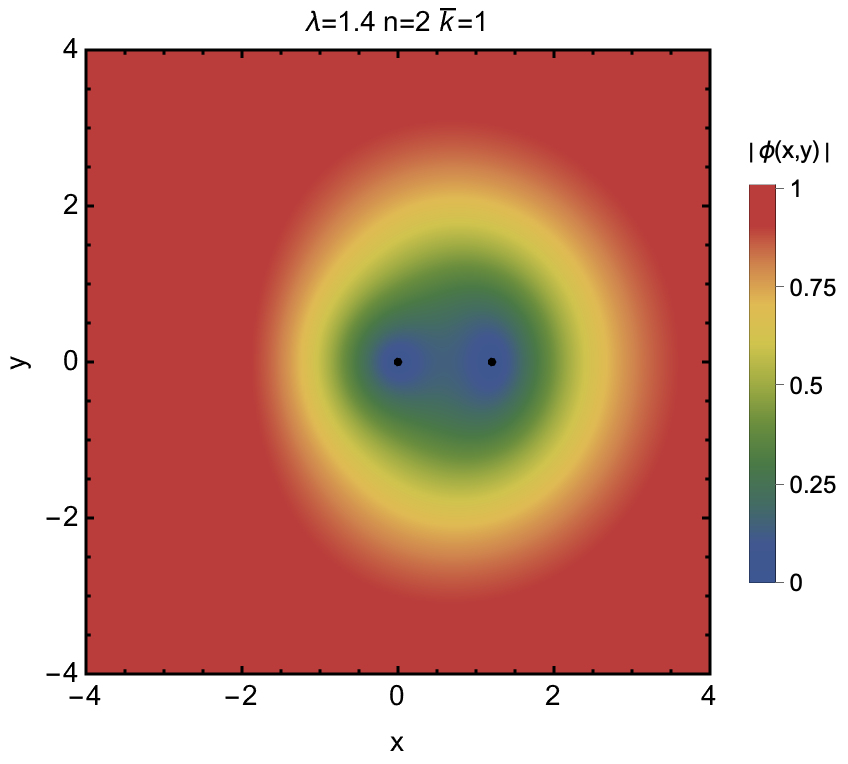}  \includegraphics[width=0.32\textwidth]{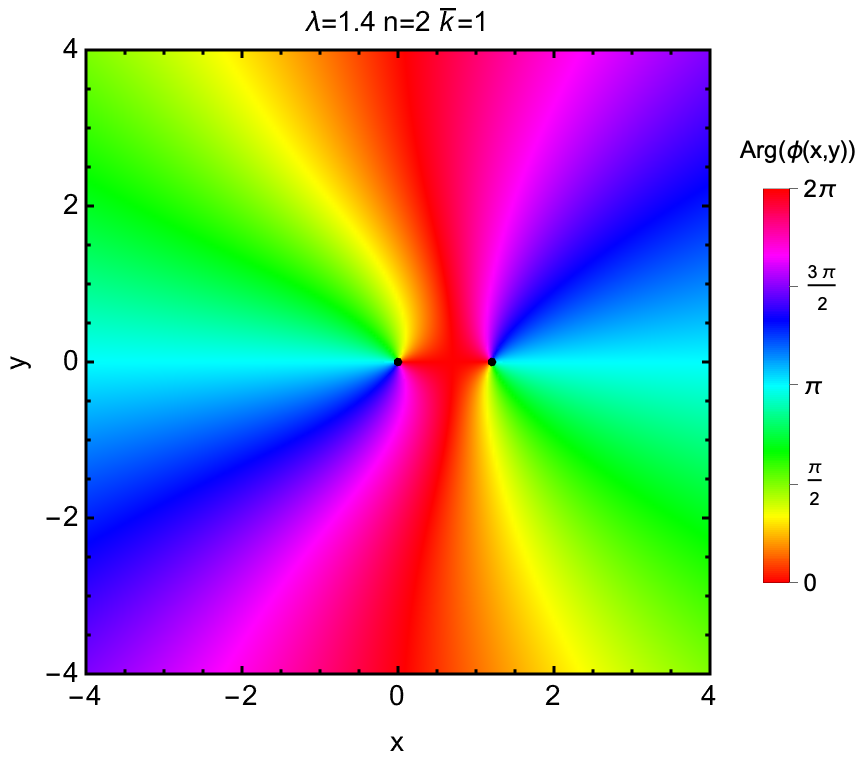}
        \raisebox{0.3\height}{\includegraphics[width=0.32\textwidth]{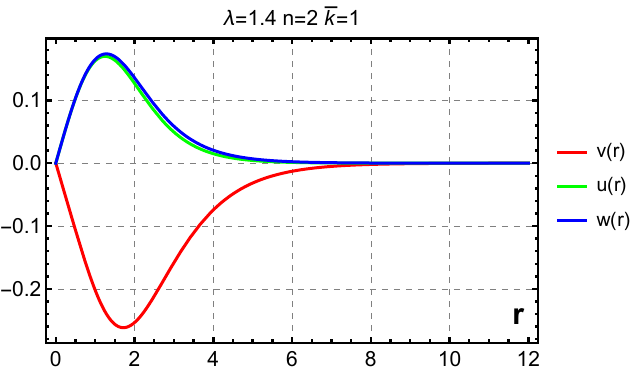}}
        \end{minipage}
 
\\

      \hline
 \multirow{3}{0.3cm}{ \rotatebox{90}{$n=2$ \hspace{2cm}$\overline{k}=2$ }} &      
          \rotatebox{90}{\hspace{-0.5cm}$\lambda=0.6$} &
        \begin{minipage}{0.5\textwidth}
        \centering
 \includegraphics[width=0.31\textwidth]{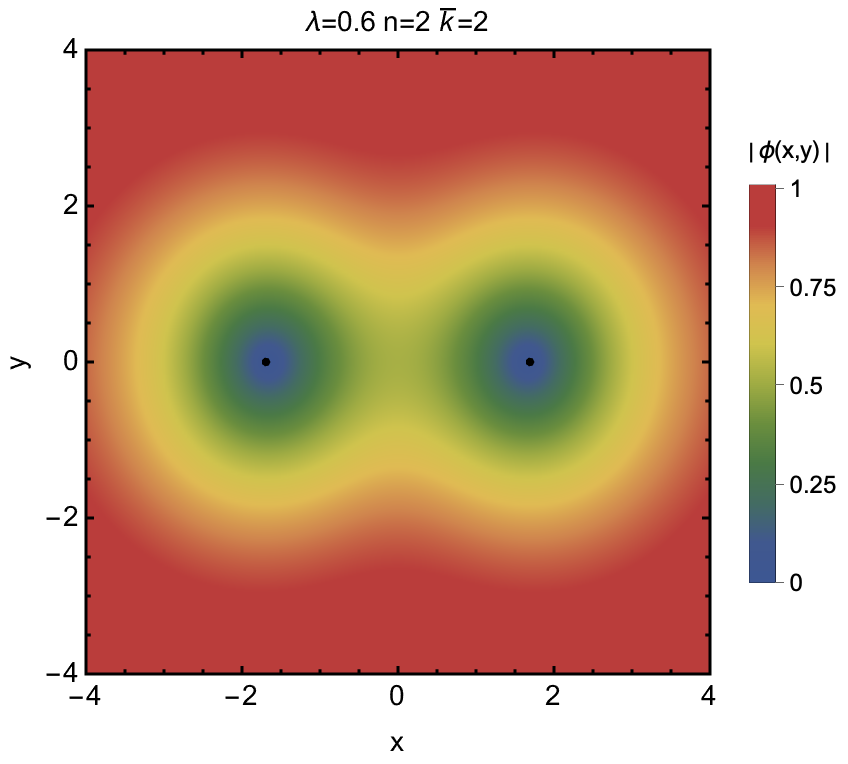}
        \includegraphics[width=0.32\textwidth]{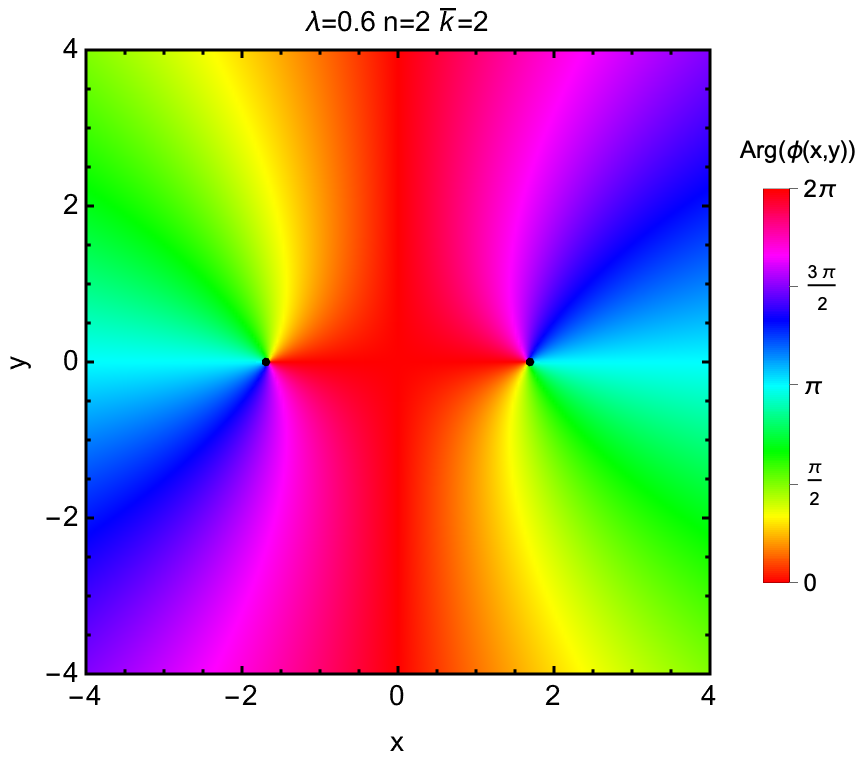}
    \raisebox{0.3\height}{\includegraphics[width=0.32\textwidth]{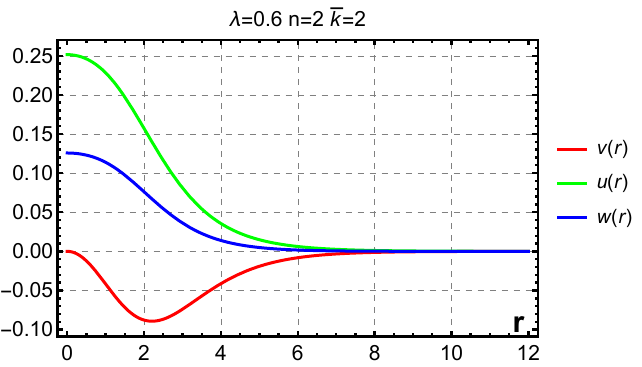}}
        \end{minipage}
\\
             &

          \rotatebox{90}{\hspace{-0.3cm}$\lambda=1$} &

        \begin{minipage}{0.5\textwidth}
        \centering
 \includegraphics[width=0.31\textwidth]{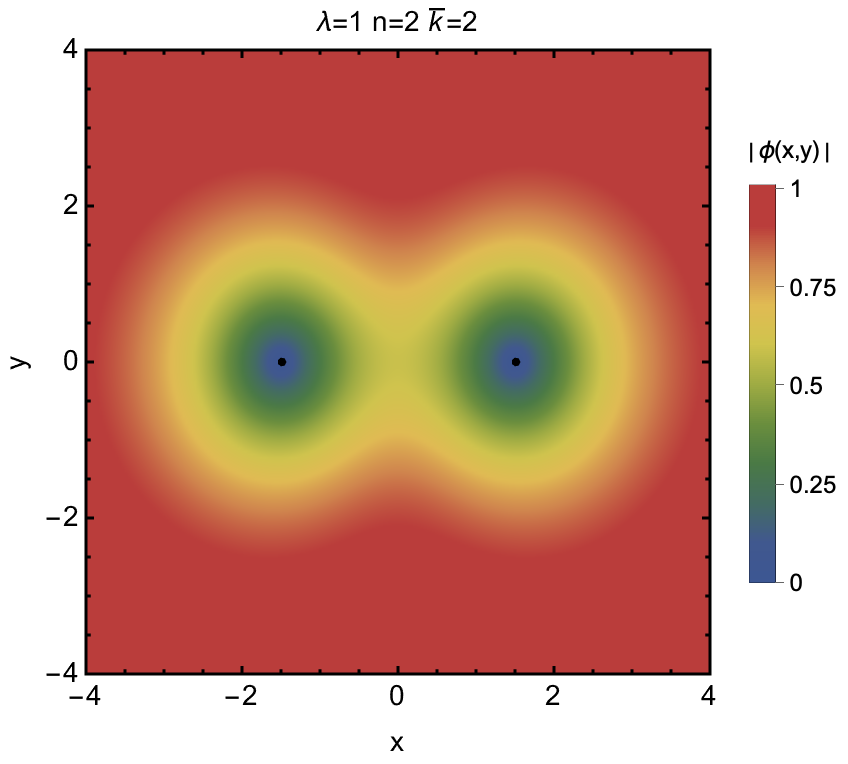}
        \includegraphics[width=0.32\textwidth]{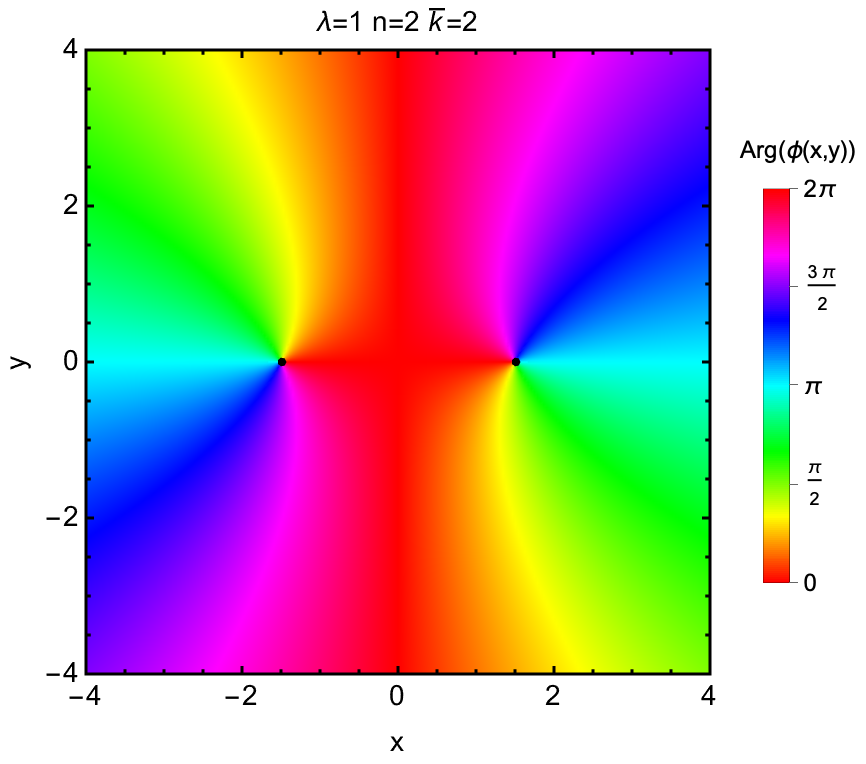}
    \raisebox{0.3\height}{\includegraphics[width=0.32\textwidth]{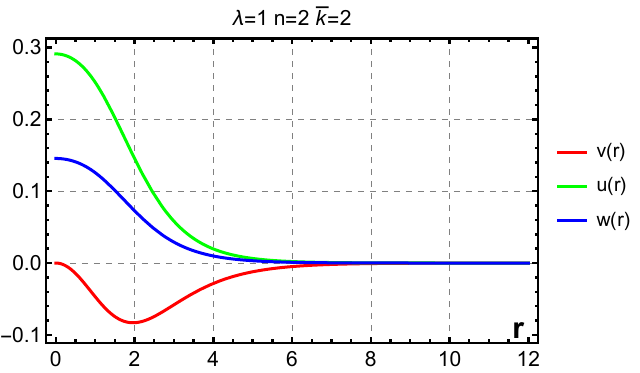}}
        \end{minipage}
\\
        &

          \rotatebox{90}{\hspace{-0.5cm}$\lambda=1.4$} &

        \begin{minipage}{0.5\textwidth}
        \centering
 \includegraphics[width=0.31\textwidth]{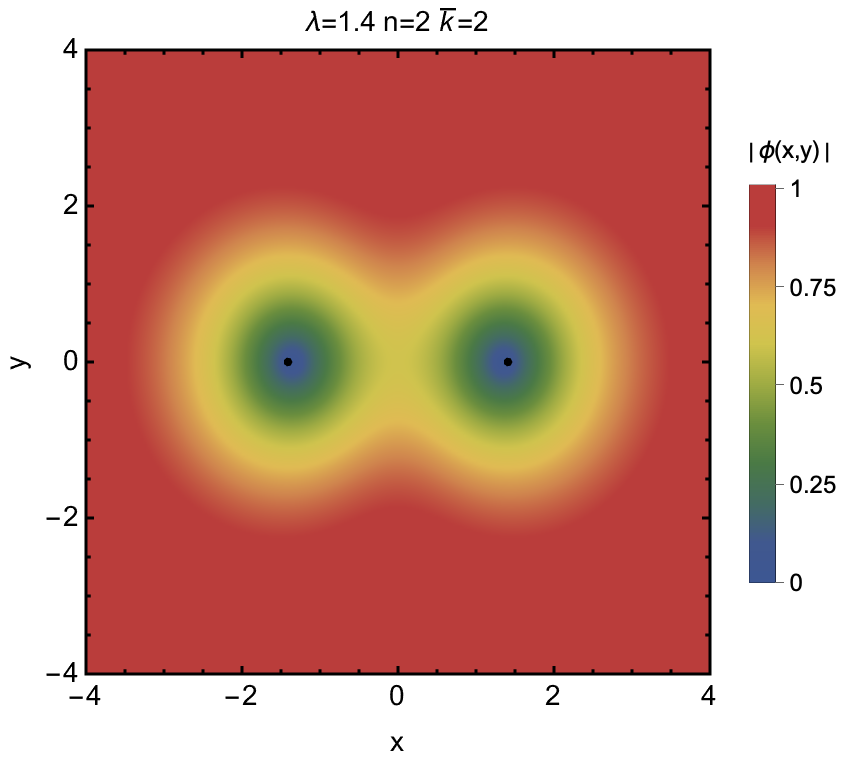}
        \includegraphics[width=0.32\textwidth]{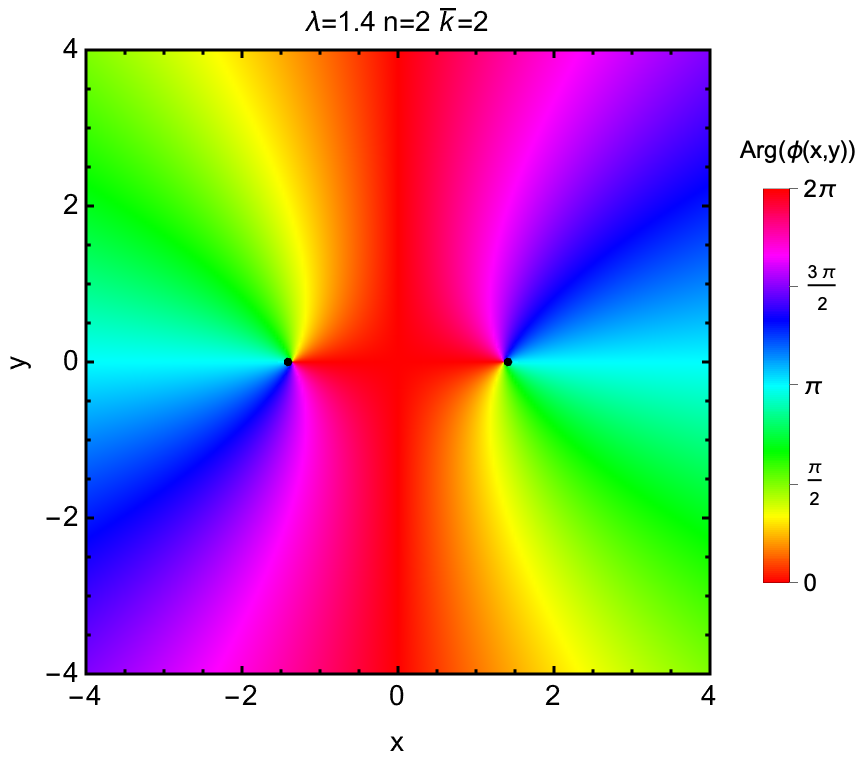}
    \raisebox{0.3\height}{\includegraphics[width=0.32\textwidth]{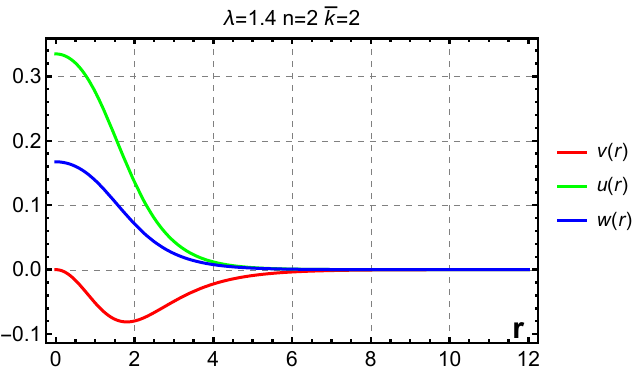}}
        \end{minipage}
\\
        \hline
    \end{tabular}
\caption{Type A internal modes for different values of $\lambda$ and $n=2$. The plots correspond to $|\phi(x,y)|$, $\mathrm{Arg}(\phi(x,y))$ once the internal modes have been added to the static vortex configurations and the profile functions $v(r)$, $u(r)$ and $w(r)$.}
    \label{Tab1:AN2}
\end{table}

\begin{table}[h!]
    \centering
        \vspace{-1cm}
    \begin{tabular}{|c|c|c|}
    \hline
 \multirow{3}{0.3cm}{ \rotatebox{90}{$n=3$ \hspace{2cm}$\overline{k}=1$ } } 
              &

      \rotatebox{90}{\hspace{-0.5cm}$\lambda=0.6$} &
        \begin{minipage}{0.5\textwidth}
        \centering
                            
        \includegraphics[width=0.31\textwidth]{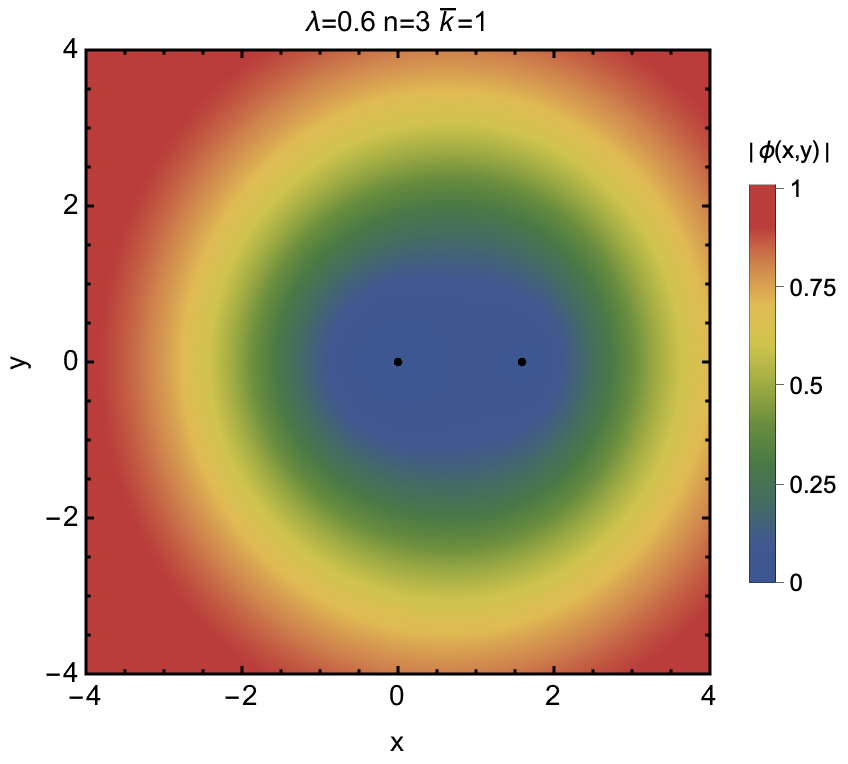}  \includegraphics[width=0.32\textwidth]{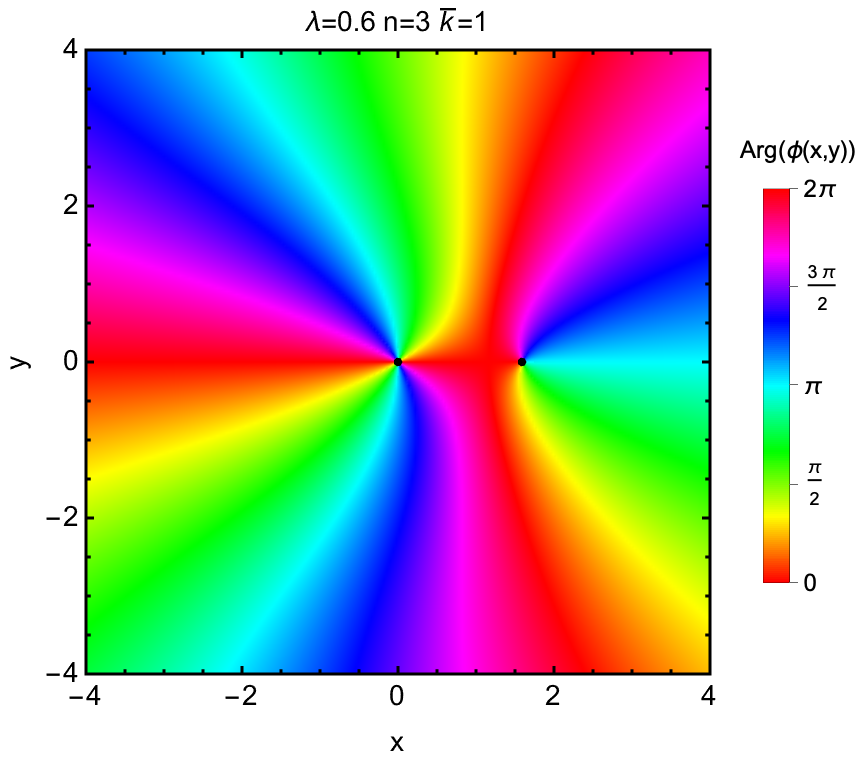}
        \raisebox{0.3\height}{\includegraphics[width=0.32\textwidth]{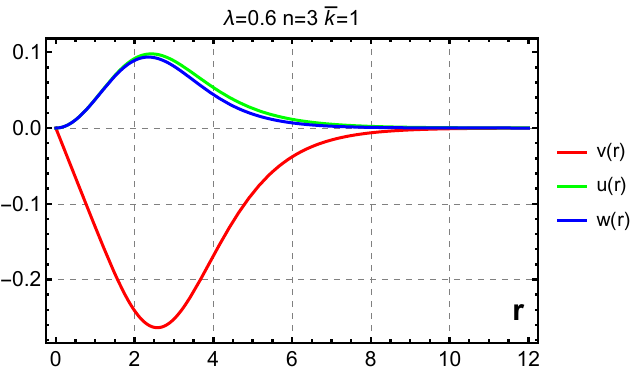}}
        \end{minipage}

\\
            & 

      \rotatebox{90}{\hspace{-0.3cm}$\lambda=1$} &
        \begin{minipage}{0.5\textwidth}
        \centering
                            
        \includegraphics[width=0.31\textwidth]{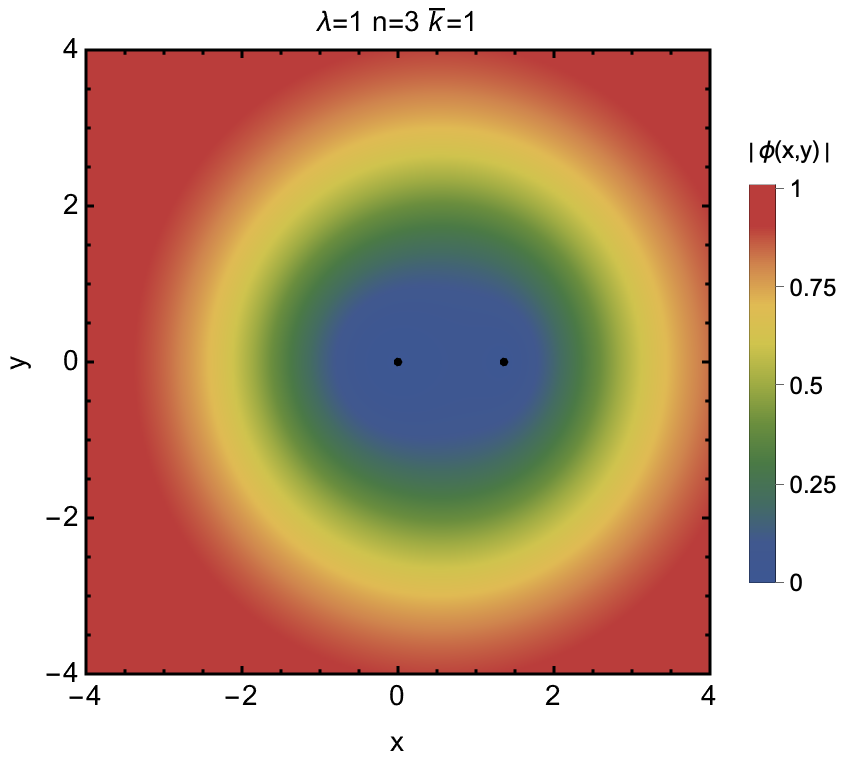}  \includegraphics[width=0.32\textwidth]{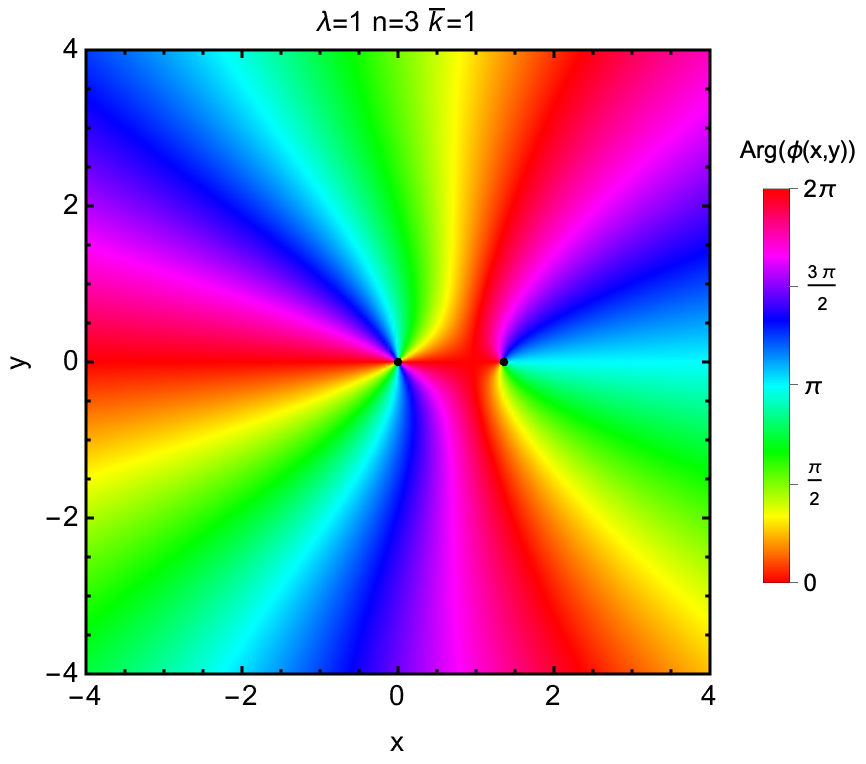}
        \raisebox{0.3\height}{\includegraphics[width=0.32\textwidth]{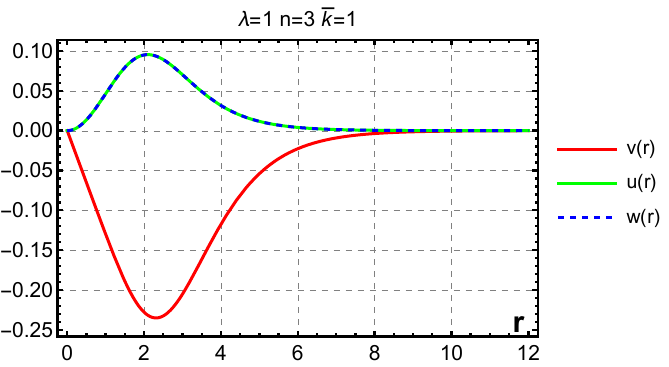}}
        \end{minipage}
 
\\
     & 

      \rotatebox{90}{\hspace{-0.5cm}$\lambda=1.4$} &
        \begin{minipage}{0.5\textwidth}
        \centering
                            
        \includegraphics[width=0.31\textwidth]{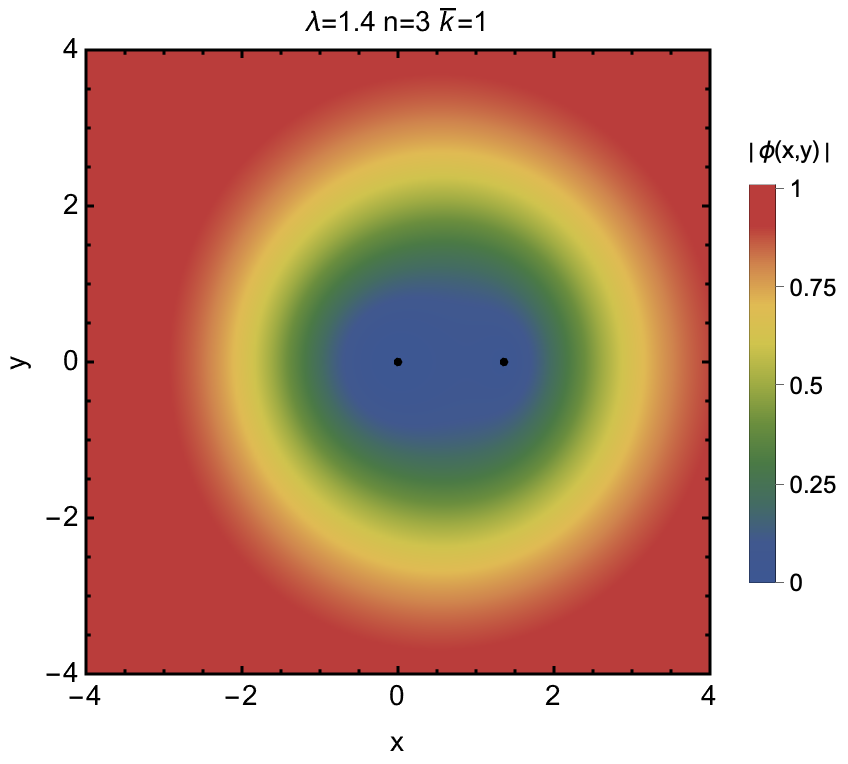}  \includegraphics[width=0.32\textwidth]{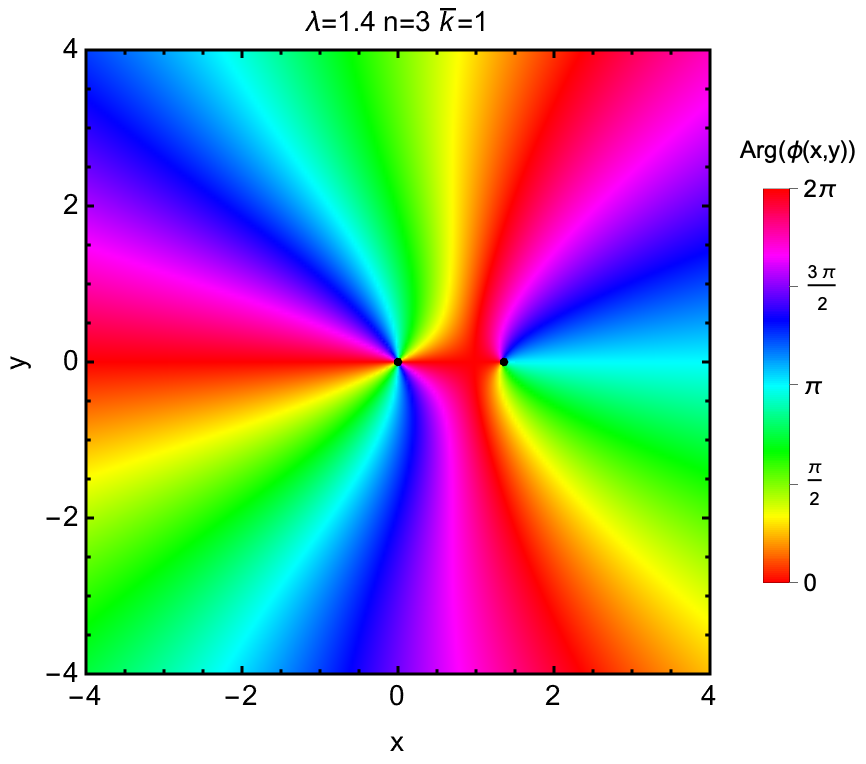}
        \raisebox{0.3\height}{\includegraphics[width=0.32\textwidth]{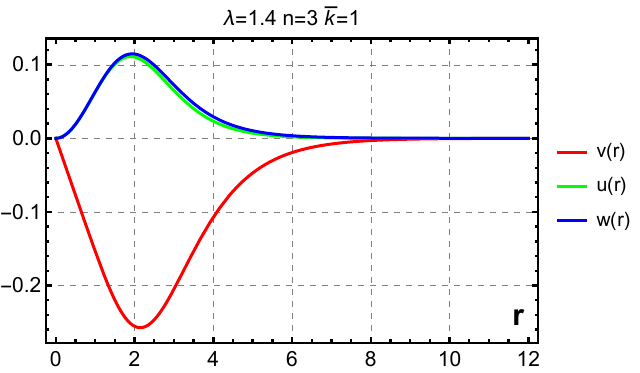}}
        \end{minipage}

\\
           \hline
 \multirow{3}{0.3cm}{ \rotatebox{90}{$n=3$ \hspace{2cm}$\overline{k}=2$ } } 
&
      \rotatebox{90}{\hspace{-0.5cm}$\lambda=0.6$} &

        \begin{minipage}{0.5\textwidth}
        \centering
 \includegraphics[width=0.31\textwidth]{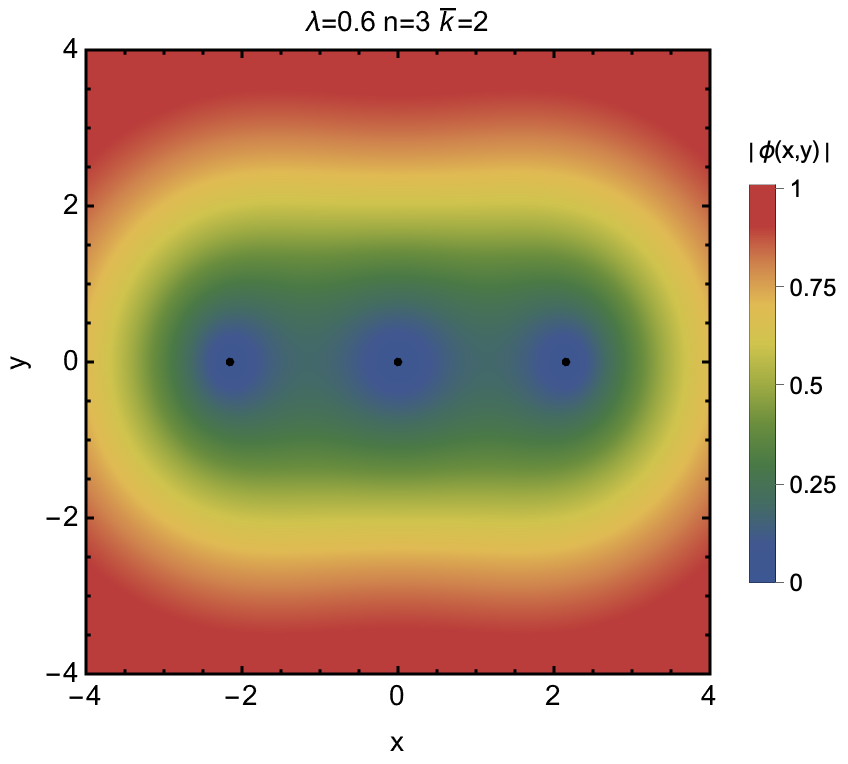}
        \includegraphics[width=0.32\textwidth]{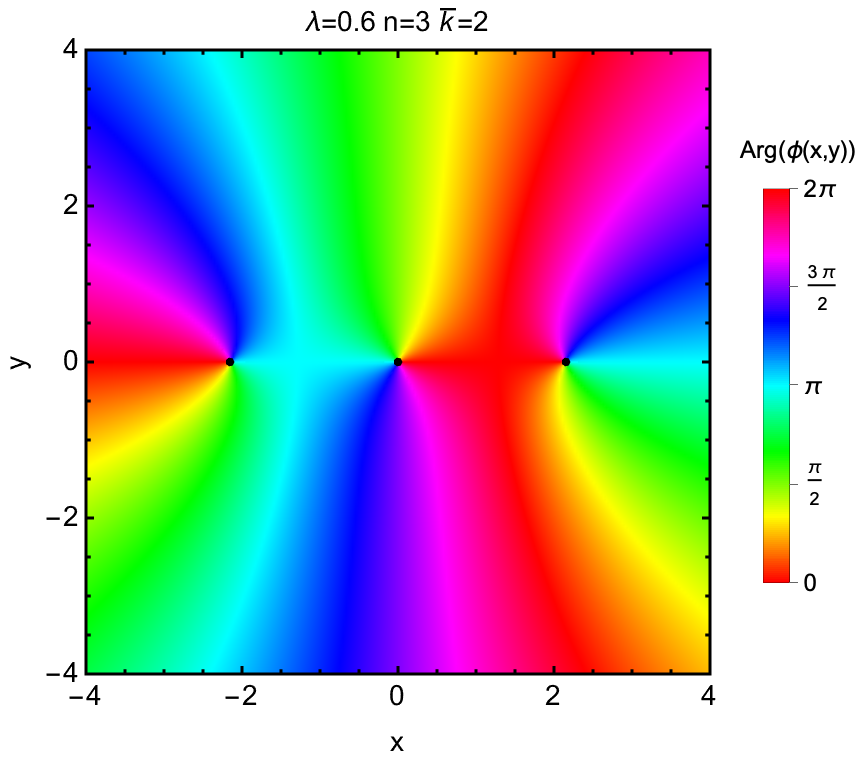}
    \raisebox{0.3\height}{\includegraphics[width=0.32\textwidth]{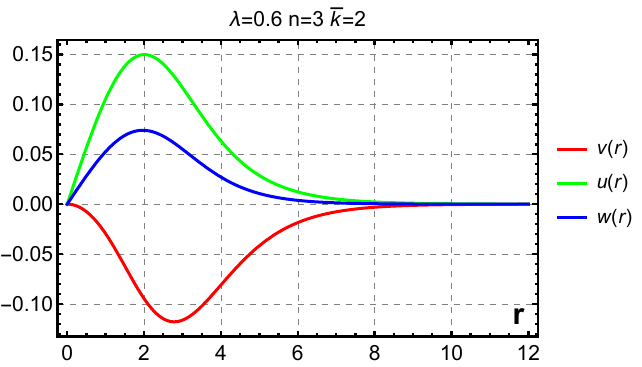}}
        \end{minipage}
\\
        &     

      \rotatebox{90}{\hspace{-0.3cm}$\lambda=1$} &

        \begin{minipage}{0.5\textwidth}
        \centering
 \includegraphics[width=0.31\textwidth]{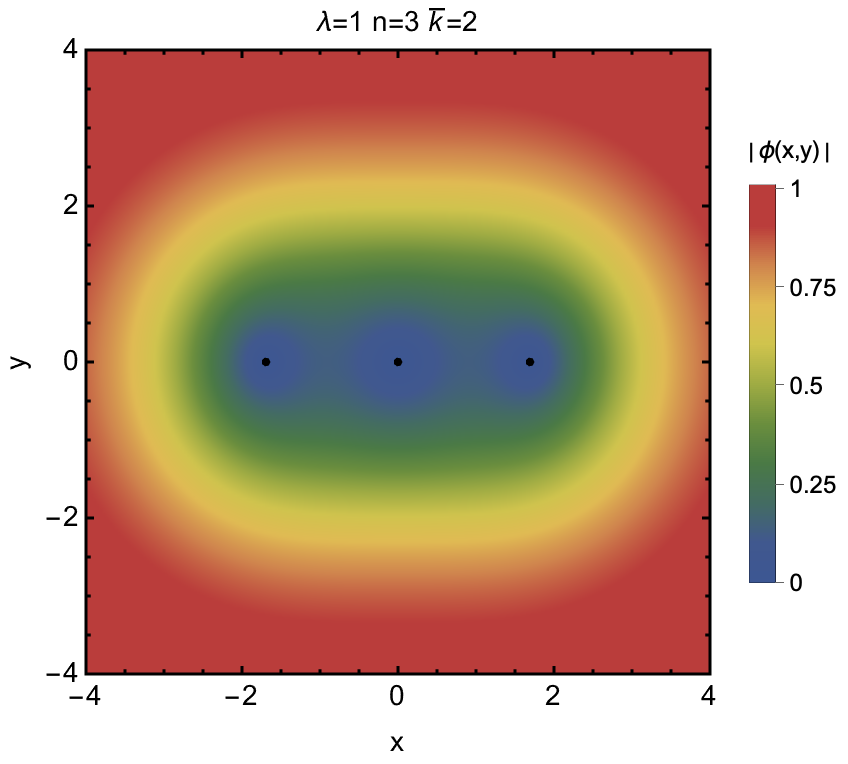}
        \includegraphics[width=0.32\textwidth]{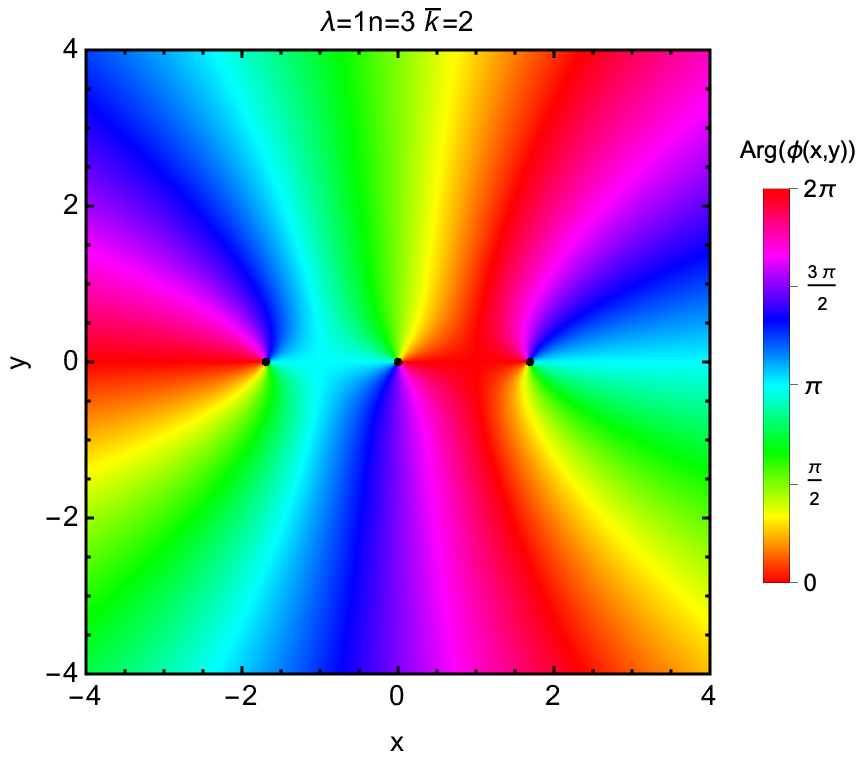}
    \raisebox{0.3\height}{\includegraphics[width=0.32\textwidth]{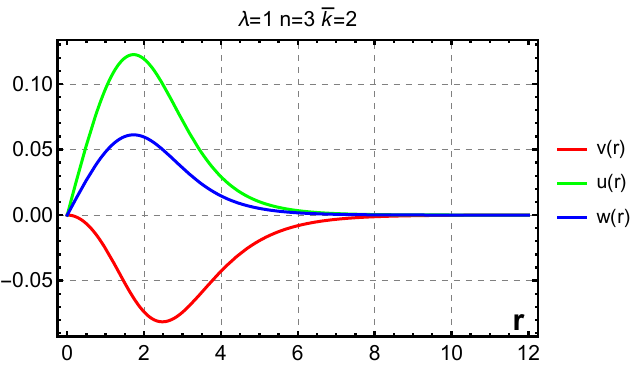}}
        \end{minipage}
\\
      & 

      \rotatebox{90}{\hspace{-0.5cm}$\lambda=1.4$} &

        \begin{minipage}{0.5\textwidth}
        \centering
 \includegraphics[width=0.31\textwidth]{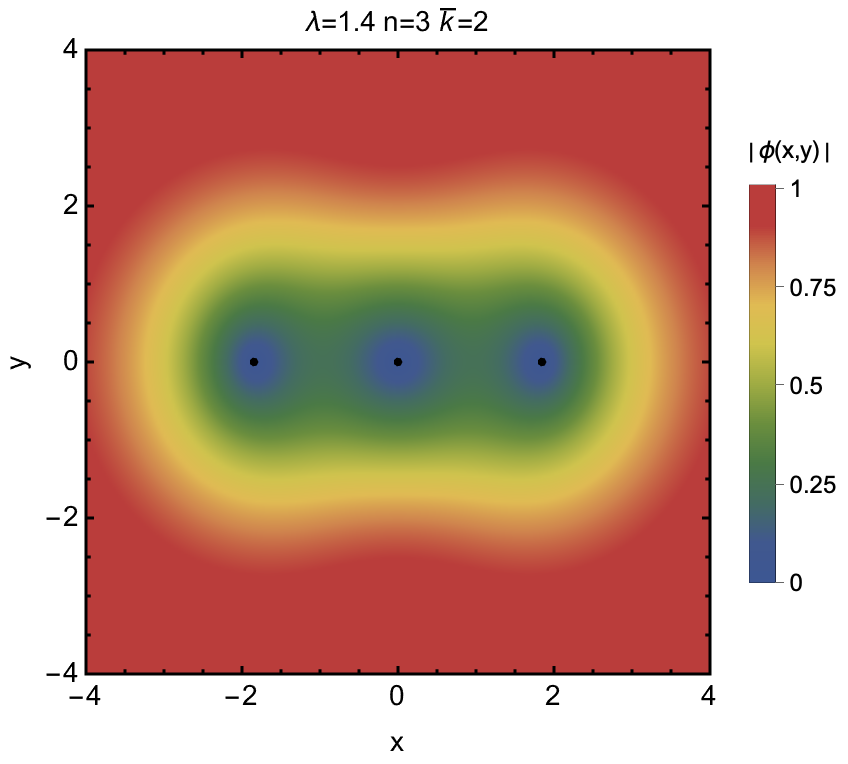}
        \includegraphics[width=0.32\textwidth]{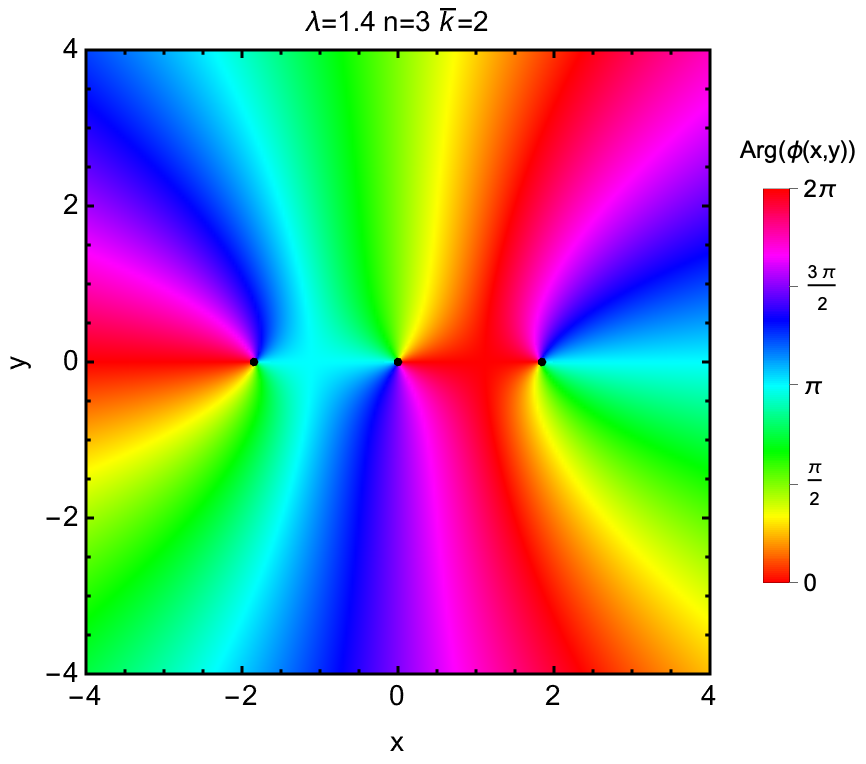}
    \raisebox{0.3\height}{\includegraphics[width=0.32\textwidth]{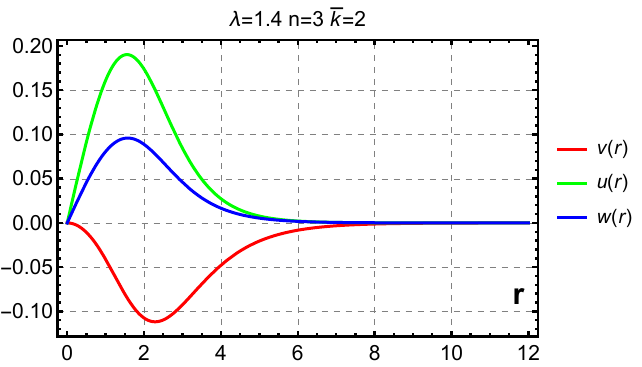}}
        \end{minipage}
\\
        \hline

 \multirow{3}{0.3cm}{ \rotatebox{90}{$n=3$ \hspace{2cm}$\overline{k}=3$ } } 
&
      \rotatebox{90}{\hspace{-0.5cm}$\lambda=0.6$} &

        \begin{minipage}{0.5\textwidth}
        \centering
 \includegraphics[width=0.31\textwidth]{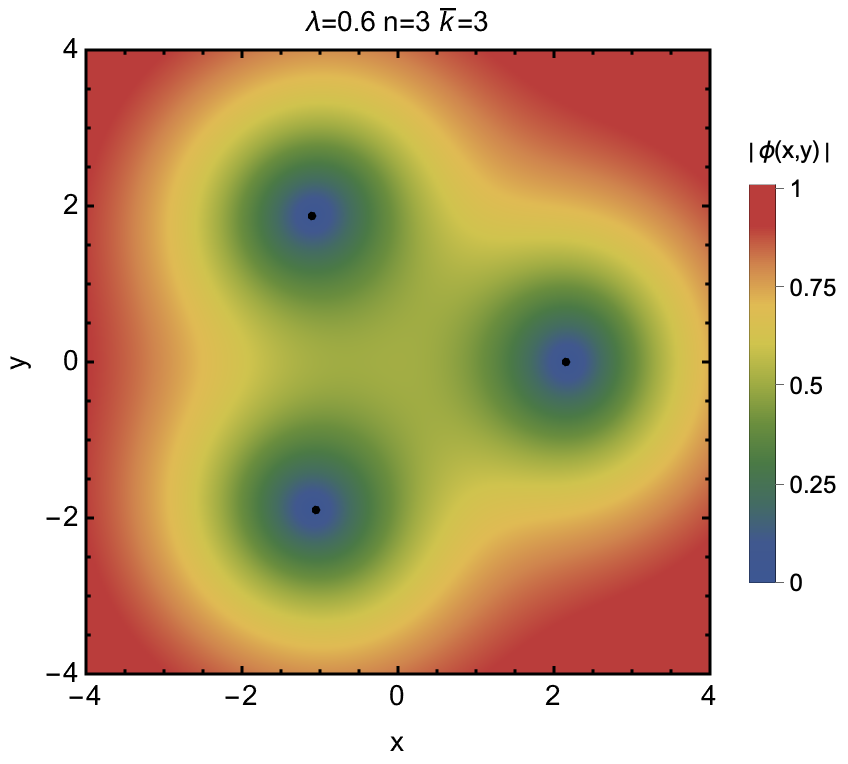}
        \includegraphics[width=0.32\textwidth]{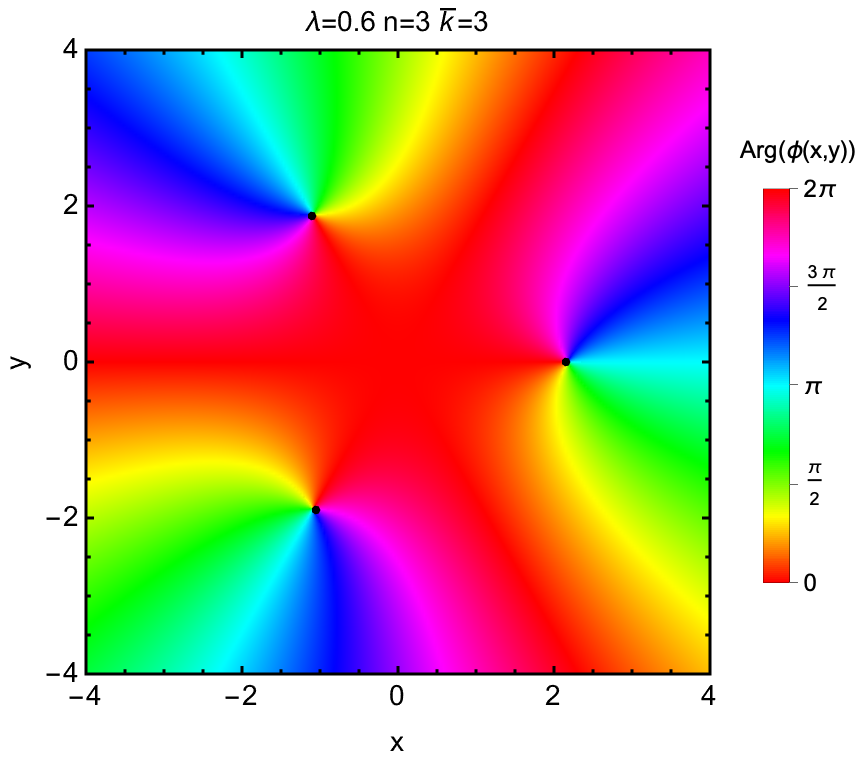}
    \raisebox{0.3\height}{\includegraphics[width=0.32\textwidth]{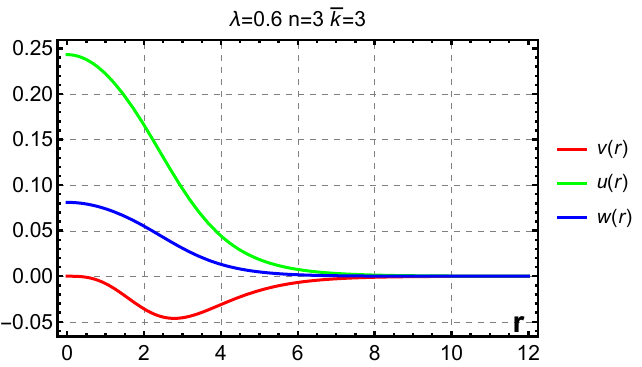}}
        \end{minipage}
\\
        &     

      \rotatebox{90}{\hspace{-0.3cm}$\lambda=1$} &

        \begin{minipage}{0.5\textwidth}
        \centering
 \includegraphics[width=0.31\textwidth]{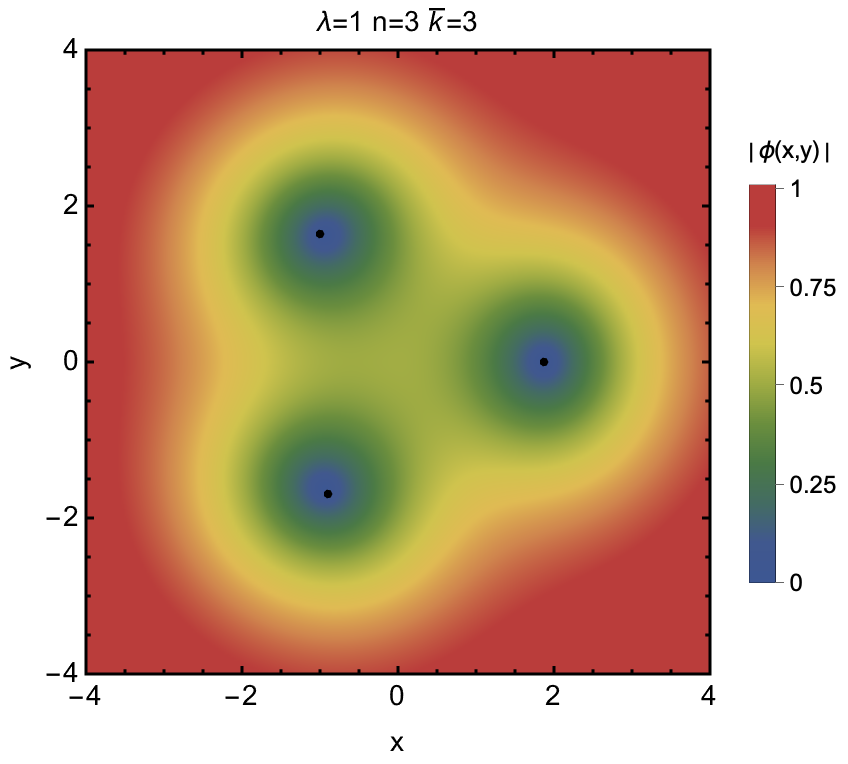}
        \includegraphics[width=0.32\textwidth]{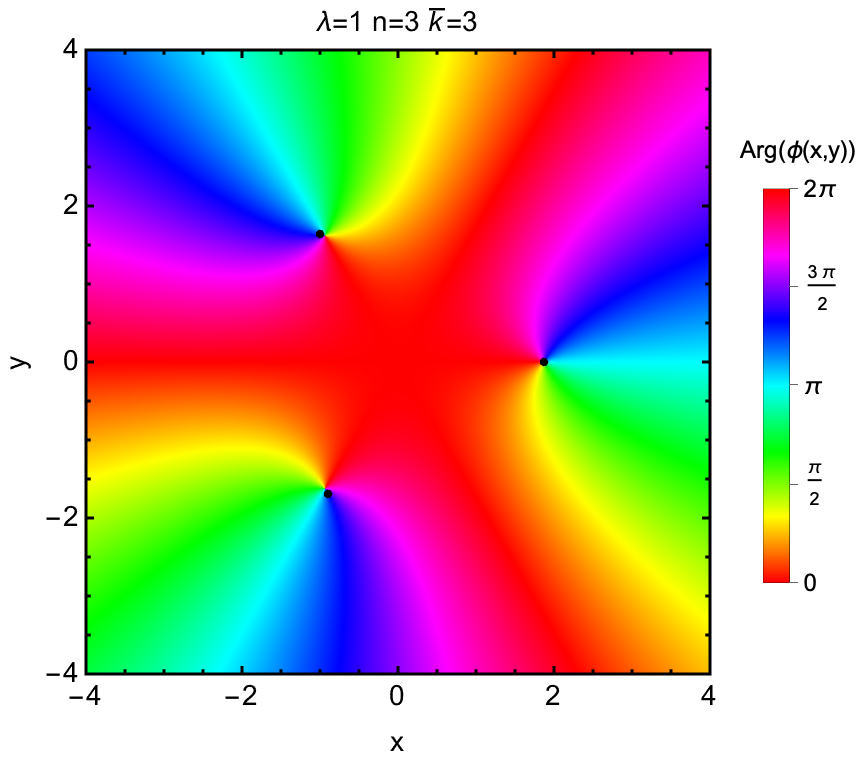}
    \raisebox{0.3\height}{\includegraphics[width=0.32\textwidth]{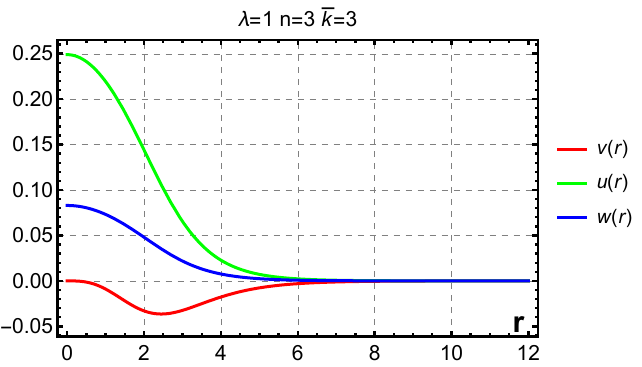}}
        \end{minipage}
\\
      & 

      \rotatebox{90}{\hspace{-0.5cm}$\lambda=1.4$} &

        \begin{minipage}{0.5\textwidth}
        \centering
 \includegraphics[width=0.31\textwidth]{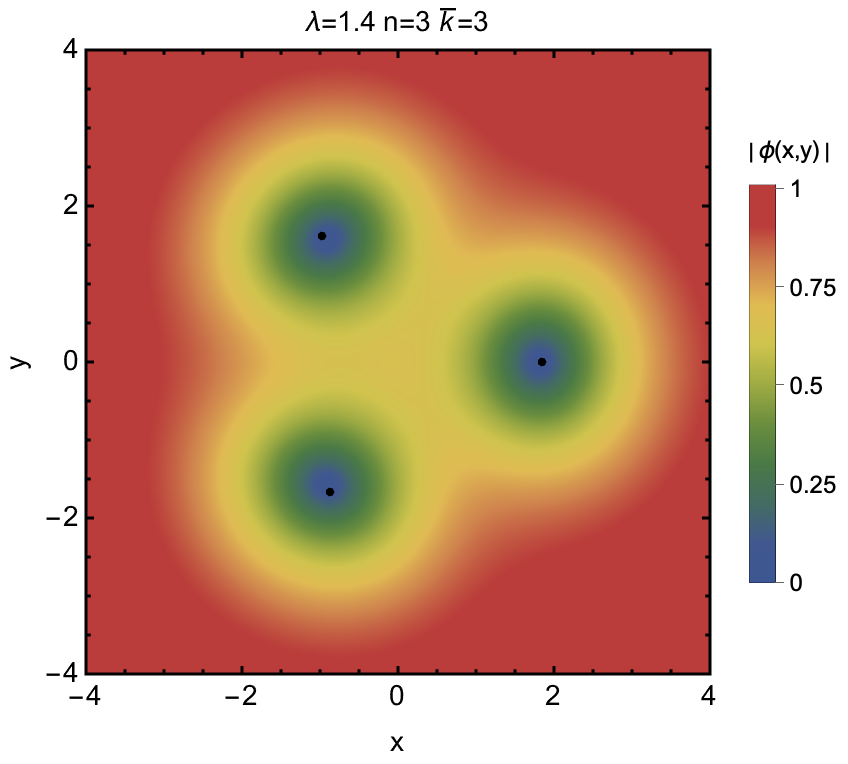}
        \includegraphics[width=0.32\textwidth]{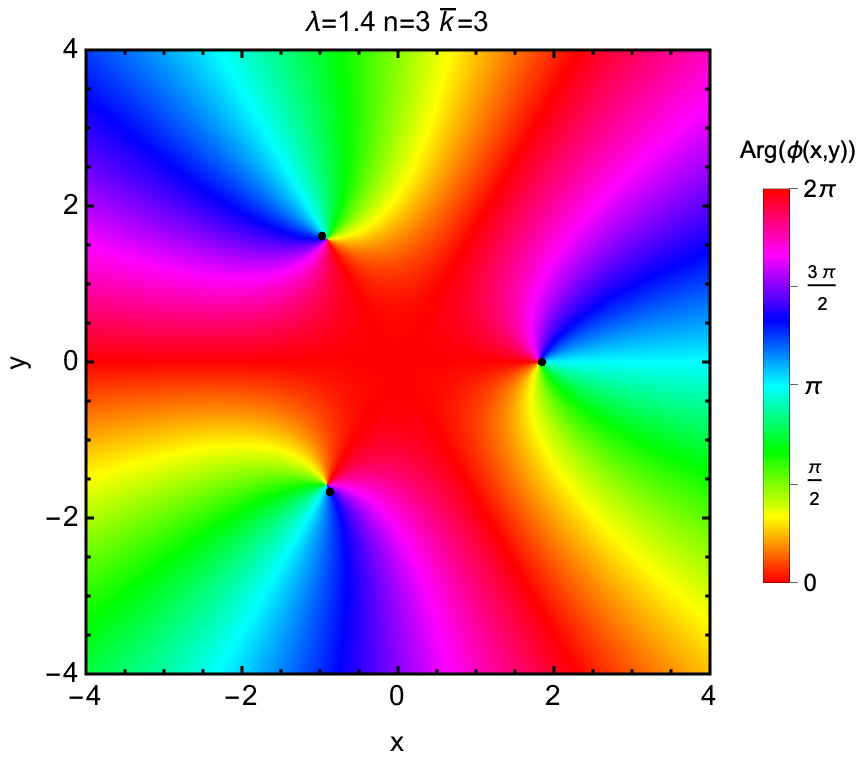}
    \raisebox{0.3\height}{\includegraphics[width=0.32\textwidth]{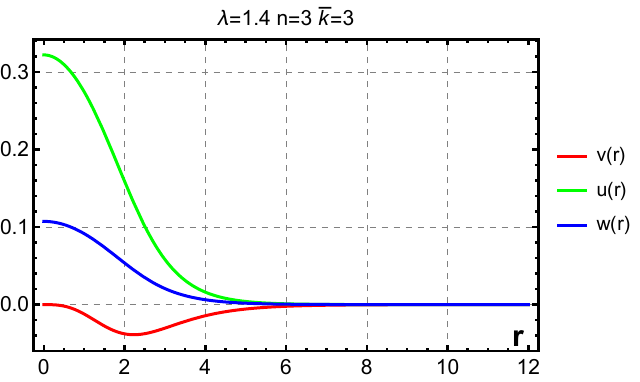}}
        \end{minipage}
\\
        \hline
    \end{tabular}
    \caption{Type A internal modes for different values of $\lambda$ and $n=3$. The plots correspond to $|\phi(x,y)|$, $\mathrm{Arg}(\phi(x,y))$ once the internal modes have been added to the static vortex configurations and the profile functions $v(r)$, $u(r)$ and $w(r)$.}
    \label{Tab2:AN3}
\end{table}

In Tables \ref{Tab1:AN2} and \ref{Tab2:AN3}, all Type A modes for $n=2,3$ and  several values of $\lambda$ are shown. In particular, we plot the functions $v(r)$, $u(r)$ and $w(r)$ as well as $|\phi(x,y)|$ and $\mathrm{Arg}(\phi(x,y))$ for the linearly excited vortex. 

As mentioned in Section \ref{non-bps}, Type A modes with $\overline{k}=1$ correspond to the translational modes of the vortex. This implies that these modes are zero modes for any value of $\lambda$. The effect of this mode on the vortex profile is just the splitting of one single vortex zero,  leaving a zero of multiplicity $n-1$ at the origin. It may seem that the translational mode does not move the entire  vortex as a whole, but, in reality, once these modes are triggered, the ``moving" center attracts the rest of them translating the entire vortex. 

Regarding the remaining Type A modes, these separate the vortex centers and are responsible for the instabilities of vortex configurations with $\lambda>1$. For example, modes with $\overline{k}=2$ separate two vortex centers leaving the rest of them at the origin, forming a line. In the  case of a 2-vortex, this mode splits the condiguration into two 1-vortices. For a 3-vortex, the effect of the mode is a symmetrical spliting in a line of three 1-vortices. 

The effect of modes with higher $\overline{k}$ is similar to the behavior previously described. For example, modes with $\overline{k}=4$ would generate a configuration in which four vortices with $n=1$ form a square leaving a vortex with vorticity $n-4$ at the center. This behaviour can be seen in Table \ref{Tab2:AN3}, where modes with $\overline{k}=3$ generate a configuration in which the vortex splits into three individual vortices forming a triangle.




\subsection{Derrick type modes and type B internal modes}

\begin{table}[h!]
    \centering
    \begin{tabular}{|c|c|c|}
 
      \hline
 \multirow{3}{0.3cm}{ \rotatebox{90}{$n=1$ \hspace{2cm}$\overline{k}=0$ } } 
              &
          \rotatebox{90}{\hspace{-0.5cm}$\lambda=0.6$} &
            \begin{minipage}{0.51\textwidth}
        \centering
                            
        \includegraphics[width=0.45\textwidth]{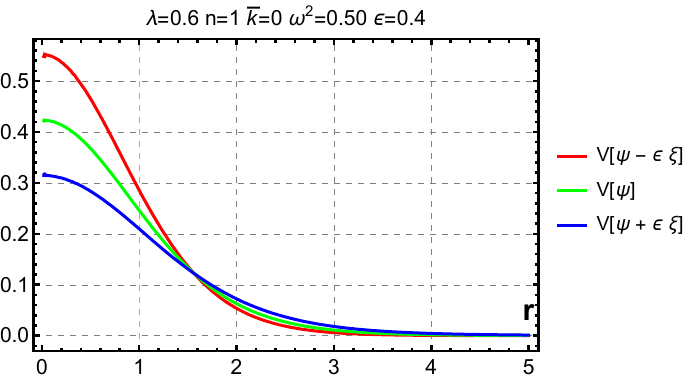}  
\includegraphics[width=0.45\textwidth]{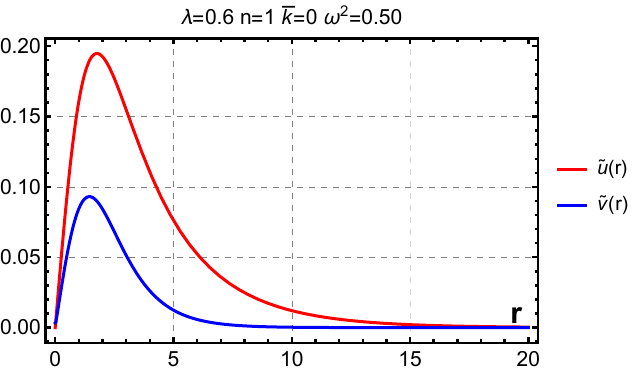}
        \end{minipage}
     
\\
   & \rotatebox{90}{\hspace{-0.3cm}$\lambda=1$}
    &

            \begin{minipage}{0.51\textwidth}
        \centering
        \includegraphics[width=0.45\textwidth]{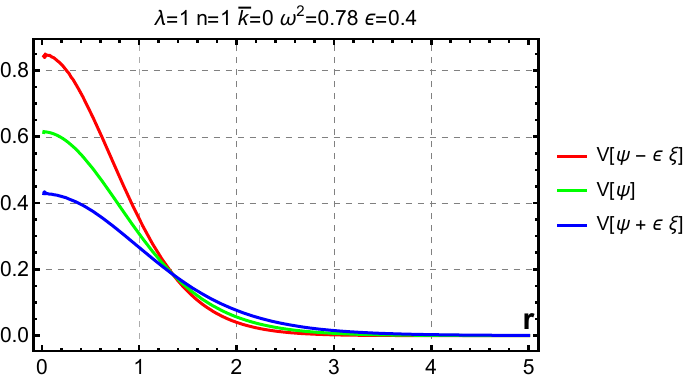}  
 \includegraphics[width=0.45\textwidth]{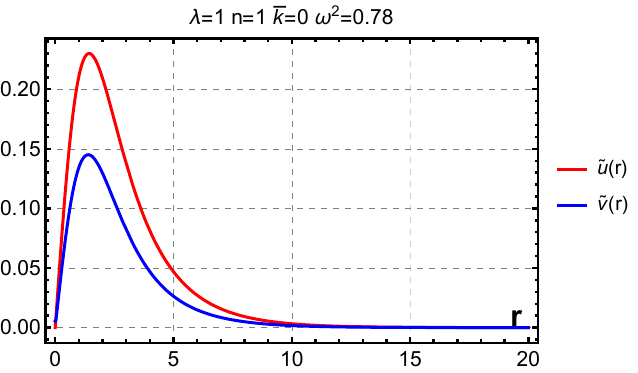}
        \end{minipage}
 
\\
   &   \rotatebox{90}{\hspace{-0.3cm}$\lambda=1.4$} &

            \begin{minipage}{0.51\textwidth}
        \centering
        \includegraphics[width=0.45\textwidth]{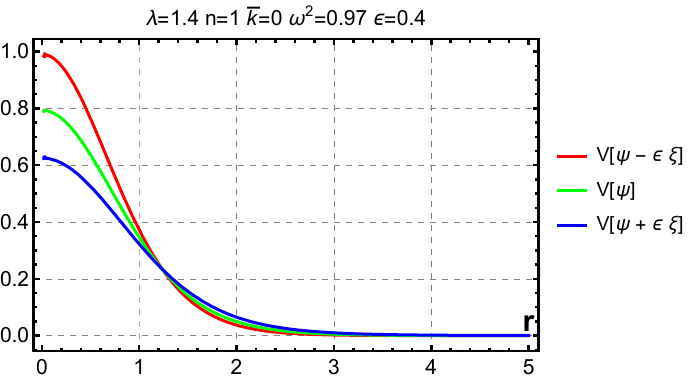}  
 \includegraphics[width=0.45\textwidth]{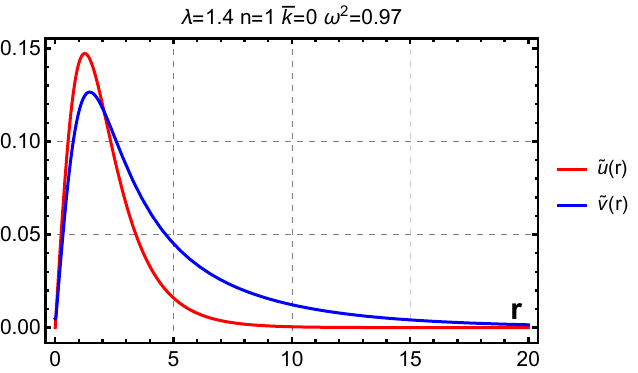}
        \end{minipage}

\\

      \hline
 \multirow{3}{0.3cm}{ \rotatebox{90}{$n=2$ \hspace{2cm}$\overline{k}=0$ }} &      \rotatebox{90}{\hspace{-0.3cm}$\lambda=0.6$} &
            \begin{minipage}{0.51\textwidth}
        \centering
        \includegraphics[width=0.45\textwidth]{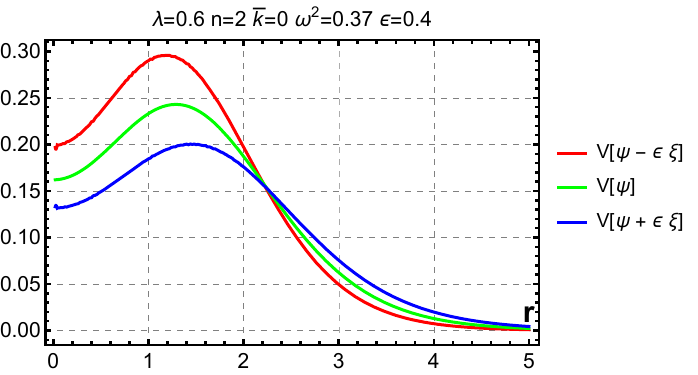}  
 \includegraphics[width=0.45\textwidth]{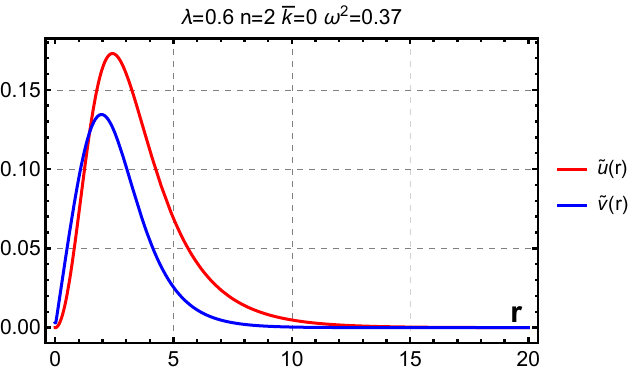}
        \end{minipage}
\\
             &  \rotatebox{90}{\hspace{-0.3cm}$\lambda=1$} &

          \begin{minipage}{0.51\textwidth}
        \centering
        \includegraphics[width=0.45\textwidth]{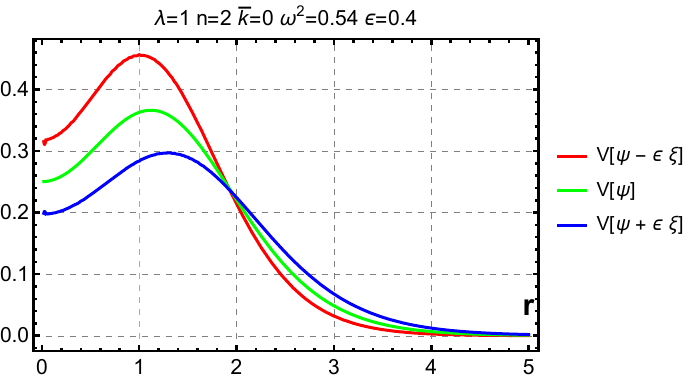}  
 \includegraphics[width=0.45\textwidth]{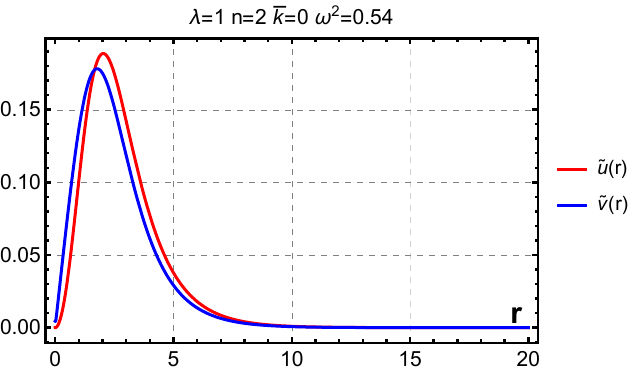}
        \end{minipage}
\\
          &
    \rotatebox{90}{\hspace{-0.3cm}$\lambda=1.4$} &

     \begin{minipage}{0.51\textwidth}
        \centering
        \includegraphics[width=0.45\textwidth]{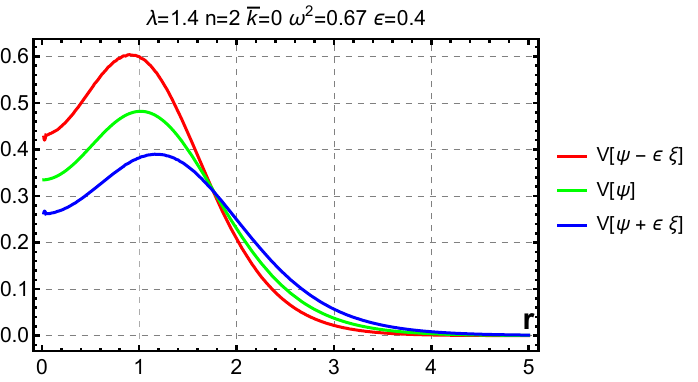}  
 \includegraphics[width=0.45\textwidth]{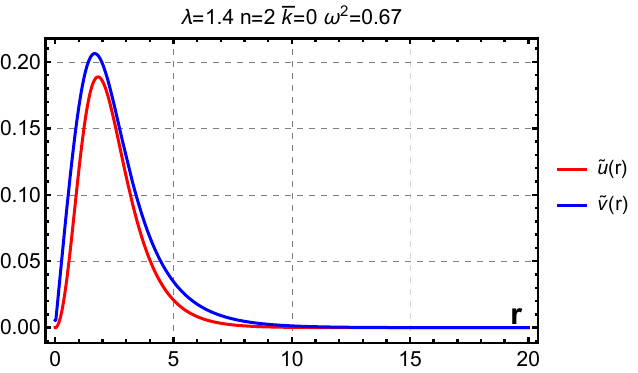}
        \end{minipage}
\\
        \hline
    \end{tabular}
\caption{Profile functions $\widetilde{v}(r)$ and $\widetilde{u }(r)$ and variation of the energy density for Derrick mode types for a vortex with $\lambda=0.6, 1, 1.4$ and vorticity $n=1,2$. }
    \label{Tab3}
\end{table}

In Table \ref{Tab3} the plots corresponding to the Derrick type modes for $n=1,2$ and several values of $\lambda$ are depicted. Additionally, we also show how the energy density changes once the internal modes have been triggered with a certain amplitude.  It can be seen that, when the corresponding eigenvalue approaches the mass threshold, profile functions exhibit a slower decay at large $r$. That is, they are less localized with respect to the vortex core. This behavior is clearly observed for $\lambda=1.4$ and $n=1$ in Table \ref{Tab4}. 

On the other hand, in Table \ref{Tab4}, the profile functions for Type B modes that arises for a vortex with $n=3$ are depicted along with the plots corresponding to the variations of the energy density once the internal mode has been triggered with a certain amplitude. It can be clearly seen that now, the variations of the energy density are not circularly symmetric; instead, the maximum deviation with respect to the unaltered vortex is located along the $x$ axis. Note that for the the mode corresponding to the form \eqref{genericform02b} these deviations would lie over the $y$ axis.

\begin{table}[h!]
        \hspace{2.9cm}
    \begin{tabular}{|c|c|}
      \hline   
   & $n=3$,  $\overline{k}=1$ \\
              \hline

      \rotatebox{90}{ \hspace{-0.3cm} $\lambda=0.6$} &
        \begin{minipage}{0.6\textwidth}
        \centering
                        
        \includegraphics[width=0.39\textwidth]{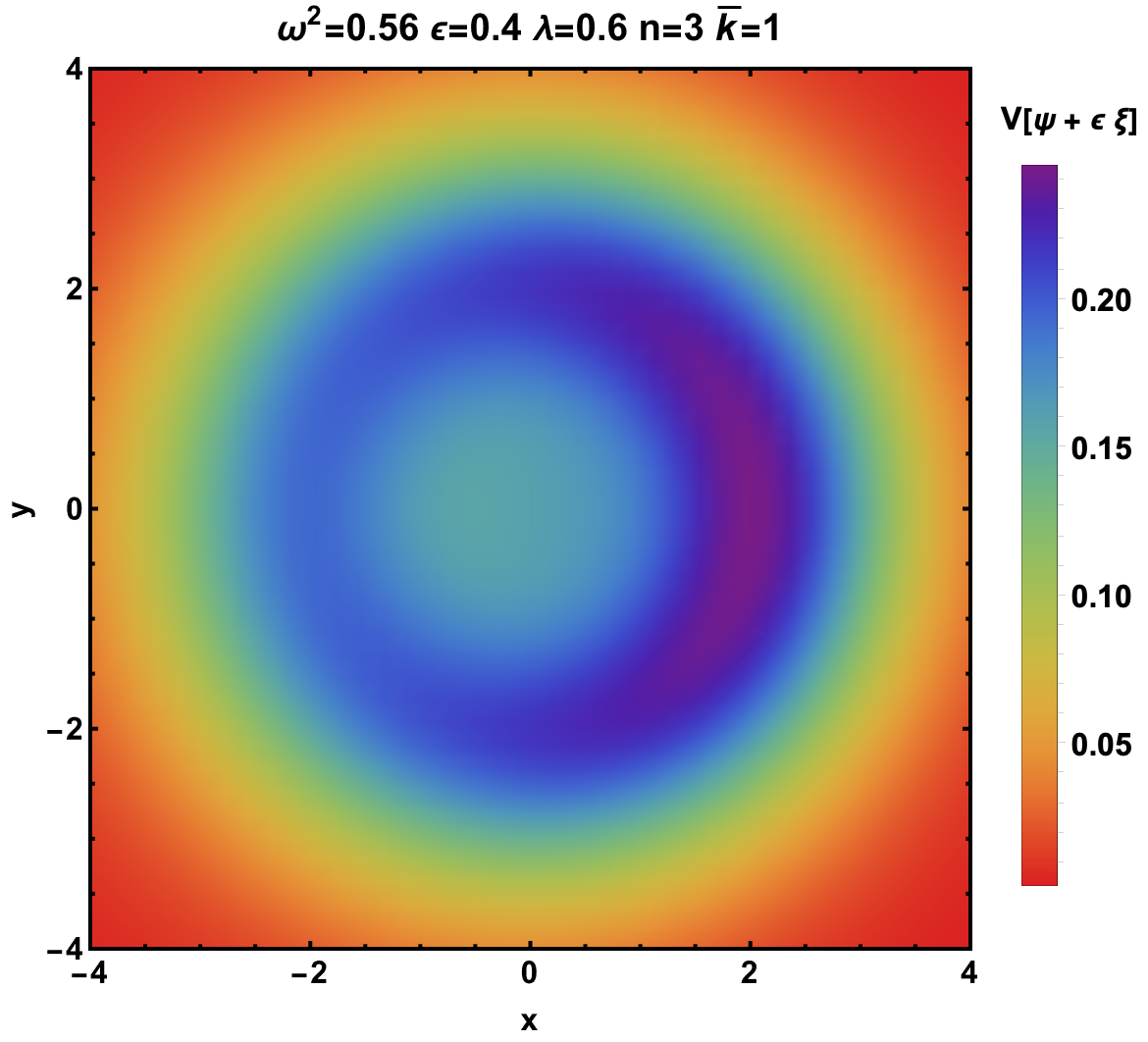}  
     \raisebox{0.25\height}{\includegraphics[width=0.45\textwidth]{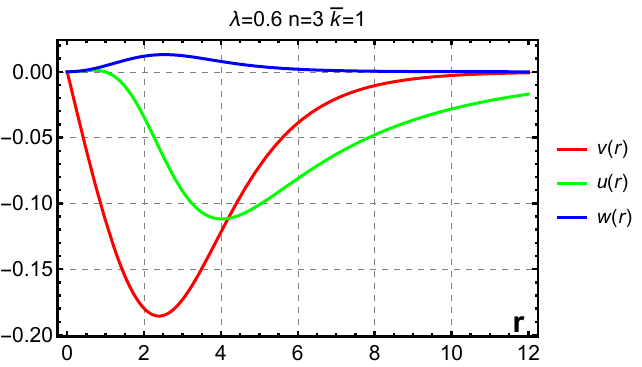}}
        \end{minipage}

\\
              \hline

      \rotatebox{90}{ \hspace{-0.3cm} $\lambda=1$} &
        \begin{minipage}{0.6\textwidth}
        \centering
                       
        \includegraphics[width=0.39\textwidth]{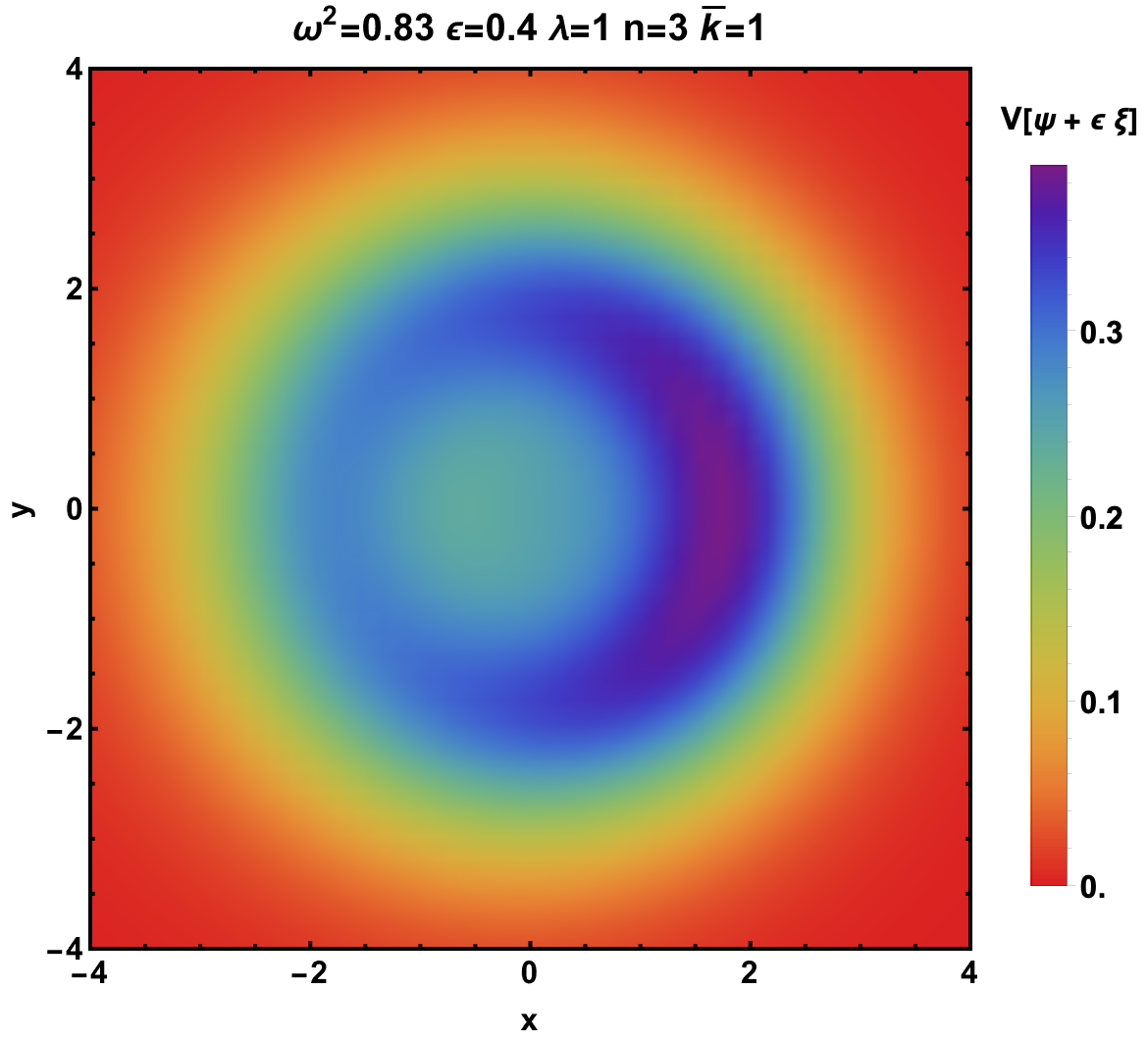}  
    \raisebox{0.25\height}{ \includegraphics[width=0.45\textwidth]{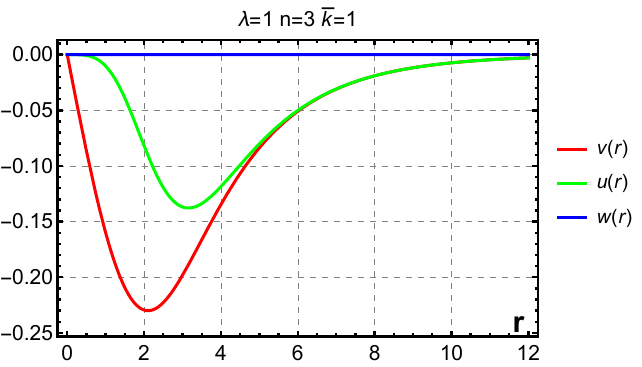}}
        \end{minipage}

\\
        \hline

       \rotatebox{90}{ \hspace{-0.5cm}$\lambda=1.2$ }&
        \begin{minipage}{0.6\textwidth}
        \centering

        \includegraphics[width=0.39\textwidth]{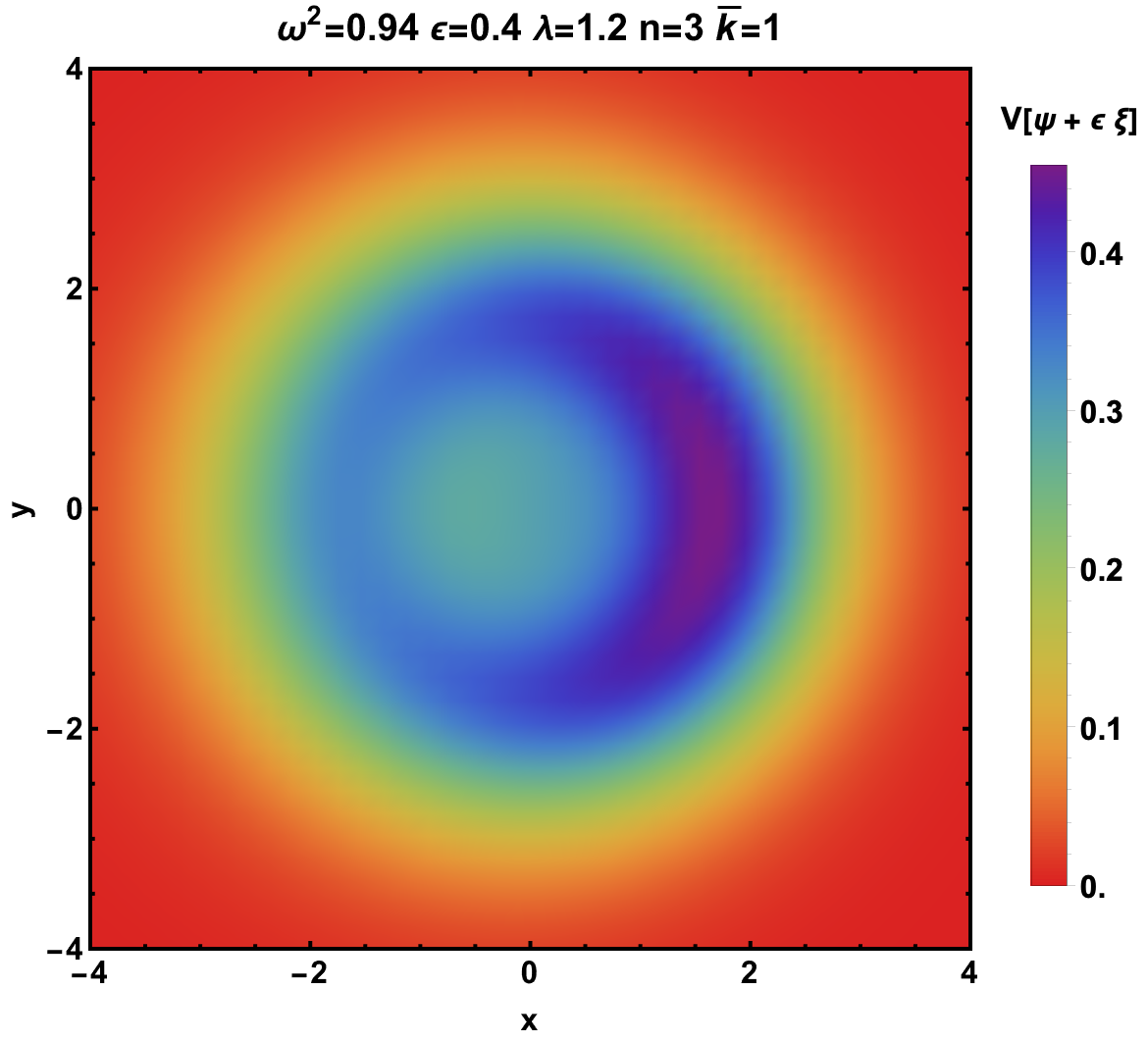}  
  \raisebox{0.25\height}{   \includegraphics[width=0.45\textwidth]{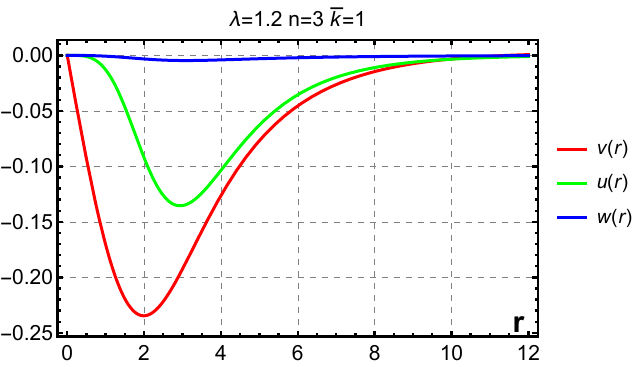}}
        \end{minipage}\\
        \hline
    \end{tabular}
     \caption{Profile functions $v(r)$, $u(r)$ and $w(r)$ for Type B multipolar modes for a vortex with $n=3$ and $\lambda=0.6,1,1.2$.}
     \label{Tab4}
\end{table}





\newpage

\section{Conclusions and Outlook}\label{conclusions}

We have studied the internal structure of higher-charge $U(1)$ symmetric non-self-dual abelian Higgs vortices in $2+1$ dimensions. We have started our discussion analyzing the second order fluctuation operator and we have reviewed the internal fluctuations of the self-dual vortices. These fluctuations are classified into two sets: the zero and shape modes. The zero modes are responsible for the translations of the individual vortices and for a vortex of charge $n$ there are $2n$ of them. The shape modes correspond to fluctuations around the vortex and their number depends nontrivially on the vortex charge.  

The spectral structure of the non-self-dual vortices is richer and depends on the vortex type. For type $I$ vortices we have found two zero modes. They correspond to the rigid translations of the vortex in two independent directions. The positive modes can be organized into two sets: the Derrick type modes and the multipolar modes. The number of Derrick type modes depends on both, the self-coupling $\lambda$ and the charge $n$. They correspond to axially symmetric fluctuations around the vortex center. 

On the other hand, the number of multipolar modes can be classified into two sets, type A and type B. The scalar and vector type A modes have opposite dependence of the parameter $\bar{k}$ near the vortex center. Their frequency decreases monotonically with the self-coupling $\lambda$. At critical coupling, type A modes become the zero modes of the BPS vortex. Therefore, there are in general $2n-2$ positive frequency type A modes and $2$ zero modes, which gives a total of $2n$ zero modes in the critical case. Type B modes exhibit the same dependence in $\bar{k}$ near the origin and their frequency increases monotonically with $\lambda$. The number of type B modes depends on both, the topological charge $n$ and the self-coupling constant $\lambda$.

 For type $II$ vortices the situation is similar. Again, the number Derrick modes depends on both, the topological charge and the self-coupling. Their frequency grows monotonically with $\lambda$, and eventually, they dissappear into the continuum. As for type $I$ vortices, the multipolar modes are continuously connected to the self-dual zero modes, but this time their frequency is negative. As a consequence, they become unstable modes that split the vortex into vortex configuration of smaller charge. 

 The zero modes (zero frequency multipolar modes) deserve a separate comment. For $n=1$ vortex, an infinitesimal excitation in the direction of the zero mode, corresponds, as expected, to a rigid translation of the vortex center. For $n>1$ vortices the situation is different. The position of the vortex is a zero of the scalar field with multiplicity $n$. An infinitesimal excitation in the direction of the zero mode only translates a zero multiplicity 1 from the origin, leaving a zero of multiplicity $n-1$ in the original position of the vortex. As a consequence, the linear excitation of the zero mode does not correspond to a rigid translation of the vortex. Instead, the translation of a zero of multiplicity $n$ requires a $n$-th order excitation in the direction of the zero mode.

Another interesting aspect, regarding the internal structure of the non-self-dual vortices in the existence of quasibound modes. As argued in \cite{AlonsoIzquierdo2024c} for large $\lambda$ the vortex core does not feel the magnetic field. This implies that the vortex behaves effectively as a global vortex. In this situation, the scalar modes of the Abelian-Higgs vortex remain bounded and approach the modes of the global vortex \cite{blanco2021, blanco2023}, while the vector modes become scattering states. As a consequence, the scalar-vector modes behave as quasibound modes. Therefore, for large $\lambda$ there is a correspondence between normal modes of the global vortex and quasibound modes of the Abelian-Higgs vortex. 

The consequences of bound and quasibound modes for non-self-dual vortex scattering,  as well as their decay, are left for a future investigation.

\section*{Acknowledgments}

The authors have been supported in part by Spanish Ministerio de Ciencia e Innovación (MCIN) with funding from the European Union NextGenerationEU (Grant No. PRTRC17.I1) and the Consejería de Educación, Junta de Castilla y León, through QCAYLE project, as well as the grant PID2023-148409NB-I00 MTM.  J.Q.  is also supported partially under the project Programa C2 from the University of Salamanca.  D.M.C. acknowledges financial support from the European Social Fund, the Operational Programme of Junta de Castilla y León and the regional Ministry of Education.

\section*{Conflict of Interest}
The authors declare that they have no conflict of interest.

\section*{Data availibility}
Data sharing is not applicable to this article as no datasets were generated or analysed during
the current study.


\appendix

\section{Numerical methods}\label{appen}

The numerical method employed in this study to solve the eigenvalue problem \eqref{radialedo03} is based on a second-order finite difference scheme, which discretizes the problem as follows:
{\small \begin{eqnarray}
	&&\hspace{-.5cm} - \frac{v_{i+1}-2 v_i+v_{i-1}}{(\Delta r)^2 } -  \frac{v_{i+1}-v_{i-1}}{2 i (\Delta r)^2}+ \Big[ \frac{\overline{k}^2}{ (i \Delta r)^2} + f_{n,i}^2 \Big] v_i+ 2\left[f_{n,i} + \frac{i}{2}(f_{n,i+1}-f_{n,i-1}) \right] w_i = \omega_n^2 v_i, \nonumber \\
	&& \hspace{-.5cm}- \frac{u_{i+1}-2 u_i+u_{i-1}}{(\Delta r)^2} -  \frac{u_{i+1}-u_{i-1}}{2 i (\Delta r)^2}+ \Big[ \frac{\overline{k}^2}{(i \Delta r)^2} + \frac{n^2(1-\beta_{n,i})^2}{(i \Delta r)^2} + \frac{3\lambda}{2} f_{n,i}^2 - \frac{\lambda}{2} \Big] u_{i}- \frac{2n(1-\beta_{n,i})(\overline{k}^2 +   (i \Delta r)^2 f_{n,i}^2)}{ (i \Delta r)^2} w_i \nonumber\\
	&& \hspace{-.5cm}\hspace{0.5cm}+ \frac{n\,f_{n,i}\,(1-\beta_{n,i})(v_{i+1}-v_{i-1})}{ i (\Delta r)^2} = \omega_n^2 u_i, \label{numeric} \\
	&& \hspace{-.5cm}- \frac{w_{i+1}-2 w_i+w_{i-1}}{(\Delta r)^2} -  \frac{w_{i+1}-w_{i-1}}{2 i (\Delta r)^2}+ \Big[ \frac{\overline{k}^2}{(i \Delta r)^2} + \frac{n^2(1-\beta_{n,i})^2}{(i \Delta r)^2} + f_{n,i}^2 + \frac{\lambda}{2} f_{n,i}^2 - \frac{\lambda}{2} \Big] w_i- \frac{2n(1-\beta_{n,i})}{(i \Delta r)^2} u_i + \nonumber\\
	&& \hspace{-.5cm}\hspace{0.5cm}+ \frac{f_{n,i+1}-f_{n,i-1} }{i (\Delta r)^2} v_i= \omega_n^2 w_i, \nonumber
\end{eqnarray}
}
where $f_{n,i}=f_n(i\Delta r)$ and $\beta_{n,i} = \beta_n(i\Delta r)$ are respectively the values of the radial profiles of the scalar and vector field at the mesh points and $v_{i}= v( i\Delta r )$,  $u_{i}=u( i\Delta r )$, $w_{i}=w(i\Delta r )$ are the eigenfunction components at these points.  $\Delta r = r_{max}/N$ denotes the spatial step. 

The eigenfunctions asymptotically vanish so that the boundary conditions $v_N=u_N=w_N=0$ will be assumed in our problem. On the other hand, we shall consider the regularity conditions $u'(0)=u'(0)=w'(0)=0$. By using the progressive numerical first derivative this leads to the relations
\begin{equation}
v_0 = \frac{4}{3} v_1 - \frac{1}{3} v_2 \hspace{0.5cm},\hspace{0.5cm}
u_0 = \frac{4}{3} u_1 - \frac{1}{3} u_2 \hspace{0.5cm},\hspace{0.5cm}
w_0 = \frac{4}{3} w_1 - \frac{1}{3} w_2 \label{regularity01}
\end{equation}
between the discretized values of the fields. The previous relations allow us to avoid the singularities at $r=0$ arising in the discretized equations (\ref{numeric}) with $i=0$ (notice that the spectral problem is well-defined at $r=0$, as analytically demonstrated in Section 4). Using (\ref{regularity01}), we can construct the discretization of the Hessian operator without explicitly considering the point $r=0$; instead, we simply take the discretization with values $v_i,u_i,w_i$, with $i=1,2,\dots,N$, replacing $v_0, u_0$ and $w_0$ with the value given above whenever necessary. For example, the first equation in this scheme reduces to 
\begin{eqnarray}
	&& - \frac{4\,v_2 }{3(\Delta r)^2} +\Big[ \frac{1}{(\Delta r)^2} \Big(\frac{4}{3}+ \overline{k}^2\Big)  +f_{n,1}^2 \Big] v_1  + 2[f_{n,1}+\frac{1}{2} (f_{n,2}-f_{n,0})]w_1 = \omega_n^2 v_1 \nonumber \\
    && - \frac{4\,u_2 }{3(\Delta r)^2} + \Big[ \frac{1}{(\Delta r)^2} \Big(\frac{4}{3}+\overline{k}^2 + n^2 (1-\beta_{n,1})^2 \Big)  + \frac{3\lambda}{2} f_{n,1}^2  -\frac{\lambda}{2} \Big] u_1  + \frac{4 n (1-\beta_{n,1})f_{n,1}}{3(\Delta r)^2} v_2  - \nonumber\\
    && \hspace{0.8cm}- \frac{4 n (1-\beta_{n,1})f_{n,1}}{3(\Delta r)^2} v_1 - 2n(1-\beta_{n,1}) \Big[ \frac{\overline{k}^2}{(\Delta r)^2} + f_{n,1}^2 \Big]w_1 = \omega_n^2 u_1 \label{numericbc} \\
    && - \frac{4\,w_2 }{3(\Delta r)^2} +\Big[ \frac{1}{(\Delta r)^2} \Big(\frac{4}{3}+ \overline{k}^2 + n^2 (1-\beta_{n,1})^2\Big) +f_{n,1}^2 + \frac{\lambda}{2} f_{n,1}^2 - \frac{\lambda}{2} \Big] w_1  + \frac{f_{n,2}-f_{n,0}}{(\Delta r)^2} v_1 - \nonumber \\
    && \hspace{0.8cm}- \frac{2n(1-\beta_{n,1})}{(\Delta r)^2} u_1 = \omega_n^2 w_1 \nonumber
\end{eqnarray}
and the rest of entries are given by (\ref{numeric}) with $i=2,3,\dots,N$ as usual. There is no dependence on $g_0$ in our discretized spectral problem. 

The choice of the regularity boundary conditions are analytically justified. It has been shown that the eigenfunction profiles $v(r)$, $u(r)$ and $w(r)$ behave as $C r^j$, with $j\in \mathbb{N}$. Then, if $j > 1$, the functions clearly satisfies the imposed condition. When some $j = 0$, the functions takes the form $C_1 + C_2 r^2$, which also satisfies the condition. When some of the functions follow the previous behavior with $j = 1$ the regularity condition is not verified, however the imposed condition is very flexible in the sense that, although the exact eigenfunction does not analytically satisfy this condition, when computed numerically, it quickly adapts to the real eigenfunction, provided that a sufficiently high resolution is used. That is, the difference between the numerical and the real eigenfunction is only noticeable in a very small range.

We have tested other boundary conditions at the origin, leveraging the analytical insights provided in the article, but the obtained eigenvalues are the same.

\end{document}